\documentclass[preprint,11pt,authoryear]{elsarticle}

\usepackage{lineno,hyperref}
\modulolinenumbers[5]

\journal{Expert Systems with Applications}

\usepackage{graphicx}
\usepackage[english]{babel,varioref}
\usepackage[utf8]{inputenc}
\usepackage[T1]{fontenc}
\usepackage{natbib}
\usepackage[a4paper, total={6in, 8.5in}]{geometry}
\usepackage{amssymb,amsthm,amsmath,dsfont}
\usepackage{setspace}
\usepackage{verbatim} 
\usepackage{fancyhdr}
\usepackage{multirow}
\usepackage{eurosym}
\usepackage{epstopdf}

\usepackage{subfigure}
\usepackage[section]{placeins}
\usepackage{lineno}
\usepackage {tikz}
\usetikzlibrary {positioning}
\definecolor {processblack}{cmyk}{0.96,0,0,0}
\usepackage[lined,algoruled,commentsnumbered, linesnumbered]{algorithm2e}

\usepackage{bibentry} 
\usepackage{soul}
\usepackage{hyperref}
\usepackage {tikz}
\usetikzlibrary{shapes,arrows}
\usetikzlibrary {positioning}
\tikzstyle{decision} = [diamond, draw, fill=green!40, 
    text width=8.5em, text badly centered, node distance=3cm, inner sep=0pt]
\tikzstyle{block} = [rectangle, draw, fill=red!20, text width=0.128cm,text centered]
    \tikzstyle{block_highlight} = [rectangle, draw, fill=red!50, 
    text centered, text width=0.15cm, minimum height=0.3cm]
\tikzstyle{block_gray} = [rectangle, draw, fill=gray!50, 
    text centered,  text width=0.15cm, minimum height=0.3cm]
 \tikzstyle{node} = [circle, draw, fill=red!20, text width=0.1cm,
    text centered, rounded corners, minimum height=0.1cm]
\tikzstyle{label} = [rectangle, rounded corners, minimum height=0.1cm]
\tikzstyle{line} = [draw, -latex']
\tikzstyle{cloud} = [draw, ellipse,fill=red!20, node distance=3cm,    minimum height=1em]
\usepackage{collcell}

\newcommand*{\MinNumber}{0.0}%
\newcommand*{\MidNumber}{0.5} %
\newcommand*{\MaxNumber}{1.0}%

\newcommand{\ApplyGradient}[1]{%
        \ifdim #1 pt > \MidNumber pt
            \pgfmathsetmacro{\PercentColor}{max(min(100.0*(#1 - \MidNumber)/(\MaxNumber-\MidNumber),100.0),0.00)} %
            \hspace{-0.4em}\colorbox{black!\PercentColor!gray!50}{#1}
        \else
            \pgfmathsetmacro{\PercentColor}{max(min(100.0*(\MidNumber - #1)/(\MidNumber-\MinNumber),100.0),0.00)} %
            \hspace{-0.4em}\colorbox{white!\PercentColor!gray!50}{#1}
        \fi
}
\newcolumntype{R}{>{\collectcell\ApplyGradient}c<{\endcollectcell}}

\newcommand*{\MinNumberP}{-1.0}%
\newcommand*{\MidNumberP}{0.0} %
\newcommand*{\MaxNumberP}{1.0}%

\newcommand{\ApplyGradientP}[1]{%
        \ifdim #1 pt > \MidNumberP pt
            \pgfmathsetmacro{\PercentColor}{max(min(100.0*(#1 - \MidNumberP)/(\MaxNumberP-\MidNumberP),100.0),0.00)} %
            \hspace{-0.33em}\colorbox{black!\PercentColor!gray!50}{#1}
        \else
            \pgfmathsetmacro{\PercentColor}{max(min(100.0*(\MidNumberP - #1)/(\MidNumberP-\MinNumberP),100.0),0.00)} %
            \hspace{-0.33em}\colorbox{white!\PercentColor!gray!50}{#1}
        \fi
}
\newcolumntype{P}{>{\collectcell\ApplyGradientP}c<{\endcollectcell}}

\begin{document}

\begin{frontmatter}

\title{An ensemble based on a bi-objective evolutionary spectral algorithm for graph clustering}
\author[l1]{Camila P.S. Tautenhain}
\ead{santos.camila@unifesp.br}
%novel

\author[l1]{Mariá C.V. Nascimento\corref{cor1}}
\ead{mcv.nascimento@unifesp.br}

\cortext[cor1]{Corresponding author.}
\address[l1]{Instituto de Ciência e Tecnologia, Universidade Federal de São Paulo\\
São José dos Campos, Brazil}

\begin{abstract}
Graph clustering is a challenging pattern recognition problem whose goal is to identify vertex partitions with high intra-group connectivity. This paper investigates a bi-objective problem that maximizes the number of intra-cluster edges of a graph and minimizes the expected number of inter-cluster edges in a random graph with the same degree sequence as the original one. The difference between the two investigated objectives is the definition of the well-known measure of graph clustering quality: the modularity.  We  introduce a spectral decomposition hybridized with an evolutionary heuristic, called \emph{MOSpecG}, to approach this bi-objective problem and an ensemble strategy to consolidate the solutions found by \emph{MOSpecG} into a final robust partition. The results of computational experiments with real and artificial LFR networks demonstrated a significant improvement in the results and performance of the introduced method in regard to another bi-objective algorithm found in the literature.  \textcolor{black}{The crossover operator based on the geometric interpretation of the modularity maximization problem   to match the communities of a pair of individuals was of utmost importance for the good performance of \emph{MOSpecG}. Hybridizing spectral graph theory and intelligent systems allowed us to define significantly high-quality community structures.}  
\end{abstract}
\end{frontmatter}

%\enlargethispage{-2.6\baselineskip}
\enlargethispage{-1.0\baselineskip}

%%%%%%%%%%%%%%%%%%%%%%%%%%%%%%%%%%%%%%%%%%%%%%%%%%%%%%
\section{Introduction}
%%%%%%%%%%%%%%%%%%%%%%%%%%%%%%%%%%%%%%%%%%%%%%%%%%%%%%

The majority of graphs that describe real networks, such as social and metabolic networks \citep{Zachary1977,Lancichinetti2011}, are characterized by vertex partitions with high intra-cluster connectivity \citep{Girvan2002}. The graph clustering problem, also known as community detection problem, aims at finding such partitions. \textcolor{black}{ \citet{Ferrara2014}, for example, developed an expert system to detect communities in mobile phone networks formed by interactions  of criminals to possibly identify criminal organizations.  \cite{Larsson2012} and \cite{Golbeck2010} applied community detection algorithms to Twitter data to classify the users' political leaning. In practice, this type of information usually benefits political campaigners.}

 \textcolor{black}{The formal definition of a graph clustering problem leans    towards the  criterion to assess the partitioning quality. Examples of optimization criteria to finding graph clusterings are the maximization of modularity \citep{Newman2004c}, the map equation minimization  \citep{Rosvall2008} and the maximization of the statistical significance of communities according to the measure introduced by \citet{Lancichinetti2011}.
In particular, the map equation measure is based on the observation on the duality between  graph clustering problems and the data compression problem described by the minimization of the path length of a random walker. The Infomap algorithm was then proposed to detect communities that minimize the map equation.  
\citet{Lancichinetti2011} studied a measure that evaluates the statistical significance of the communities in a network by calculating their probability of existing  in a random graph with the same degree sequence as the original one. The authors introduced a solution method to find a partitioning of the vertices that maximizes these probabilities named  Order Statistics Local Optimization Method (OSLOM).}

\textcolor{black}{Despite Infomap and OSLOM being considered state-of-the-art methods, the optimization criteria they employ have not been  properly explored by other algorithms yet. Modularity maximization, on the other hand,} is one of the most popular optimization criteria to define graph clusterings. The modularity of a partition is the difference between the number of edges in the same groups (first term) and the expected number of edges within the groups in a random graph with the same vertex degree sequence as the original graph (second term) \citep{Newman2004c}. However, many studies in the literature point out that by simply defining the measure as the difference between these two terms, without scaling them, may be a poor way to evaluate graph clusterings \citep{Fortunato2007,Reichardt2006}.

As an attempt to mitigate the scaling problem of modularity, \citet{Reichardt2006} suggested multiplying the second term of the modularity measure by a parameter called resolution parameter. A few studies approaching this modified modularity have shown interesting results \citep{Santos2016a,Carvalho2014}. \citet{Carvalho2014}, for example, introduced a supervised method that automatically adjusts the resolution parameter based on the graph topology. The method was later employed in the consensus algorithm proposed by \citet{Santos2016a}. In spite of the potential of the strategies, they require labeled data for defining a training set of the supervised algorithm.
\citet{Berry2011, DeMeo2013} and \citet{Khadivi2011} introduced pre-processing strategies to change the edge weights of a graph in order to diminish the negative effects of the resolution limit without the prior knowledge of the resolution parameter.

Another approach that explores the duality between the first and second terms of the modularity measure was introduced by \citet{Shi2012}. The authors introduced an evolutionary algorithm called MOCD for solving the bi-objective problem that maximizes the first term of modularity and minimizes the second term of modularity. \textcolor{black}{The studies in \citep{Pizzuti2012,Gong2012,Gong2014} also investigate bi-objective problems by optimizing different criteria. In particular,} MOCD achieved good quality partitions when compared to the other  evolutionary bi-objective clustering algorithm, known as Moga-Net \citep{Pizzuti2012}. 

This paper investigates a weighted aggregation method for solving the bi-objective problem that optimizes the first and second terms of modularity. The resulting problem is here called weighted aggregate modularity and is equivalent to solving the problems that maximize the modularity with different resolution parameter values, as demonstrated in this paper. To solve the weighted aggregate modularity, we propose a multi-objective evolutionary algorithm whose fitness function is the spectral relaxation of the weighted aggregate modularity matrix. 
In addition,  we explore the close relationship between multi-objective clusterings and ensemble clusterings by introducing an ensemble of the approximation of the Pareto solutions that adjusts the edge weights of the graph. \textcolor{black}{To the best of our knowledge, ensemble or consensus clustering strategies have not been applied to  solutions of the studied bi-objective graph clustering problem.} The proposed algorithm deals with the resolution limit by combining both the edge weighting and resolution parameter strategies, without the need of pre-defining the resolution parameter. Furthermore, we estimate an upper bound to the number of clusters in advance, which might contribute to further reductions of the negative effects of the resolution limit according to the computational experiments performed by \citet{Darst2014}.

Computational experiments were carried out using real and LFR networks \citep{Lancichinetti2008}. \textcolor{black}{We contrasted the results achieved by the proposed algorithm with those found by Moga-Net, a reference multi-objective method.
Moreover, we  compared the results with  OSLOM and Infomap.}
The proposed algorithm outperformed the multi-objective algorithm Moga-Net in all the networks and was from 6 to 64 times faster in the LFR networks. Despite the slightly better results achieved by the reference mono-objective algorithms   OSLOM  \citep{Lancichinetti2011} and Infomap \citep{Rosvall2008} in most of the LFR networks, the proposed algorithm outperformed them in the LFR networks with large mixture coefficients.

The rest of this paper is organized as follows: 
Section \ref{sec:related} presents a brief literature review of multi-objective and ensemble graph clustering algorithms; Section \ref{sec:spectral} thoroughly describes the studied spectral decomposition of the weighted aggregate modularity; Section \ref{sec:method} introduces the multi-objective evolutionary algorithm proposed in this paper; Section \ref{sec:experiments} discusses the computational experiments carried out with the  algorithm in question along with the analysis of the results; and, to sum up, Section \ref{sec:conclusion} brief summarizes the contributions of the paper and outlines further works.

 

%%%%%%%%%%%%%%%%%%%%%%%%%%%%%%%%%%%%%%%%%%%%%%%%%%%%%%
\section{Related Works}\label{sec:related}
%%%%%%%%%%%%%%%%%%%%%%%%%%%%%%%%%%%%%%%%%%%%%%%%%%%%%%

\textcolor{black}{This section presents a concise literature review focusing on multi-objective optimization and consensus clustering. As earlier mentioned, both types of strategies are approached in this paper to mitigate the bias of algorithms that optimize a single quality measure.}

\subsection{Multi-objective graph clustering methods}

Multi-objective optimization involves solving problems with two or more conflicting objective functions. The existence of trade-offs amongst objective functions is the reason why a single solution cannot optimize all the functions simultaneously;
instead, a number of  efficient solutions, known as Pareto solutions, describes the best solutions for adequate decision-making. 
In a multi-objective problem, a solution is called efficient when it is not possible to improve the value of any objective function without worsening the value of another function.

Because of the computational challenges involved in graph partitioning problems, especially in large-scale networks, \textcolor{black}{the existing multi-objective solution methods  are heuristics.} In particular,  \textcolor{black}{the overwhelming majority of multi-objective graph clustering solution} methods are evolutionary algorithms \citep{Pizzuti2012,Gong2012,Shi2012,Amiri2013,Shang2016,Zalik2018,Cheng2018,Zou2019}, due to the set of evolved solutions provided by their population-based structure.  \textcolor{black}{Methods based on particle swarm optimization \citep{Gong2014,Chen2016,Pourkazemi2017,Rahimi2018} and other nature- or human-inspired algorithms \citep{Gong2011,Li2012,Gong2013,Xu2015,Zhou2016_cuckoo, Amiri2011,Amiri2013} were also proposed to solve multi-objective graph clustering problems.}

\subsubsection{Optimization of the modularity terms}

\textcolor{black}{As mentioned in the earlier section of this paper, \citet{Shi2012} introduced MOCD to  optimize the two terms of the modularity measure.
 \citet{Li2012} and \citet{Zalik2018} also optimized the two terms of the modularity measure using multi-objective evolutionary algorithms.}

 \textcolor{black}{For this, \citet{Li2012} applied a multi-objective harmony search clustering algorithm called SCAH-MOHSA to the matrix of eigenvectors of the normalized adjacency matrix. It is worth pointing out that  \citet{Li2012} have suggested a spectral-based algorithm. 
 %The authors apply an intelligent system to the eigenvectors of the normalized adjacency matrix.
 This strategy of detecting communities in networks by finding clusters in an eigenvector matrix which is the solution of the spectral relaxation of graph partitioning problems is widely employed in the literature.  However, clustering algorithms based on this strategy are known to not scale well since they work with a non-sparse matrix. In this context, there is a dearth in the literature on efficient spectral-based methods to optimize multi-objective graph clustering problems.}

  \textcolor{black}{\cite{Zalik2018} introduced CM-Net as a combination of problem-specific genetic operators with a multi-objective algorithm based on the Non-dominated Sorting Genetic Algorithm II (NSGA-II) \citep{Deb2002}. SCAH-MOHSA and CM-Net outperformed an algorithm found in the literature -- known as Moga-Net, which is discussed in the next section -- in artificial networks proposed by \citep{Girvan2002}. 
On the one hand, both SCAH-MOHSA and CM-Net found partitions with higher modularity values than Moga-Net in real networks.
On the other hand, when contrasting the partitions obtained by SCAH-MOHSA and by Moga-Net with the expected partitions, the algorithms were competitive\footnote{\cite{Zalik2018} did not contrast the partitions obtained by CM-Net with the expected partitions.}.}

\textcolor{black}{In the next section, we briefly present studies about multi-objective graph clustering algorithms that employ criteria different from modularity to optimize.}

\subsubsection{Other optimization criteria}

\citet{Pizzuti2009,Pizzuti2012} introduced a bi-objective genetic algorithm, also  based on NSGA-II, which the authors named Moga-Net,  \textcolor{black}{to detect communities by maximizing the so-called community score \citep{Pizzuti2008} and minimizing a function named community fitness \citep{Lancichinetti2008}. On the one hand, the community score is based on the evaluation of the number of edges  inside  communities. On the other, the community fitness relies on the assessment of the number of edges between vertices from different communities.}  In computational experiments with large real networks, the modularity values of the best modularity valued partitions from the Pareto sets found by Moga-Net were worse than those found by a mono-objective spectral clustering algorithm  in the literature. \textcolor{black}{The studies performed in \citep{Gong2011,Amiri2011,Amiri2012,Amiri2013} approached the same bi-objective problem and presented heuristic methods  competitive with Moga-Net.}

\textcolor{black}{\citet{Gong2012} suggested a bi-objective problem that aims at maximizing the ratio association \citep{Angelini2007} and  minimizing the ratio cut \citep{Wei1991}. The ratio association and ratio cut assess the sum of the internal and external degrees, respectively, of the subgraphs induced by the communities of the graph. Both  measures are normalized by the number of vertices in each community. Other authors also studied these measures in the literature e.g. in \citep{Zhou2016_cuckoo,Chen2016,Shang2016,Pourkazemi2017,Zou2017,Cheng2018,Zhu2008}.}

\subsubsection{Solution selection for the decision-making}

\textcolor{black}{It is worth mentioning that solution selection strategies can be used in applications which require a single solution from multi-objective community detection algorithms that return a Pareto set approximation.
One of the most common strategies  is to select from the  set the partition with the highest modularity value \citep{Pizzuti2009,Pizzuti2012,Shi2012,Gong2012,Gong2013,Ghaffaripour2016,Pourkazemi2017}. 
\citet{Shi2012}, in addition to this strategy, suggested considering the minimum standard deviation of the Pareto solutions from those obtained to a graph generated randomly with the same degree sequence as the graph under study.  The selected Pareto solution is the one whose minimum standard deviation is the largest among all Pareto solutions. \citet{Zalik2018} suggested ranking the partitions according to their non-domination level measured by the  crowding distance, as suggested in  NSGA-II}.

\textcolor{black}{Another form to return a single partition from a given  set of solutions is by consensus clustering strategies.  \citet{Kanawati2015} suggested using different consensus and ensemble strategies to obtain a  partition from outputs of a graph clustering algorithm. Nevertheless, the method of \citet{Kanawati2015} was designed only to find  clusters of target nodes in a distributed form. In this paper, we propose a  consensus strategy to define a partition from the Pareto solutions, instead of employing the measure-based strategies introduced in the literature that are biased to a single evaluation metric. By using the consensus clustering, our goal is to capture the core communities of the Pareto set and to weight the joint relation between vertices to define their final communities. }

\textcolor{black}{In this context, the next section briefly reviews ensemble and consensus clustering methods for graph clustering.}

\subsection{Consensus clustering}

Ensemble and consensus clustering are both solution methods that combine algorithms, partitions or models to perform the clustering task. These methods have been intensively studied in the last decades \citep{Nascimento2008,Lancichinetti2012,Santos2016a}. They  tend to be more robust than those that optimize a single criterion.

The ensemble algorithms for graph clustering related to the study performed in this paper belong to the class of  consensus methods that combines partitions from a set of diverse partitions in order to determine a consensus partition. The strategy to define such consensus partitions relies on observing whether a pair of vertices is in the same group in most of the partitions in the set. Studies \citep{Lancichinetti2012},  \citep{Liang2014} and \citep{Santos2016a} obtained good results using these methods.

\textcolor{black}{The consensus strategy proposed by \citet{Lancichinetti2012} achieved better  results than  ensemble algorithms based on the modularity maximization using the majority rule. \citet{Liang2014} combined a consensus strategy with a label propagation (LP) algorithm \citep{Raghavan2007} to obtain better partitions than LP. As previously mentioned, although \citet{Kanawati2015} approached a graph clustering problem which does not find a partitioning, one of the  strategies the author employed is founded on the definition of a consensus matrix, similar to the strategy that we introduce in this paper.}

In their consensus clustering, \citet{Santos2016a}  identified a consensual partition from a set of partitions obtained by an algorithm that aims to maximize the modularity adjusted for different values of the resolution parameter. The consensual partition is obtained by assigning the same community to vertices that are in the same community on at least half of the partitions of the set. 

 %%%%%%%%%%%%%%%%%%%%%%%%%%%%%%%%%%%%%%%%%%%%%%%%%%%%%%
\section{Weighted Aggregate Modularity}\label{sec:spectral}
%%%%%%%%%%%%%%%%%%%%%%%%%%%%%%%%%%%%%%%%%%%%%%%%%%%%%%

This section discusses the spectral decomposition of the weighted aggregate modularity.
Throughout this paper, let $G=(V,E)$ be an undirected graph, where $V$ is its set of $n$ vertices and $E$ is its set of $m$ edges. The edges of $G$ are unordered pairs of distinct adjacent vertices $(i,j)$, where $i,j\in V$. Let $A=[a_{ij}] \in \mathds{N}^{n \times n}$ be the adjacency matrix of $G$, i.e., $a_{ij}$ is $1$ if $(i,j)\in E$, and 0 otherwise. The degree of a vertex $i$, $d_i$, is given by $\sum_{j \in V} a_{ij}$.
A vertex partition with $k$ clusters (groups or communities) is here defined as $P = \{C_1, C_2, \dots, C_k \}$, where $\bigcup_{t=1}^k C_t=V$ and $C_t \cap C_{t'}=\emptyset$, $\forall t\neq t' \in \{1,2,\ldots, k\}$. The label of a cluster $C_t$ is $t$ and, for ease of notation, we refer to cluster $C_t$ as the cluster with label $t$ \textcolor{black}{and to the label of the cluster of a vertex $i$ in a partition $P$ as $\mathcal{C}_P(i)$}.

Modularity is a measure that assesses the difference between the number of edges within clusters and its expected number in a random graph with the same degree sequence as the graph under consideration. 
Equation \eqref{eq:fo_modQ_orig} presents a way to calculate the modularity measure originally introduced by \citet{Newman2004c}.

 \begin{equation}
 Q(P) = \frac{1}{2m} \sum_{i,j \in V} \left( a_{ij}-\frac{d_i d_j}{2m}  \right)\delta_{\mathcal{C}_P(i),\mathcal{C}_P(j)}
 \label{eq:fo_modQ_orig}
 \end{equation}
 
 In Equation~\eqref{eq:fo_modQ_orig}, $\delta_{\mathcal{C}_P(i),\mathcal{C}_P(j)}$ is an indicator function that assumes value 1 if $\mathcal{C}_P(i) = \mathcal{C}_P(j)$, and 0 otherwise. The resolution parameter, as suggested by \citet{Reichardt2006}, is a scalar $\gamma$ that multiplies the term $\frac{d_i d_j}{2m}$ in Equation \eqref{eq:fo_modQ_orig}.

 Equation~\eqref{eq:fo_modQ_orig} shows that in order to maximize the modularity, the first term, i.e. $a_{ij}$, must be maximized and the second term, i.e. $\frac{d_i d_j}{2m}$, has to be minimized.
 On the one hand, the higher the number of edges within clusters, the higher the first term. On the other, the lower the number of edges within clusters, the lower the expected number of edges within clusters and consequently, the lower the second term. These two terms, therefore, are conflicting and result in a trade-off in the modularity measure \citep{Brandes2008}.
 
 As discussed earlier in this paper, \citet{Shi2012} have approached the bi-objective problem that optimizes the two terms of modularity. Equations \eqref{eq:fo_modQ} and \eqref{eq:fo_modNull} present the pair of objective functions of the bi-objective problem. 
 
 \begin{equation}
 \max_{P} Q^{IN}(P) = \frac{1}{2m} \sum_{i,j \in V} a_{ij} \delta_{\mathcal{C}_P(i),\mathcal{C}_P(j)}
 \label{eq:fo_modQ}
 \end{equation}

 \begin{equation}
 \min_{P} Q^{NULL}(P) = \frac{1}{2m} \sum_{i,j \in V} \frac{d_i d_j}{2m} \delta_{\mathcal{C}_P(i),\mathcal{C}_P(j)}
 \label{eq:fo_modNull}
 \end{equation}

 Consider the weighted aggregation of the objective functions $Q^{IN}(P)$ and $Q^{NULL}(P)$ as presented in Equation~\eqref{eq:mod_aggr_2}. The objective function \eqref{eq:fo_modNull} can be transformed into a maximization function without loss of generality by multiplying the function by -1.
 
 \begin{equation}
 QW(P) =\frac{1}{2m} \sum_{i,j \in V} \left [ \gamma_1 a_{ij} - \gamma_2 \frac{d_i d_j}{2m} \right ] \delta_{\mathcal{C}_P(i),\mathcal{C}_P(j)},
 \label{eq:mod_aggr_2}
 \end{equation}
 \noindent where $\gamma_1, \gamma_2 \in \mathds{R}, \gamma_1+ \gamma_2=1$.
 
 The set of solutions for the weighted aggregation problem for different values of $\gamma_1$ and $\gamma_2$ are efficient \citep{Ehrgott2005}, and thereby provide an approximation to the Pareto frontier of the bi-objective problem. 
 Moreover, as $\gamma_1$ and $\gamma_2$ are both scalars, when $\gamma_1>0$ the optimization problem $\max_{P}QW$ is equivalent to $\frac{1}{\gamma_1}\max_{P}QW$, which is exactly the adjusted modularity maximization problem. Therefore, the solutions of the modularity maximization problem with different values of resolution parameter are also efficient Pareto solutions for the bi-objective problem \eqref{eq:fo_modQ}-\eqref{eq:fo_modNull}.  In particular, the maximization of Equation \eqref{eq:mod_aggr_2} for $\gamma_1=\gamma_2=0.5$ is equivalent to the classical modularity maximization problem.
 
We also say that a partition $P_a$  dominates a partition $P_b$ if and only if $Q^{IN}(P_a)>Q^{IN}(P_b)$ and $Q^{NULL}(P_a) \leq Q^{NULL}(P_b)$ or if and only if $Q^{IN}(P_a) \geq Q^{IN}(P_b)$ and $Q^{NULL}(P_a)$ $<$ $Q^{NULL}(P_b)$.
 
%%%%%%%%%%%%%%%%%%%%%%%%%%%%%%%%%%%%%%%%%%%%%%%%%%%%%%
\subsection{Spectral decomposition}\label{sec:spec_dec}
%%%%%%%%%%%%%%%%%%%%%%%%%%%%%%%%%%%%%%%%%%%%%%%%%%%%%%

This section presents the spectral decomposition of the weighted aggregation of modularity provided in Equation \eqref{eq:mod_aggr_2}. It is strongly based on the spectral decomposition proposed by \citet{Newman2006}. 
Let us first define in Equation \eqref{eq:modMatrix} the weighted aggregate matrix $BW=[bw_{ij}] \in \mathds{R}^{n \times n}$.
 
 \begin{equation}
 bw_{ij} = \gamma_1 a_{ij} - \gamma_2 \frac{d_i d_j}{2m}
 \label{eq:modMatrix}
 \end{equation}

Note that the modularity matrix is $B=\frac{1}{\gamma_1} BW,$ where $\gamma = \frac{\gamma_2}{\gamma_1}=1$.
Consider the sequence of eigenvalues of matrix $BW$, $\lambda_1,$ $\lambda_2,$ $\dots,$ $\lambda_n$, sorted in the decreasing order of absolute value, that is, $|\lambda_1| \geq |\lambda_2| \geq \dots \geq |\lambda_n|$. Let $U \in \mathds{R}^{n \times n}$ be a matrix such that its $j$-th column, referred to as column $u_j$, is an eigenvector of $BW$ associated with eigenvalue $\lambda_j$. $BW$ is symmetric and thus admits an eigen-decomposition: $BW = U \Lambda U^T$, where $\Lambda=[\Lambda_{ij}] \in \mathds{R}^{n \times n}$ is a diagonal matrix such that $\Lambda_{ii} = \lambda_i$.

Let $S=[s_{it}] \in \mathds{N}^{n \times k}$ be a binary matrix associated with a solution of the graph clustering problem. Element $s_{it}$ receives 1 if vertex $i$ belongs to cluster $C_t$, and $0$ otherwise. Therefore, $\delta_{\mathcal{C}_P(i),\mathcal{C}_P(j)} = \sum_{t=1}^{k} s_{it}s_{jt}$. Equation \eqref{eq:mod_aggr_2} can hence be rewritten as indicated in Equation \eqref{eq:modEsp2}.
 
 \begin{equation}
 QW(P) = \frac{1}{2m} \sum_{i,j \in V} \sum_{t=1}^{k} bw_{ij} s_{it}s_{jt} = \frac{1}{2m}Tr(S^T BW S)
 \label{eq:modEsp2}
 \end{equation}
 
Any given vertex belongs to exactly and only one cluster, which implies that $\sum_{t=1}^{k} s_{it} = 1, i=1,\dots,n$, and $Tr(S^TS)= n$.  Knowing that $U$ is an orthogonal matrix, we can rewrite Equation \eqref{eq:modEsp2} as Equation \eqref{eq:modEsp4}.
 \begin{equation}
  QW(P) = \frac{1}{2m} Tr[S^T U \Lambda U^T S] 
 	= \frac{1}{2m} \sum_{j=1}^{n} \sum_{t=1}^{k} \lambda_j ( \sum_{i=1}^n u_{ij} s_{it})^2
  \label{eq:modEsp4}
 \end{equation}

Since Equation~ \eqref{eq:modEsp4} shows that only positive eigenvalues increase the value of $QW$, \citet{Newman2006} suggested approximating Equation \eqref{eq:modEsp4} using only the first largest positive eigenvalues. Nonetheless, 
\citet{Newman2006} also demonstrated that negative eigenvalues are important to indicate vertices that  decrease the $QW(P)$ in case they are clustered together. 
This paper takes into account the negative eigenvalues by selecting the first $p$ eigenvalues sorted in decreasing order of absolute value.

Consider $\mathcal{E}$ the set of the first $p$ eigenvalues of $BW$; let $\mathcal{E}p=\{j | \lambda_j \in \mathcal{E} \mbox{ such that } \lambda_j \geq 0\}$ and $\mathcal{E}n=\{j | \lambda_j \in \mathcal{E} \mbox{ such that } \lambda_j < 0\}$ be the positive and negative eigenvalue indices, respectively. 
Moreover, let $rp^i \in \mathds{R}^{p}$ and $rn^i \in \mathds{R}^{p}$ be the vectors regarding vertex $i$ whose components are defined by Equations \eqref{eq:rpi} and \eqref{eq:rni}, respectively. 
Also, in this paper, $rp^i$ is called positive vector of vertex $i$, whereas $rn^i$ is referred to as negative vector of vertex $i$.

 \begin{equation}
 rp^i_j =  \left \{
 \begin{split}
  { \sqrt{\lambda_j } u_{ij} } & \quad , \mbox{ if } j\in \mathcal{E}_p  \\
  {0} & \quad ,\mbox{ if } j\in \mathcal{E}_n
  \end{split}
  \right .
  \label{eq:rpi}
 \end{equation}
 \begin{equation}
 rn^i_j =  \left \{
 \begin{split}
  {\sqrt{-\lambda_j} u_{ij} } & \quad , \mbox{ if } j\in \mathcal{E}_n  \\
  {0} & \quad ,\mbox{ if } j\in \mathcal{E}_p
  \end{split}
  \right .
  \label{eq:rni}
 \end{equation}
  
%--------------------------------------------------

Equation \eqref{eq:modEsp5} approximates Equation \eqref{eq:modEsp4} using the $p$ largest eigenvalues in absolute value.

 \begin{equation}
 \begin{split}
  QW(P) \simeq & \frac{1}{2m} \sum_{\lambda_j \in \mathcal{E}} \sum_{t=1}^{k} {\left [ \sum_{i=1}^{n} \sqrt{|\lambda_j|} u_{ij} s_{it}\right ]}^2\\
	= & \frac{1}{2m}  \sum_{t=1}^{k} \sum_{j=1}^{p} \left [
	{ \left(\sum_{i \in C_t} rp^i_j \right ) }^2
	- 
	{ \left ( \sum_{i \in C_t} rn^i_j \right ) }^2 
	\right ] \\
	= & \frac{1}{2m} \sum_{t=1}^{k} ( ||Rp^t||^2 - ||Rn^t||^2)	
 \end{split}
\label{eq:modEsp5}
 \end{equation}

 \noindent where $\forall j\in \{1,\ldots,p\}$, $Rp^t_j = \sum_{i \in C_t} rp^i_j$ and $Rn^t_j = \sum_{i \in C_t} rn^i_j$.
 
Furthermore, $Rp^t=[Rp^t_j]_{j=1\ldots p}$ and $Rn^t=[Rn^t_j]_{j=1\ldots p}$ are referred to as vectors of cluster $C_t$. In this paper, $Rp^t$ is called positive vector of cluster $C_t$, whereas $Rn^t$ is referred to as negative vector of cluster $C_t$.

Similarly to the results of the approximation with positive eigenvalues carried out by \citet{Newman2006}, we have reduced the weighted aggregate modularity maximization problem into a vector partitioning problem.
The goal of the vector partitioning problem is to find a vertex partition by maximizing the terms $Rp^t$ and minimizing the terms $Rn^t$, for $t=1,2,\dots,k$. 

It is well-known that the number of groups has a direct impact on the number of eigenvectors required to determine graph clusterings. Thereby, most spectral heuristics must first define the number of groups, which is generally not known in advance. 

%%%%%%%%%%%%%%%%%%%%%%%%%%%%%%%%%%%%%%%%%%%%%%%%%%%%%% 
\subsection{Defining the number of clusters}\label{sec:alg_number_k}
%%%%%%%%%%%%%%%%%%%%%%%%%%%%%%%%%%%%%%%%%%%%%%%%%%%%%% 

In this paper, we adapted the strategy to identify the number of clusters presented by \citet{Krzakala2013}, who constructed a matrix called non-backtracking matrix from the adjacency matrix of a given graph and estimated the number of clusters through the eigenvalues of this matrix.

The adaptation proposed here consists in estimating the number of clusters based on the weighted aggregate modularity matrix $BW$.
Let $\chi$ be the largest (leading) eigenvalue of $BW$. The proposed algorithm sets $k'$ as the number of eigenvalues of $BW$ higher than $\sqrt{\chi}$. In this paper, we estimate the number of clusters, $k$, to be $\lfloor1.25k' \rfloor$.
This estimation is an upper bound to the number of clusters because the proposed algorithm might leave one or more clusters empty.

\begin{figure}[!htb] 
  \centering
 \includegraphics[width=0.49\textwidth]{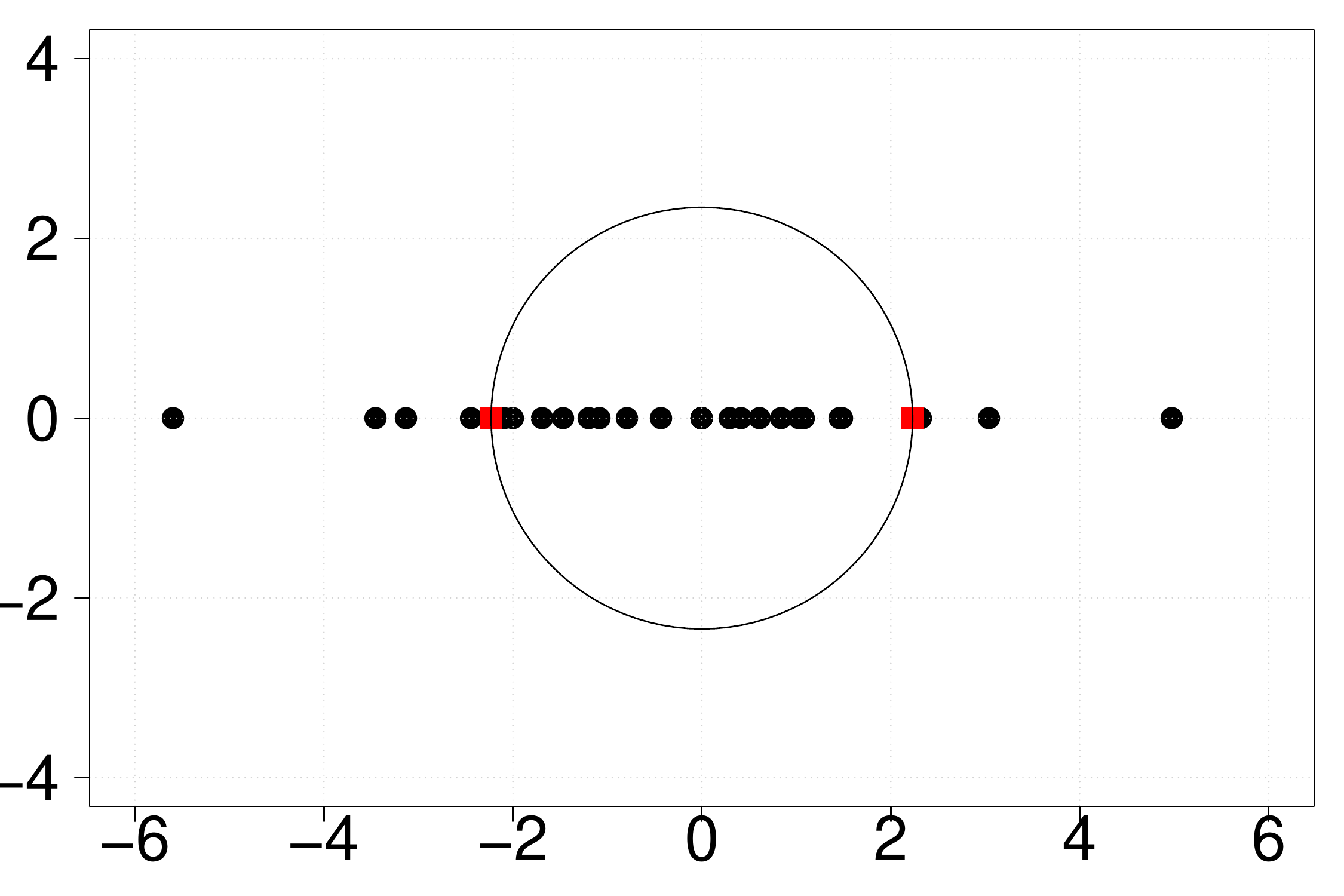}
 \caption{Distribution of the eigenvalues of the Karate network.}
 \label{fig:example_eigenvalues}
 \end{figure}

Figure \ref{fig:example_eigenvalues} displays an example of the proposed strategy by depicting the eigenvalues of the Karate network \citep{Zachary1977}, whose largest eigenvalue is $4.977$. The red squares in this figure indicate the points $(-\sqrt{\chi},0)$ and $(\sqrt{\chi},0)$ and a circumference of radius $\sqrt{\chi}$ centered at the origin of the Cartesian plane. Most of the black dots, which correspond to the eigenvalues of matrix $BW$, are enclosed by the circumference. The proposed algorithm estimates $k'$ to be the number of eigenvalues higher than $\sqrt{\chi}$, i.e., the number of positive points outside the circumference, which is $3$. Therefore, the upper bound estimation to the number of clusters is $k=3$.

%%%%%%%%%%%%%%%%%%%%%%%%%%%%%%%%%%%%%%%%%%%%%%%%%%%%%% 
\subsection{Geometric interpretation}\label{sec:geometric}
%%%%%%%%%%%%%%%%%%%%%%%%%%%%%%%%%%%%%%%%%%%%%%%%%%%%%% 

\begin{figure}[!htb]
 \subfigure[fig:geometricP5pos][Positive vertex and cluster vectors]{\includegraphics[width=0.49\textwidth]{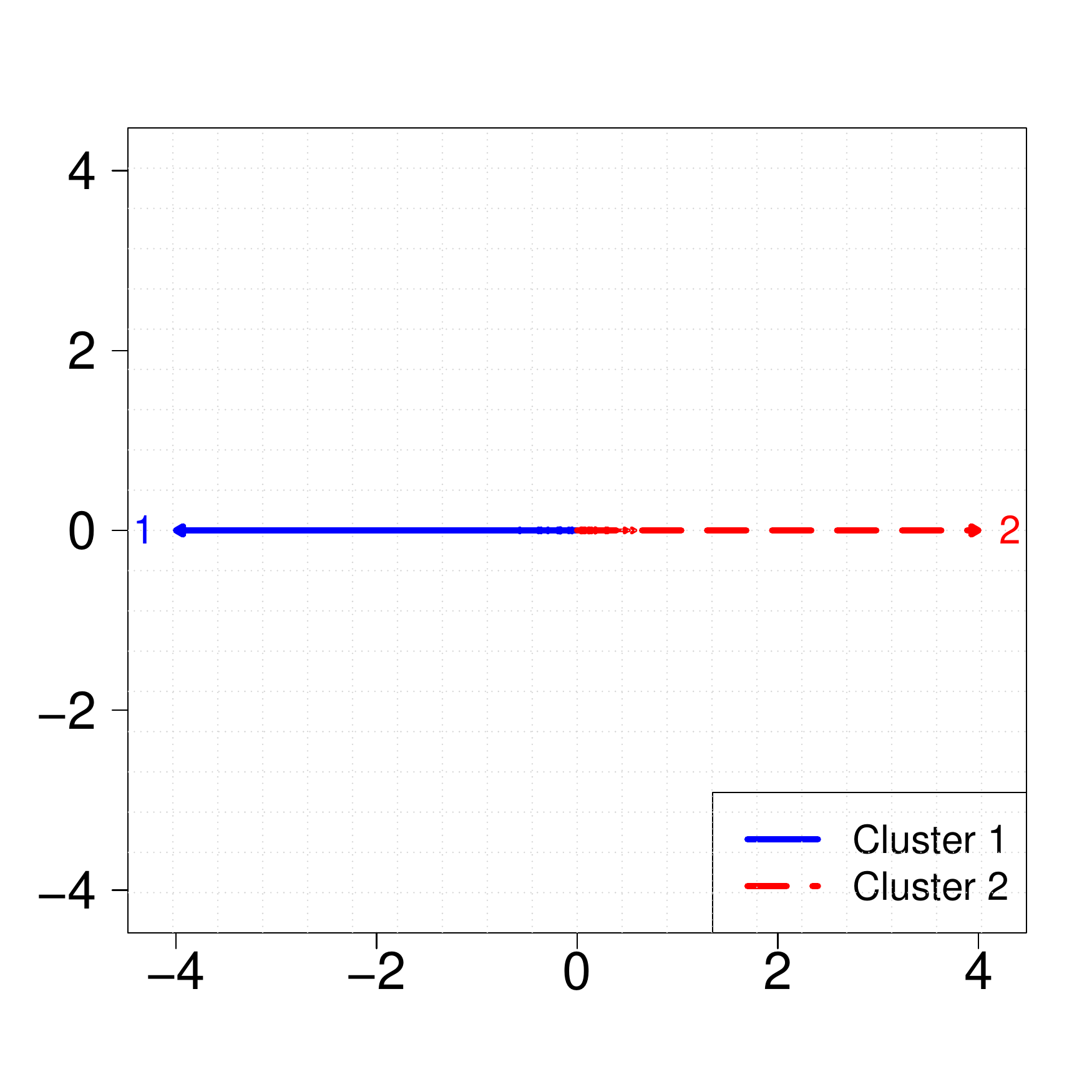}\label{fig:geometric_P5_pos}}
 \subfigure[fig:geometricP5neg][Negative vertex and cluster vectors]{\includegraphics[width=0.49\textwidth]{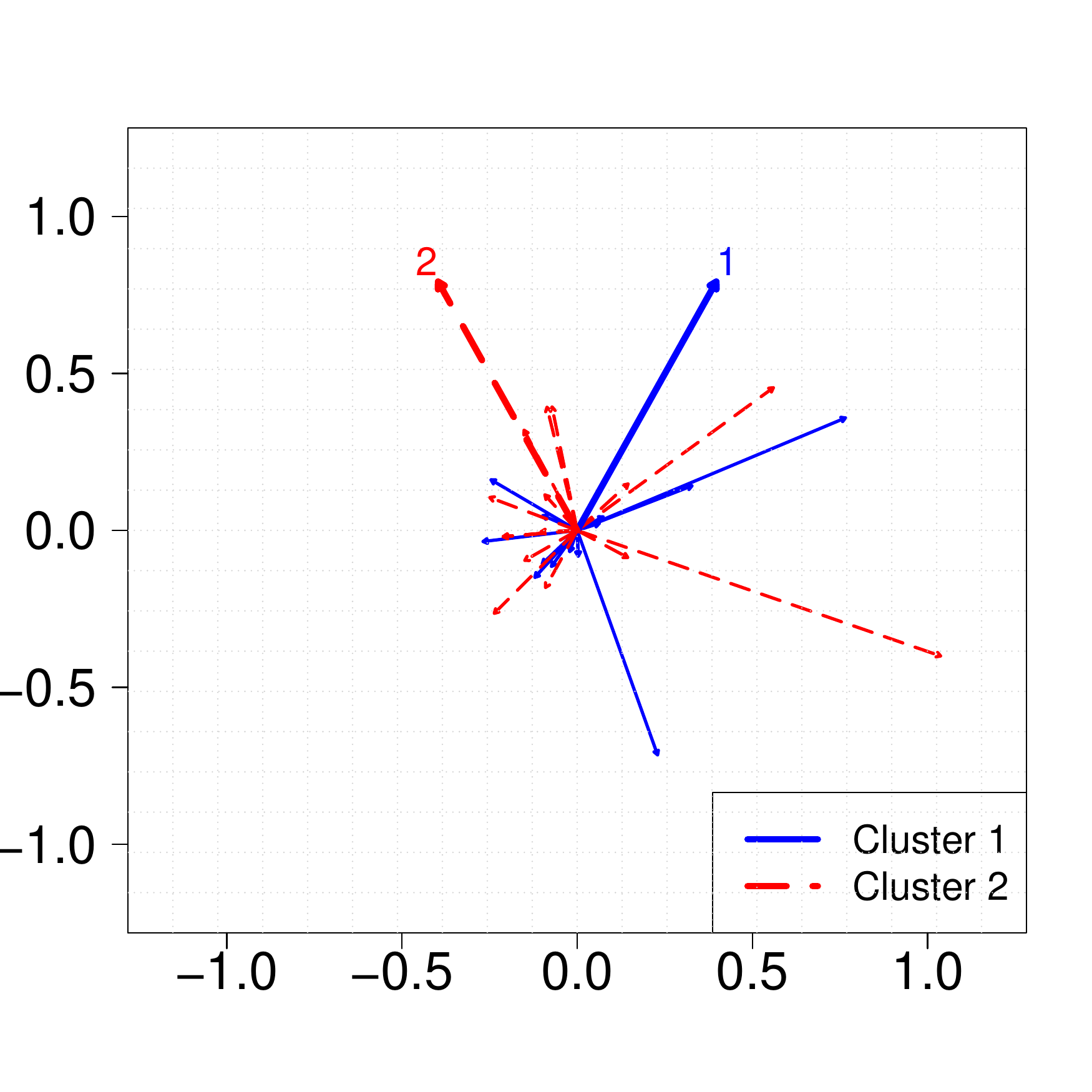}\label{fig:geometric_P5_neg}}
 \caption{Vectors of the vertices and clusters of the bipartition found by the algorithm proposed in this paper to maximize the weighted aggregate modularity with $\gamma_1=\gamma_2=0.5$ applied to the Karate network.}
 \label{fig:geometric_P5}
 \end{figure}

Figure \ref{fig:geometric_P5} illustrates, for a given bipartition $P$ of the benchmark Karate network, the geometric interpretation of all the vectors of vertices and clusters. This network has $34$ vertices. 
The positive and negative vectors are shown in Figures \ref{fig:geometric_P5_pos} and \ref{fig:geometric_P5_neg}, respectively. In these figures, the vectors of clusters are identified by their labels and the solid and dashed lines distinguish the vertex vectors regarding clusters 1 and 2, respectively.

The cluster vectors are the sum of the vertex vectors that compose the clusters. The higher ${Rp^t}^T rp^i$ and the lower ${Rn^t}^T rn^i$, $\forall i \in V$ and $t=\mathcal{C}_P(i)$, the higher the modularity. 
On the one hand, the obvious choice to maximize the magnitude of the positive cluster vectors in Figure \ref{fig:geometric_P5_pos} is to select the vertices whose positive vertex vectors point to the same direction. On the other hand, to minimize the magnitude of the negative cluster vectors in Figure \ref{fig:geometric_P5_neg}, the vertices whose negative vertex vectors point to opposite directions should be selected.
By comparing Figures \ref{fig:geometric_P5_pos} and \ref{fig:geometric_P5_neg}, it is possible to observe that the magnitude of the positive vectors of the clusters is approximately $4.51$ times higher than the magnitude of the negative vectors of the clusters.

%%%%%%%%%%%%%%%%%%%%%%%%%%%%%%%%%%%%%%%%%%%%%%%%%%%%%% 
\subsection{Moving vertices between clusters} \label{sec:move_vert}
%%%%%%%%%%%%%%%%%%%%%%%%%%%%%%%%%%%%%%%%%%%%%%%%%%%%%% 

Given a partition $P$ at hand, some procedures attempt to enhance its quality, which can be evaluated using a fitness function. One way of performing this task is by moving vertices from one cluster to another so that the modified partition has better quality than the previous one.
Many studies that employ this type of strategy can be found in the literature, e.g. \citep{Newman2006} and \citep{Zhang2015}.

Moving a vertex $i$ from a cluster $C_b$ to a cluster $C_t$ modifies the fitness function value, i.e., the weighted aggregate modularity. 
Let the vectors of clusters $C_b$ and $C_t$, disregarding the contribution of vertex $i$, be defined by $Rp^{b} = \sum_{v \in C_b, v \neq i} rp^v$, $Rn^{b} = \sum_{v \in C_b, v \neq i} rn^v$, $Rp^{t} = \sum_{v \in C_t} rp^v$ and $Rn^{t} = \sum_{v \in C_t} rn^v$. 
 
On the one hand, before moving $i$ to cluster $C_t$, the vectors of clusters $C_b$ are given by $Rp'^{b} = Rp^{b} + rp^i$ and $Rn'^{b} = Rn^{b} + rn^i$, respectively. On the other hand, before any movement, the vectors of cluster $C_t$ are $Rp'^{t}=Rp^{t}$ and $Rn'^{t}=Rn^{t}$. After moving $i$ from cluster $C_b$ to $C_t$, the vectors of the clusters are $Rp''^{b} = Rp^{b}$, $Rp''^{t}=Rp^{t} + rp^i$, $Rn''^{b} = Rn^{b}$ and $Rn''^{t}=Rn^{t} + rn^i$. Equation \eqref{eq:mod_delta} presents the change in the weighted aggregate modularity of partition $P$ after moving a vertex $i$ from a cluster $C_b$ to a cluster $C_t$. 

\begin{equation}
 \begin{split}
  \Delta QW(i,C_b,C_t) = & \frac{1}{2m} 
  \left[\right. ||Rp''^{b}||^2- ||Rn''^{b}||^2 + ||Rp''^{t}||^2 - ||Rn''^{t}||^2 \\
	 -&( ||Rp'^{b}||^2 - ||Rn'^{b}||^2 + ||Rp'^{t}||^2 - ||Rn'^{t}||^2 )\left.\right]\\
	 =& \frac{1}{m}\left[ {Rp^{t}}^T rp^i - {Rn^{t}}^T rn^i - {Rp^{b}}^T rp^i 
	 + {Rn^{b}}^T rn^i\right]	 	 
  \end{split}
  \label{eq:mod_delta}
 \end{equation}

From Equation \eqref{eq:mod_delta}, it is possible to see that $\Delta QW(i,b,t) \geq 0$ if $({Rp^{t}}^T rp^i - {Rn^{t}}^T rn^i) \geq ({Rp^{b}}^T rp^i - {Rn^{b}}^T rn^i)$.
 
Recently, \citet{Zhang2015} presented a spectral greedy heuristic to solve the vector partitioning problem considering only positive eigenvalues. In this heuristic, starting from an initial group of vectors, at each iteration, the algorithm moves a vertex $i$ to the cluster $C_{t^*}$ that results in the largest positive gain in modularity.
Concerning both positive and negative eigenvalues, a simple greedy heuristic consists of moving vertex $i$ to the cluster $C_{t^*}$ that results in the largest value for ${Rp^{t^*}}^{T} rp^i - {Rn^{t^*}}^T rn^i$. Equation \eqref{eq:modMovMax} defines the choice of ${t^*}$.
 
 \begin{equation}
 {t^*} = \arg\max_{t\in\{1,\dots, k\}} \left \{
 \begin{split}
 {Rp^{t}}^T rp^i - {Rn^{t}}^T rn^i & \quad , \mbox{ if } \mathcal{C}_P(i) \neq t \\
  {0} & \quad ,\mbox{ if } \mathcal{C}_P(i) = t
  \end{split}
  \right .
  \label{eq:modMovMax}
 \end{equation}

If ${t^*} = \mathcal{C}_P(i)$, vertex $i$ will remain in its original cluster.

%%%%%%%%%%%%%%%%%%%%%%%%%%%%%%%%%%%%%%%%%%%%%%%%%%%%%%
\section{Proposed \textcolor{black}{Spectral-}evolutionary Hybrid Multi-objective Algorithm}\label{sec:method}
%%%%%%%%%%%%%%%%%%%%%%%%%%%%%%%%%%%%%%%%%%%%%%%%%%%%%%

This section thoroughly describes the \textcolor{black}{spectral-}evolutionary \textcolor{black}{hybrid} multi-objective algorithm proposed in this paper and called \emph{MOSpecG}. \emph{MOSpecG} is \textcolor{black}{an iterative two-phase algorithm. At the first phase,   the weighted aggregate modularity matrix is updated and its eigen-decomposition is performed. At the second phase, a memetic algorithm works based on the information of the vertex vectors -- discussed in the earlier section. To a better understanding of the method, Algorithm~\ref{alg:mo_alg} presents a pseudocode of \emph{MOSpecG}.}

 \begin{algorithm}[!htb]
 {
 \caption{\emph{MOSpecG}}
 \label{alg:mo_alg}
 \SetKwInOut{Input}{Input}
 \SetKwInOut{Output}{Output}
   \Input{$G$, $N\mathcal{F}$, $N\mathcal{G}$, $N\mathcal{P}$, $N\mathcal{O}$, $p$ and $IT$}
   \Output{$\mathcal{F}$}

   $\mathcal{F} = \emptyset$
   
   $inc = \frac{1}{N\mathcal{F}-1}$
   
   \For{$\gamma_1=0$ to $1$, $\gamma_1=\gamma_1+inc$ }{
    $\gamma_2 = 1-\gamma_1$
    
    Construct the weighted aggregate modularity matrix $BW$ with weights $\gamma_1$ and $\gamma_2$
    
    $\Lambda, U :=$ Eigen-decomposition of $BW$ regarding the $p$ largest eigenvalues in absolute value

    $\chi := \max_{\Lambda_{ii},\forall i}( \Lambda_{ii})$ 
    
    $k' :=$ number of eigenvalues of $BW$ with value larger than or equal to $\sqrt{\chi}$
    
    $k :=\lfloor 1.25 k'\rfloor$
    
    Define vertex vectors $rp^i$ and $rn^i$, $\forall i \in V$
    
     $P$:=\emph{Memetic Algorithm}($N\mathcal{G},\mathcal{N}{P}, \mathcal{N}O, IT, k, rp^i, rn^i, \forall i \in V$) 
    
    $\mathcal{F}:= \mathcal{F} \cup P $ 
   }
 }
 \end{algorithm}

According to Algorithm \ref{alg:mo_alg},  \emph{MOSpecG} has as input: an undirected unweighted graph $G$; the size of the Pareto frontier, $N\mathcal{F}$; the number of generations, $N\mathcal{G}$; the number of solutions in the population, $N\mathcal{P}$; the percentage of solutions from the offspring, $N\mathcal{O}$; the number of eigenvalues and eigenvectors to be computed, $p$; and the number of iterations of the local search procedure, $IT$. 

In line 1 of Algorithm \ref{alg:mo_alg}, set $\mathcal{F}$ is initialized as empty. \textcolor{black}{Consider that the possible values of $\gamma_1$ and $\gamma_2$ are defined in a grid to ensure a good spreading of the solutions in the Pareto frontier approximation. Therefore, the} grid spacing is dependent on the number of solutions of the resulting Pareto frontier. In line 2, the grid spacing is assigned to variable $inc$ in order to define values for $\gamma_1$.  In the sequence, weight $\gamma_2$ is calculated taking $\gamma_1$ as  reference, in line 4. From lines 5 to 11, the proposed heuristic creates a new solution to the approximation of the Pareto frontier by optimizing $QW$ with the current values of $\gamma_1$ and $\gamma_2$.

In particular, in line 5, the algorithm constructs matrix $BW$ with weights $\gamma_1$ and $\gamma_2$ according to Equation \eqref{eq:modMatrix}. \textcolor{black}{In line 6, the largest $p$ eigenvalues and the associated eigenvectors that compose $\Lambda$ and $U$ are computed using the \textit{implicitly restarted Arnoldi method} from ARPACK++ library \citep{Arpack}}.
In line 7, the leading eigenvalue is assigned to $\chi$. In lines 8 and 9, \emph{MOSpecG} estimates the number of clusters, $k$, according to Section \ref{sec:alg_number_k}.
In line 10, vertex vectors $rp^i$ and $rn^i$, $\forall i \in V$, are defined according to Equations \eqref{eq:rpi} and \eqref{eq:rni}, respectively. In line 11, \emph{MOSpecG} calls the \emph{Memetic Algorithm} function presented in Algorithm \ref{alg:memetic} to optimize $QW$ with weights $\gamma_1$ and $\gamma_2$. The resulting partition $P$ is included in  the Pareto frontier approximation $\mathcal{F}$ in line 12. At the end, Algorithm \ref{alg:mo_alg} returns $\mathcal{F}.$

\textcolor{black}{In the next section, the \emph{Memetic Algorithm} employed in line 11 of Algorithm~\ref{alg:mo_alg} is comprehensively discussed.}

%%%%%%%%%%%%%%%%%%%%%%%%%%%%%%%%%%%%%%%%%%%%%%%%%%%%%%
 \subsection{\emph{Memetic Algorithm}}
%%%%%%%%%%%%%%%%%%%%%%%%%%%%%%%%%%%%%%%%%%%%%%%%%%%%%%
\textcolor{black}{Before going into detail on the algorithm, let us briefly introduce the notations employed in this section.}

Let the population of the $g$-ith generation be defined by $\mathcal{P}^{g}=\{P^g_1,P^g_2,\dots,P^g_{N\mathcal{P}}\}$, where $g \in \{1,2,\dots,N\mathcal{G}\}$. The individuals from the population of the $g$-ith generation are the partitions $P^g_h$, $h \in \{1,2,\dots,N\mathcal{P}\}$.

Algorithm \ref{alg:memetic} presents the proposed \emph{Memetic Algorithm}, whose inputs are: $N\mathcal{G}$; $\mathcal{N}{P}$; $\mathcal{N}O$; $IT$; the number of clusters, $k$; and the vertex vectors $rp^i$ and $rn^i$, $\forall i \in V$.
In line 1 of Algorithm \ref{alg:memetic}, the initial population, i.e., individuals from the first generation, is constructed using the strategy suggested by \citet{Zhang2015}. This strategy selects $k$ vertices and assigns each of them to a different cluster ($k$ singletons). Then, the vectors of the selected clusters $Rp^t$ and $Rn^t, t=1,\dots,k$, are updated. The remaining vertices are assigned to clusters $C_{t^*}$, where $t^*$ is chosen according to Equation \eqref{eq:modMovMax}.

 \begin{algorithm}[!htb]
  {
  \caption{\emph{Memetic Algorithm}}
  \label{alg:memetic}
  \SetKwInOut{Input}{Input}
  \SetKwInOut{Output}{Output}
      \Input{$N\mathcal{G}$,  $N\mathcal{P}$, $N\mathcal{O}$, $IT$,   $k$, $rp^i$ and $rn^i$, $\forall i \in V$}
    \Output{Fittest individual $P^*$}
    
    $P_{s}^1, s=1, \dots,N\mathcal{P} :=$ construct solution using vertex vectors as directions
    
    \For{$g=1$ to $N\mathcal{G}$}{
        
        $\mathcal{O}:=Crossover(\mathcal{P}^g,k,rp^i,rn^i, \forall i \in V)$
        
        $\mathcal{O} :=Mutation(\mathcal{O},k,rp^i,rn^i, \forall i \in V)$
        
        $\mathcal{O}:=LocalSearch(\mathcal{O},IT,k,rp^i,rn^i, \forall i \in V)$
        
        $\mathcal{P}^{g+1}:=$ Update population $\mathcal{P}^g$ using $\mathcal{O}$
    }

    $P^*:=$ the fittest individual from $\mathcal{P}^{N\mathcal{G}}$
  }
  \end{algorithm}
  
Figure \ref{fig:example_initial} shows an example of an initial partition of the Karate network. To calculate $QW$, we considered $\gamma_1=\gamma_2=0.5$. In this figure, each square identifies the cluster label of a vertex of the network.

\begin{figure}[htb]
\begin{center}
\footnotesize
\begin{tikzpicture}[node distance = 0.6cm, auto]
\node [block] (v1) {$1$};
\node [block, right=0cm of v1] (v2) {$2$};
\node [block, right=0cm of v2] (v3) {$1$};
\node [block, right=0cm of v3] (v4) {$1$};
\node [block, right=0cm of v4] (v5) {$1$};
\node [block, right=0cm of v5] (v6) {$1$};
\node [block, right=0cm of v6] (v7) {$1$};
\node [block, right=0cm of v7] (v8) {$1$};
\node [block, right=0cm of v8] (v9) {$1$};
\node [block, right=0cm of v9] (v10) {$1$};
\node [block, right=0cm of v10] (v11) {$1$};
\node [block, right=0cm of v11] (v12) {$1$};
\node [block, right=0cm of v12] (v13) {$1$};
\node [block, right=0cm of v13] (v14) {$1$};
\node [block, right=0cm of v14] (v15) {$2$};
\node [block, right=0cm of v15] (v16) {$2$};
\node [block, right=0cm of v16] (v17) {$1$};
\node [block, right=0cm of v17] (v18) {$1$};
\node [block, right=0cm of v18] (v19) {$2$};
\node [block, right=0cm of v19] (v20) {$1$};
\node [block, right=0cm of v20] (v21) {$2$};
\node [block, right=0cm of v21] (v22) {$1$};
\node [block, right=0cm of v22] (v23) {$2$};
\node [block, right=0cm of v23] (v24) {$2$};
\node [block, right=0cm of v24] (v25) {$2$};
\node [block, right=0cm of v25] (v26) {$2$};
\node [block, right=0cm of v26] (v27) {$2$};
\node [block, right=0cm of v27] (v28) {$2$};
\node [block, right=0cm of v28] (v29) {$2$};
\node [block, right=0cm of v29] (v30) {$2$};
\node [block, right=0cm of v30] (v31) {$2$};
\node [block, right=0cm of v31] (v32) {$2$};
\node [block, right=0cm of v32] (v33) {$2$};
\node [block, right=0cm of v33] (v34) {$2$};
\node [label, left=0.2cm of v1] (lb1) {$P_s^1$};
\node [label, below=0.2cm of v1] (lb2) {$QW(P_s^1)=0.1443$};

\end{tikzpicture}
\caption{Example of a solution for the Karate network when $\gamma_1=\gamma_2=0.5$.}
\label{fig:example_initial}
\end{center}
\end{figure}

In line 3, the \emph{Memetic Algorithm} constructs the offspring of generation $g$, $\mathcal{O}$, by applying the genetic operator crossover (Algorithm \ref{alg:crossover}) to the current population $\mathcal{P}^g$. 
In lines 4 and 5, the genetic operator mutation (Algorithm \ref{alg:mutation}) and a local search procedure (Algorithm~\ref{alg:local_search}) update the offspring population. \textcolor{black}{The  population of the next generation, $\mathcal{P}^{g+1}$, is the population $\mathcal{P}^{g}$ but with the $N\mathcal{O}\%$ fittest individuals from the offspring $\mathcal{O}$ replacing the $N\mathcal{O}\%$ least fit individuals from $\mathcal{P}^{g}$. }
In line 8, the algorithm returns the fittest individual from $\mathcal{P}^{N\mathcal{G}}$, i.e., individual $P^*$ such that $P^*=\arg\max_{P \in \mathcal{P}^{N\mathcal{G}}}QW(P)$.

%%%%%%%%%%%%%%%%%%%%%%%%%%%%%%%%%%%%%%%%%%%%%%%%%%%%%% 
\subsubsection{\textcolor{black}{Crossover}}
%%%%%%%%%%%%%%%%%%%%%%%%%%%%%%%%%%%%%%%%%%%%%%%%%%%%%% 
 
Algorithm \ref{alg:crossover} presents the one-way crossover procedure of the \emph{Memetic Algorithm}, which has as input $\mathcal{P}^g$, $k$, $rp^i$ and $rn^i$ , $\forall i \in V$. At each iteration $f$, the crossover constructs a new solution for the offspring population, $\mathcal{O}$, by combining two solutions from the current population $\mathcal{P}^g$.
In line 2, the fitness proportionate roulette method selects two individuals $P^g_b$ and $P^g_d$, $b,d \in \{1,2,\dots,{N\mathcal{P}}\}, b \neq d$, to  perform the crossover.
In line 3, the algorithm creates an offspring individual $W$ as a copy of $P^g_d$. 
In line 4, the method randomly selects a vertex $vs$ and, in line 5, $ls$ stores the label of the cluster of $vs$ in individual $P^g_b$.

  \begin{algorithm}[!htb]
  {
  \caption{\emph{Crossover}}
  \label{alg:crossover}
  \SetKwInOut{Input}{Input}
  \SetKwInOut{Output}{Output}
      \Input{$\mathcal{P}^g$, $k$, $rp^i$ and $rn^i$, $\forall i \in V$}
    \Output{$\mathcal{O}=\{O_1,O_2,\dots,O_{N\mathcal{P}} \}$}

    \For{$f=1$ to $N\mathcal{P}$}{
     Pick randomly $P^g_b$ and $P^g_d$, $b,d \in \{1,2,\dots,{N\mathcal{P}}\}, b \neq d$, from $\mathcal{P}^g$ with probability distribution $pr(P^g_h)=\frac{QW(P^g_h)}{\sum_{P^g_j \in \mathcal{P}} QW(P^g_j)}, h \in \{1,2,\dots,{N\mathcal{P}}\}$.

    $W := P^g_d$
    
     Randomly select  a vertex $vs$ from $V$
    
    $ls := P^g_b(vs)$
    
    $ld^* :=$ choose according to Equation \eqref{eq:crossover_ld}
    
    Move $vd$ to cluster $C_{ld^*}$ in individual $W$, $\forall vd \in V$ such that  $\textcolor{black}{\mathcal{C}_{P^g_b}}(vd)=ls$ and $\textcolor{black}{\mathcal{C}_W}(vd) \neq ld^*$
    
    Update $QW$ and cluster vectors 
    
    $\mathcal{O}f := W$
    
    }

  }
  \end{algorithm}

In line 6, the crossover procedure selects the cluster with label $ld^*$ from individual $P_d^g$ as the cluster whose sum of the inner products ${Rp^{b}_{ls}}^T Rp^{d}_{ld}$ and ${Rn^{b}_{ls}}^T Rn^{d}_{ld}$ is the maximum amongst all $ld \in \{1,\dots,k\}$, according to Equation \eqref{eq:crossover_ld}.

 \begin{equation}
 ld^* = {\arg\max}_{ld \in \{1,\dots,k\}} ( {Rp^{b}_{ls}}^T Rp^{d}_{ld} + {Rn^{b}_{ls}}^T Rn^{d}_{ld})
 \label{eq:crossover_ld}
 \end{equation}

\begin{figure}[!htb]
 \subfigure[fig:][Positive cluster vectors]{\includegraphics[width=0.49\textwidth]{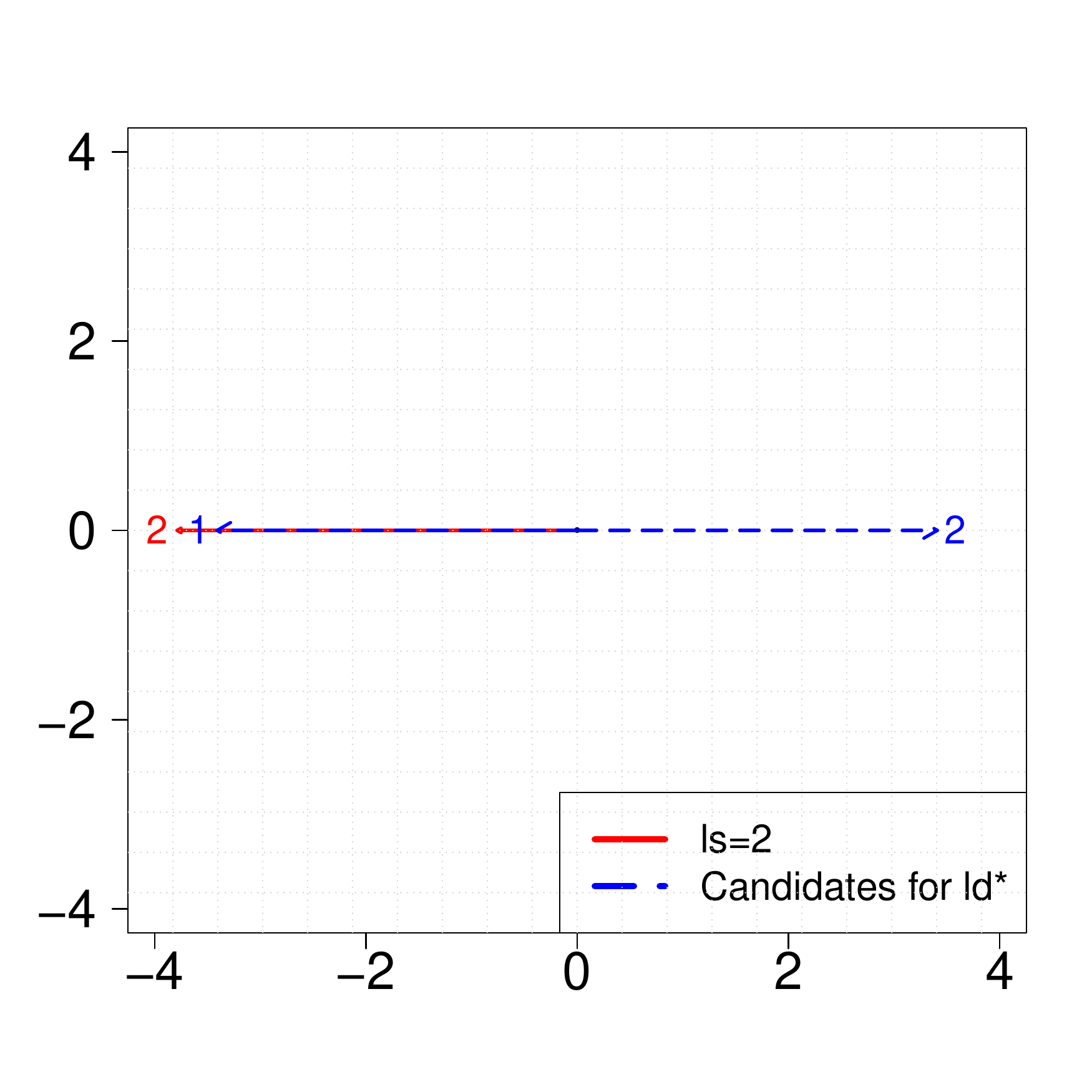}\label{fig:geometric_P5_crossover_pos}}
 \subfigure[fig:][Negative cluster vectors]{\includegraphics[width=0.49\textwidth]{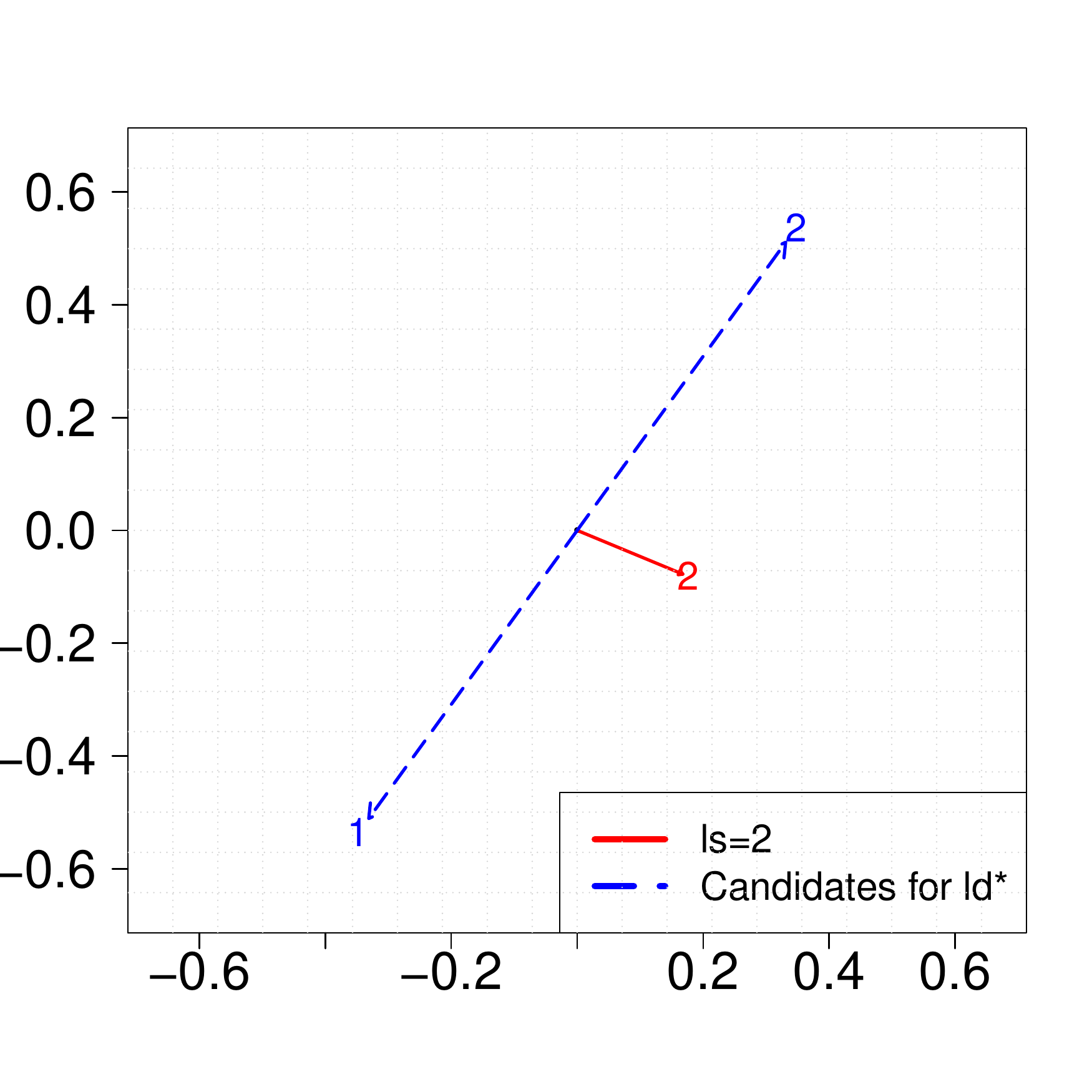}\label{fig:geometric_P5_crossover_neg}}
 \caption{Example of the selection of $ld^*$ performed by the crossover procedure in the Karate network.}
 \label{fig:geometric_P5_crossover}
 \end{figure}

Figure \ref{fig:geometric_P5_crossover} shows an example of the selection performed in line 6 of Algorithm \ref{alg:crossover}. It illustrates the cluster vector with label $ls=2$ in individual $P_b^g$, as a solid red line, and the cluster vectors with labels $ld \in \{1,2\}$ in individual $P_d^g$ -- candidates to $ld^*$ -- as dashed lines. The positive and negative vectors are identified by the label of the clusters and are shown in Figures \ref{fig:geometric_P5_crossover_pos} and \ref{fig:geometric_P5_crossover_neg}, respectively. The conjecture that justifies the selection choice is that the clusters whose vectors point to the same direction have more vertices in common. In this example, the cluster with label $ld^*=1$ from individual $P_d^g$ is selected because \textcolor{black}{${Rp^{b}_{ls=2}}^T Rp^{d}_{ld^*=1} + {Rn^{b}_{ls=2}}^T Rn^{d}_{ld^*=1}$
 is higher than ${Rp^{b}_{ls=2}}^T Rp^{d}_{ld=2} + {Rn^{b}_{ls=2}}^T Rn^{d}_{ld=2}$ in individual $P_b^g$.}

In line 7, the method moves the vertices $vd$ in the cluster labeled $ls$ in individual $P_b^g$ to cluster  labeled  $ld^*$ in individual $W$. For all $vd \in V$ already belong to the cluster labeled $ld^*$, nothing is done.
After each movement, line 8 of Algorithm \ref{alg:crossover} updates: (i) the weighted aggregate modularity $QW$, according to Equation \eqref{eq:mod_delta}, and (ii) the vectors of the clusters involved in the vertex moves in individual $W$, according to Section \ref{sec:move_vert}. After setting $W$ as the offspring individual $\mathcal{O}_f$ in line 9, the crossover returns the offspring population $\mathcal{O}=\{\mathcal{O}_1,\mathcal{O}_2,\dots,\mathcal{O}_{N\mathcal{P}}\}$.

Figure \ref{fig:example_crossover} gives an example of the crossover procedure when $\gamma_1=\gamma_2=0.5$. In the example, the offspring individual $W$ had a higher weighted aggregate modularity value than the parents $P_b^1$ and $P_d^1$. Let $ls=2$; the selection of $ld^{*}=1$ was illustrated in Figure \ref{fig:geometric_P5_crossover}. The vertices whose cluster label is $ls=2$ on individual $P_b^1$ are in bold on the partitions. At the offspring individual, which is initially a copy of $P_d^1$, these vertices are moved to the group  labeled $ld^{*}=1$, if they are not yet in this group.

\begin{figure}[htb]
\begin{center}
\small
\begin{tikzpicture}[node distance = 0.6cm, auto]
    % Place nodes\node [block] (v1) {$0$};
\node [block] (v1) {$1$};
\node [block_highlight, right=0cm of v1] (v2) {$\mathbf{2}$};
\node [block, right=0cm of v2] (v3) {$1$};
\node [block_highlight, right=0cm of v3] (v4) {$\mathbf{2}$};
\node [block_highlight, right=0cm of v4] (v5) {$\mathbf{2}$};
\node [block_highlight, right=0cm of v5] (v6) {$\mathbf{2}$};
\node [block_highlight, right=0cm of v6] (v7) {$\mathbf{2}$};
\node [block_highlight, right=0cm of v7] (v8) {$\mathbf{2}$};
\node [block, right=0cm of v8] (v9) {$1$};
\node [block, right=0cm of v9] (v10) {$1$};
\node [block_highlight, right=0cm of v10] (v11) {$\mathbf{2}$};
\node [block_highlight, right=0cm of v11] (v12) {$\mathbf{2}$};
\node [block_highlight, right=0cm of v12] (v13) {$\mathbf{2}$};
\node [block_highlight, right=0cm of v13] (v14) {$\mathbf{2}$};
\node [block, right=0cm of v14] (v15) {$1$};
\node [block, right=0cm of v15] (v16) {$1$};
\node [block_highlight, right=0cm of v16] (v17) {$\mathbf{2}$};
\node [block_highlight, right=0cm of v17] (v18) {$\mathbf{2}$};
\node [block, right=0cm of v18] (v19) {$1$};
\node [block_highlight, right=0cm of v19] (v20) {$\mathbf{2}$};
\node [block, right=0cm of v20] (v21) {$1$};
\node [block_highlight, right=0cm of v21] (v22) {$\mathbf{2}$};
\node [block, right=0cm of v22] (v23) {$1$};
\node [block, right=0cm of v23] (v24) {$1$};
\node [block, right=0cm of v24] (v25) {$1$};
\node [block, right=0cm of v25] (v26) {$1$};
\node [block, right=0cm of v26] (v27) {$1$};
\node [block, right=0cm of v27] (v28) {$1$};
\node [block, right=0cm of v28] (v29) {$1$};
\node [block, right=0cm of v29] (v30) {$1$};
\node [block, right=0cm of v30] (v31) {$1$};
\node [block, right=0cm of v31] (v32) {$1$};
\node [block, right=0cm of v32] (v33) {$1$};
\node [block, right=0cm of v33] (v34) {$1$};
\node [label, left=0.2cm of v1] (lb1) {$P_b^1$};
\node [label, below=0.2cm of v1] (lb2) {$QW(P_b^1)=0.1835$};

\node [block, below=0.2cm of lb2] (v35) {$1$};
\node [block_highlight, right=0cm of v35] (v36) {$\mathbf{2}$};
\node [block, right=0cm of v36] (v37) {$1$};
\node [block_highlight, right=0cm of v37] (v38) {$\mathbf{1}$};
\node [block_highlight, right=0cm of v38] (v39) {$\mathbf{1}$};
\node [block_highlight, right=0cm of v39] (v40) {$\mathbf{1}$};
\node [block_highlight, right=0cm of v40] (v41) {$\mathbf{1}$};
\node [block_highlight, right=0cm of v41] (v42) {$\mathbf{1}$};
\node [block, right=0cm of v42] (v43) {$1$};
\node [block, right=0cm of v43] (v44) {$1$};
\node [block_highlight, right=0cm of v44] (v45) {$\mathbf{1}$};
\node [block_highlight, right=0cm of v45] (v46) {$\mathbf{1}$};
\node [block_highlight, right=0cm of v46] (v47) {$\mathbf{1}$};
\node [block_highlight, right=0cm of v47] (v48) {$\mathbf{1}$};
\node [block, right=0cm of v48] (v49) {$2$};
\node [block, right=0cm of v49] (v50) {$2$};
\node [block_highlight, right=0cm of v50] (v51) {$\mathbf{1}$};
\node [block_highlight, right=0cm of v51] (v52) {$\mathbf{1}$};
\node [block, right=0cm of v52] (v53) {$2$};
\node [block_highlight, right=0cm of v53] (v54) {$\mathbf{1}$};
\node [block, right=0cm of v54] (v55) {$2$};
\node [block_highlight, right=0cm of v55] (v56) {$\mathbf{1}$};
\node [block, right=0cm of v56] (v57) {$2$};
\node [block, right=0cm of v57] (v58) {$2$};
\node [block, right=0cm of v58] (v59) {$2$};
\node [block, right=0cm of v59] (v60) {$2$};
\node [block, right=0cm of v60] (v61) {$2$};
\node [block, right=0cm of v61] (v62) {$2$};
\node [block, right=0cm of v62] (v63) {$2$};
\node [block, right=0cm of v63] (v64) {$2$};
\node [block, right=0cm of v64] (v65) {$2$};
\node [block, right=0cm of v65] (v66) {$2$};
\node [block, right=0cm of v66] (v67) {$2$};
\node [block, right=0cm of v67] (v68) {$2$};
\node [label, left=0.2cm of v35] (lb3) {$P_d^1$};
\node [label, below=0.2cm of v35] (lb4) {$QW(P_d^1)=0.1443$};

\node [label, below=0.4cm of lb4] (lbls) {$ls=$};
\node [block_highlight, right=0cm of lbls] (vls) {$\mathbf{2}$};
\node [label, right=0.4cm of vls] (lbld) {$ld^*=$};
\node [block_highlight, right=0cm of lbld] (vld) {$\mathbf{1}$};

\node [block, below=0.4cm of lbls] (v69) {$1$};
\node [block_highlight, right=0cm of v69] (v70) {$\mathbf{1}$};
\node [block, right=0cm of v70] (v71) {$1$};
\node [block_gray, right=0cm of v71] (v72) {$\mathbf{1}$};
\node [block_gray, right=0cm of v72] (v73) {$\mathbf{1}$};
\node [block_gray, right=0cm of v73] (v74) {$\mathbf{1}$};
\node [block_gray, right=0cm of v74] (v75) {$\mathbf{1}$};
\node [block_gray, right=0cm of v75] (v76) {$\mathbf{1}$};
\node [block, right=0cm of v76] (v77) {$1$};
\node [block, right=0cm of v77] (v78) {$1$};
\node [block_gray, right=0cm of v78] (v79) {$\mathbf{1}$};
\node [block_gray, right=0cm of v79] (v80) {$\mathbf{1}$};
\node [block_gray, right=0cm of v80] (v81) {$\mathbf{1}$};
\node [block_gray, right=0cm of v81] (v82) {$\mathbf{1}$};
\node [block, right=0cm of v82] (v83) {$2$};
\node [block, right=0cm of v83] (v84) {$2$};
\node [block_gray, right=0cm of v84] (v85) {$\mathbf{1}$};
\node [block_gray, right=0cm of v85] (v86) {$\mathbf{1}$};
\node [block, right=0cm of v86] (v87) {$2$};
\node [block_gray, right=0cm of v87] (v88) {$\mathbf{1}$};
\node [block, right=0cm of v88] (v89) {$2$};
\node [block_gray, right=0cm of v89] (v90) {$\mathbf{1}$};
\node [block, right=0cm of v90] (v91) {$2$};
\node [block, right=0cm of v91] (v92) {$2$};
\node [block, right=0cm of v92] (v93) {$2$};
\node [block, right=0cm of v93] (v94) {$2$};
\node [block, right=0cm of v94] (v95) {$2$};
\node [block, right=0cm of v95] (v96) {$2$};
\node [block, right=0cm of v96] (v97) {$2$};
\node [block, right=0cm of v97] (v98) {$2$};
\node [block, right=0cm of v98] (v99) {$2$};
\node [block, right=0cm of v99] (v100) {$2$};
\node [block, right=0cm of v100] (v101) {$2$};
\node [block, right=0cm of v101] (v102) {$2$};
\node [label, left=0.2cm of v69] (lb5) {$W$};
\node [label, below=0.2cm of v69] (lb6) {$QW(W)=0.1868$};

\end{tikzpicture}
\caption{Example of the crossover procedure in a partition of the Karate network for $\gamma_1=\gamma_2=0.5$.}
\label{fig:example_crossover}
\end{center}
\end{figure}
%- partitions from $v=\{10,11,\dots,20\}

%%%%%%%%%%%%%%%%%%%%%%%%%%%%%%%%%%%%%%%%%%%%%%%%%%%%%%
\subsubsection{\textcolor{black}{Mutation}}
%%%%%%%%%%%%%%%%%%%%%%%%%%%%%%%%%%%%%%%%%%%%%%%%%%%%%%

Algorithm \ref{alg:mutation} presents the mutation procedure  whose inputs are: the offspring population, $\mathcal{O}$; $k$; $rp^i$ and $rn^i$, $\forall i \in V$.
In line 1, a random integer number in the interval $[1, \lfloor\frac{n}{2}\rfloor]$ is assigned to $count$, which indicates the number of mutations. In line 2, an individual $O_d$ is randomly selected from $\mathcal{O}$. In line 3, the algorithm picks  $count$ vertices from $V$ at random to define the set of vertices to be mutated, $V^{'}$. 
 Each vertex $vd \in V^{'}$ is assigned to a cluster $C_{r}$ chosen randomly from individual $O_d$, in lines 5 and 6. Note that if a cluster $C_t$ is empty, $vd$ will be assigned to a new cluster.
After each movement of a vertex $vd$, both $QW$ and the vectors of cluster $C_r$ are updated in line 8. The mutation procedure halts when all vertices of $V^{'}$ have been mutated and, then, returns the updated offspring $\mathcal{O}$.

 \begin{algorithm}[!htb]
 {
 \caption{\emph{Mutation}}
 \label{alg:mutation}
 \SetKwInOut{Input}{Input}
 \SetKwInOut{Output}{Output}
 
\Input{$\mathcal{O}$, $k$, $rp^i$ and $rn^i$, $\forall i \in V$}
\Output{Updated $\mathcal{O}$}
  
   Randomly choose an integer number $count$ from interval $[1, \lfloor\frac{n}{2}]\rfloor$ with    uniform probability distribution 

    Randomly select an individual $\mathcal{O}_d$ from $\mathcal{O}$

    Randomly pick $count$ distinct elements from $V$ and assign them to $V^{'}$
    
    \While{$V^{'}\neq \emptyset$ }{
        
        Randomly choose $C_{r}$ from the $k$ possible clusters of $\mathcal{O}_d$
        
       Move a $vd \in V^{'}$ to cluster $C_{r}$, if  $\textcolor{black}{\mathcal{C}_{O_d}}(vd) \neq r$
       
       $V^{'} := V^{'} \backslash \{vd\}$

        Update $QW$ and vectors of cluster $C_r$
    }
  }
  \end{algorithm}

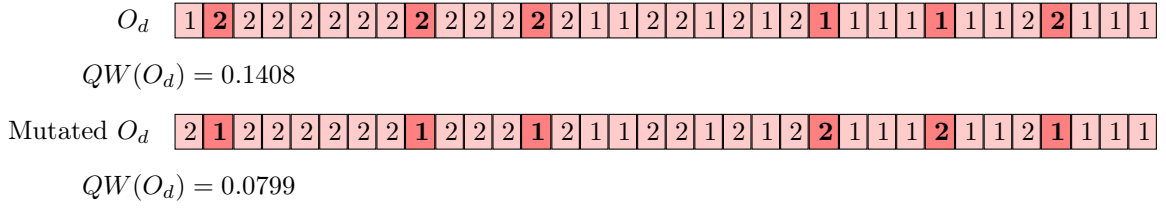
\begin{figure}[htb]
\begin{center}
\small
\begin{tikzpicture}[node distance = 0.6cm, auto]
    % Place nodes
    
\node [block] (v1) {$1$};
\node [block_highlight, right=0cm of v1] (v2) {$\mathbf{2}$};
\node [block, right=0cm of v2] (v3) {$2$};
\node [block, right=0cm of v3] (v4) {$2$};
\node [block, right=0cm of v4] (v5) {$2$};
\node [block, right=0cm of v5] (v6) {$2$};
\node [block, right=0cm of v6] (v7) {$2$};
\node [block, right=0cm of v7] (v8) {$2$};
\node [block_highlight, right=0cm of v8] (v9) {$\mathbf{2}$};
\node [block, right=0cm of v9] (v10) {$2$};
\node [block, right=0cm of v10] (v11) {$2$};
\node [block, right=0cm of v11] (v12) {$2$};
\node [block_highlight, right=0cm of v12] (v13) {$\mathbf{2}$};
\node [block, right=0cm of v13] (v14) {$2$};
\node [block, right=0cm of v14] (v15) {$1$};
\node [block, right=0cm of v15] (v16) {$1$};
\node [block, right=0cm of v16] (v17) {$2$};
\node [block, right=0cm of v17] (v18) {$2$};
\node [block, right=0cm of v18] (v19) {$1$};
\node [block, right=0cm of v19] (v20) {$2$};
\node [block, right=0cm of v20] (v21) {$1$};
\node [block, right=0cm of v21] (v22) {$2$};
\node [block_highlight, right=0cm of v22] (v23) {$\mathbf{1}$};
\node [block, right=0cm of v23] (v24) {$1$};
\node [block, right=0cm of v24] (v25) {$1$};
\node [block, right=0cm of v25] (v26) {$1$};
\node [block_highlight, right=0cm of v26] (v27) {$\mathbf{1}$};
\node [block, right=0cm of v27] (v28) {$1$};
\node [block, right=0cm of v28] (v29) {$1$};
\node [block, right=0cm of v29] (v30) {$2$};
\node [block_highlight, right=0cm of v30] (v31) {$\mathbf{2}$};
\node [block, right=0cm of v31] (v32) {$1$};
\node [block, right=0cm of v32] (v33) {$1$};
\node [block, right=0cm of v33] (v34) {$1$};
\node [label, left=0.2cm of v1] (lb1) {$O_d$};
\node [label, below=0.2cm of v1] (lb2) {$QW(O_d)=0.1408$};

\node [block, below=0.2cm of lb2] (v35) {$2$};
\node [block_highlight, right=0cm of v35] (v36) {$\mathbf{1}$};
\node [block, right=0cm of v36] (v37) {$2$};
\node [block, right=0cm of v37] (v38) {$2$};
\node [block, right=0cm of v38] (v39) {$2$};
\node [block, right=0cm of v39] (v40) {$2$};
\node [block, right=0cm of v40] (v41) {$2$};
\node [block, right=0cm of v41] (v42) {$2$};
\node [block_highlight, right=0cm of v42] (v43) {$\mathbf{1}$};
\node [block, right=0cm of v43] (v44) {$2$};
\node [block, right=0cm of v44] (v45) {$2$};
\node [block, right=0cm of v45] (v46) {$2$};
\node [block_highlight, right=0cm of v46] (v47) {$\mathbf{1}$};
\node [block, right=0cm of v47] (v48) {$2$};
\node [block, right=0cm of v48] (v49) {$1$};
\node [block, right=0cm of v49] (v50) {$1$};
\node [block, right=0cm of v50] (v51) {$2$};
\node [block, right=0cm of v51] (v52) {$2$};
\node [block, right=0cm of v52] (v53) {$1$};
\node [block, right=0cm of v53] (v54) {$2$};
\node [block, right=0cm of v54] (v55) {$1$};
\node [block, right=0cm of v55] (v56) {$2$};
\node [block_highlight, right=0cm of v56] (v57) {$\mathbf{2}$};
\node [block, right=0cm of v57] (v58) {$1$};
\node [block, right=0cm of v58] (v59) {$1$};
\node [block, right=0cm of v59] (v60) {$1$};
\node [block_highlight, right=0cm of v60] (v61) {$\mathbf{2}$};
\node [block, right=0cm of v61] (v62) {$1$};
\node [block, right=0cm of v62] (v63) {$1$};
\node [block, right=0cm of v63] (v64) {$2$};
\node [block_highlight, right=0cm of v64] (v65) {$\mathbf{1}$};
\node [block, right=0cm of v65] (v66) {$1$};
\node [block, right=0cm of v66] (v67) {$1$};
\node [block, right=0cm of v67] (v68) {$1$};
\node [label, left=0.2cm of v35] (lb3) {Mutated $O_d$};
\node [label, below=0.2cm of v35] (lb4) {$QW(O_d)=0.0799$};

    %\path[line] (v1) --node{$v$ was retweeted by $u$ } (v2);
    %path[line, above=2cm] (v1) --node{$v$ was mentioned by $z$ } (v3);
\end{tikzpicture}
\caption{Example of the mutation procedure on an individual that decodes a solution for the Karate network when  $\gamma_1=\gamma_2=0.5$.}
\label{fig:example_mutation}
\end{center}
\end{figure}

Figure \ref{fig:example_mutation} presents an example of the mutation procedure on an individual $O_d$ generated to decode a solution for the Karate network. The mutated individual $O_d$ is a result of the modification of the labels of $6$ randomly selected vertices.

%%%%%%%%%%%%%%%%%%%%%%%%%%%%%%%%%%%%%%%%%%%%%%%%%%%%%%
\subsubsection{\textcolor{black}{Local search}}
%%%%%%%%%%%%%%%%%%%%%%%%%%%%%%%%%%%%%%%%%%%%%%%%%%%%%%

Algorithm \ref{alg:local_search} shows the local search procedure of the introduced \emph{Memetic Algorithm}, which has as input $\mathcal{O}$, $IT$, $k$, $rp^i$ and $rn^i$, $\forall i \in V$.  \textcolor{black}{Each iteration  of the local search  attempts to improve the modularity value of individuals of offspring $\mathcal{O}$ by moving vertices to different communities.}
In line 4, for each individual $O_d \in \mathcal{O}$ and each vertex $vi\in V$, the local search selects the label $t^*$ such that the relocation of $vi$ to cluster $C_{t^*}$ will result in the largest modularity gain. In line 5, $vi$ is assigned to cluster $C_{t^*}$, if it does not belong to it yet. After moving $vi$, in line 6 of the algorithm, $QW$, the vectors of cluster $C_{t^*}$ and the vector of the cluster where $vi$ was found before being moved are updated.   Algorithm \ref{alg:local_search} returns the improved offspring $\mathcal{O}$.

 \begin{algorithm}[!htb]
  {
  \caption{Local Search}
  \label{alg:local_search}
  \SetKwInOut{Input}{Input}
  \SetKwInOut{Output}{Output}
      \Input{$\mathcal{O}$, $IT$, $k$, $rp^i$ and $rn^i$, $\forall i \in V$}
    \Output{$\mathcal{O}$}
    
    \For{$it=1$ to $IT$}{
    \For{$O_d \in \mathcal{O}$}{
        \For{$vi \in V$}{
        
          Select $t^*$ according to Equation \eqref{eq:modMovMax}
            
        Move $vi$ to cluster $C_{t^*}$ if  $\textcolor{black}{\mathcal{C}_{O_d}}(vi) \neq t^*$
        
        Update $QW$ and the vectors of cluster $C_{t^*}$
        }
    }
    }
  }
  \end{algorithm}

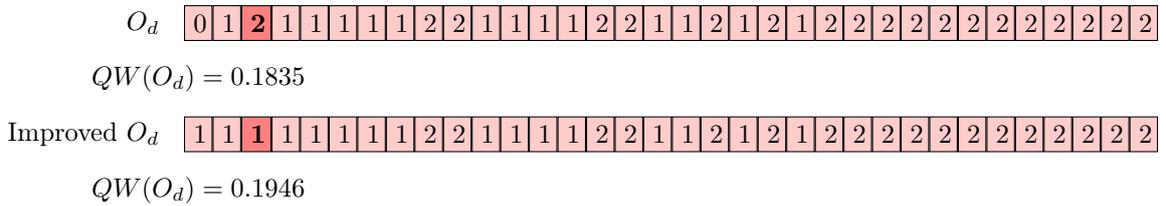
\begin{figure}[htb]
\begin{center}
\small
\begin{tikzpicture}[node distance = 0.6cm, auto]
\node [block] (v1) {$0$};
\node [block, right=0cm of v1] (v2) {$1$};
\node [block_highlight, right=0cm of v2] (v3) {$\mathbf{2}$};
\node [block, right=0cm of v3] (v4) {$1$};
\node [block, right=0cm of v4] (v5) {$1$};
\node [block, right=0cm of v5] (v6) {$1$};
\node [block, right=0cm of v6] (v7) {$1$};
\node [block, right=0cm of v7] (v8) {$1$};
\node [block, right=0cm of v8] (v9) {$2$};
\node [block, right=0cm of v9] (v10) {$2$};
\node [block, right=0cm of v10] (v11) {$1$};
\node [block, right=0cm of v11] (v12) {$1$};
\node [block, right=0cm of v12] (v13) {$1$};
\node [block, right=0cm of v13] (v14) {$1$};
\node [block, right=0cm of v14] (v15) {$2$};
\node [block, right=0cm of v15] (v16) {$2$};
\node [block, right=0cm of v16] (v17) {$1$};
\node [block, right=0cm of v17] (v18) {$1$};
\node [block, right=0cm of v18] (v19) {$2$};
\node [block, right=0cm of v19] (v20) {$1$};
\node [block, right=0cm of v20] (v21) {$2$};
\node [block, right=0cm of v21] (v22) {$1$};
\node [block, right=0cm of v22] (v23) {$2$};
\node [block, right=0cm of v23] (v24) {$2$};
\node [block, right=0cm of v24] (v25) {$2$};
\node [block, right=0cm of v25] (v26) {$2$};
\node [block, right=0cm of v26] (v27) {$2$};
\node [block, right=0cm of v27] (v28) {$2$};
\node [block, right=0cm of v28] (v29) {$2$};
\node [block, right=0cm of v29] (v30) {$2$};
\node [block, right=0cm of v30] (v31) {$2$};
\node [block, right=0cm of v31] (v32) {$2$};
\node [block, right=0cm of v32] (v33) {$2$};
\node [block, right=0cm of v33] (v34) {$2$};
\node [label, left=0.2cm of v1] (lb1) {$O_d$};
\node [label, below=0.2cm of v1] (lb2) {$QW(O_d)=0.1835$};

\node [block, below=0.2cm of lb2] (v35) {$1$};
\node [block, right=0cm of v35] (v36) {$1$};
\node [block_highlight, right=0cm of v36] (v37) {$\mathbf{1}$};
\node [block, right=0cm of v37] (v38) {$1$};
\node [block, right=0cm of v38] (v39) {$1$};
\node [block, right=0cm of v39] (v40) {$1$};
\node [block, right=0cm of v40] (v41) {$1$};
\node [block, right=0cm of v41] (v42) {$1$};
\node [block, right=0cm of v42] (v43) {$2$};
\node [block, right=0cm of v43] (v44) {$2$};
\node [block, right=0cm of v44] (v45) {$1$};
\node [block, right=0cm of v45] (v46) {$1$};
\node [block, right=0cm of v46] (v47) {$1$};
\node [block, right=0cm of v47] (v48) {$1$};
\node [block, right=0cm of v48] (v49) {$2$};
\node [block, right=0cm of v49] (v50) {$2$};
\node [block, right=0cm of v50] (v51) {$1$};
\node [block, right=0cm of v51] (v52) {$1$};
\node [block, right=0cm of v52] (v53) {$2$};
\node [block, right=0cm of v53] (v54) {$1$};
\node [block, right=0cm of v54] (v55) {$2$};
\node [block, right=0cm of v55] (v56) {$1$};
\node [block, right=0cm of v56] (v57) {$2$};
\node [block, right=0cm of v57] (v58) {$2$};
\node [block, right=0cm of v58] (v59) {$2$};
\node [block, right=0cm of v59] (v60) {$2$};
\node [block, right=0cm of v60] (v61) {$2$};
\node [block, right=0cm of v61] (v62) {$2$};
\node [block, right=0cm of v62] (v63) {$2$};
\node [block, right=0cm of v63] (v64) {$2$};
\node [block, right=0cm of v64] (v65) {$2$};
\node [block, right=0cm of v65] (v66) {$2$};
\node [block, right=0cm of v66] (v67) {$2$};
\node [block, right=0cm of v67] (v68) {$2$};
\node [label, left=0.2cm of v35] (lb3) {Improved $O_d$};
\node [label, below=0.2cm of v35] (lb4) {$QW(O_d)=0.1946$};

\end{tikzpicture}
\caption{Example of the local search applied to a solution for the Karate network when $\gamma_1=\gamma_2=0.5$.}
\label{fig:example_local_search}
\end{center}
\end{figure}

Figure \ref{fig:example_local_search} illustrates the local search procedure on an individual of the offspring. In this figure, the procedure improved the weighted aggregate modularity of an individual $O_d$ by moving a single vertex to a different cluster.

%%%%%%%%%%%%%%%%%%%%%%%%%%%%%%%%%%%%%%%%%%%%%%%%%%%%%%
\subsection{Ensemble algorithm}
%%%%%%%%%%%%%%%%%%%%%%%%%%%%%%%%%%%%%%%%%%%%%%%%%%%%%%

This section  introduces an ensemble algorithm that uses information of  partitions obtained by \emph{MOSpecG} to find a single partition that best captures the community structure of a network. Algorithm \ref{alg:ensemble} presents the proposed ensemble algorithm, called \emph{SpecG-EC}. The algorithm has as input $G$; $N\mathcal{G}$; $N\mathcal{P}$; $N\mathcal{O}$; $p$; $IT$; a set of partitions, $\mathcal{F}$, and a required threshold, $\tau$. A partition from $\mathcal{F}$ is identified by $\mathcal{F}_i, i=\{1,\dots,N\mathcal{F}\}$, and represents the solution achieved by \emph{MOSpecG} for $\gamma_1=(i-1)\frac{1}{N\mathcal{F}-1}, \gamma_2=1-\gamma_1$.

\begin{algorithm}[!htb]
{
 \caption{\emph{SpecG-EC}}
  \label{alg:ensemble}
  \SetKwInOut{Input}{Input}
  \SetKwInOut{Output}{Output}
      \Input{$G$, $N\mathcal{G}$, $N\mathcal{P}$,  $N\mathcal{O}$, $p$, $IT$, $\mathcal{F}$ and $\tau$}
    \Output{Ensemble partition $EP$}
    
    $\mathcal{F}' := \mathcal{F}\backslash\{\mathcal{F}_1,\mathcal{F}_{N\mathcal{F}}\}$

    $e_{ij} := $ number of times that $\mathcal{F}_a'(i)=\mathcal{F}_b'(j), \forall \mathcal{F}_a',\mathcal{F}'_b \in \mathcal{F}', \forall i,j \in V$
    
    $\mathcal{E} := \frac{\mathcal{E}}{|\mathcal{F^{'}}|}$
      
     $e_{ij} = 0$, if $e_{ij} < \tau$, $\forall i,j\in V$ 
     
     $A := A + \mathcal{E}$
  
    $\Lambda \mathcal{E}, U\mathcal{E} :=$ Eigen-decomposition of $B$ regarding the $p$ largest eigenvalues in absolute value
    
    $\chi := \max_{\Lambda  \mathcal{E}_{ii},\forall i}( \Lambda  \mathcal{E}_{ii})$ 
 
    $k': =$ number of eigenvalues of $B$ with values larger than or equal to $\sqrt{\chi}$

    $k :=\lfloor 1.25 k'\rfloor$

         Define vertex vectors $rp^i$ and $rn^i$, $i \in V$ 
    
    $EP :=$ \emph{Memetic Algorithm}($N\mathcal{G},\mathcal{N}{P}, \mathcal{N}O, IT, k, rp^i, rn^i, \forall i \in V$) 
  }
\end{algorithm}
 
Line 1 of Algorithm \ref{alg:ensemble} assigns to $\mathcal{F}'$ every solution from $\mathcal{F}$ except those by \emph{MOSpecG} for the pair of values $\gamma_1=0,\gamma_2=1$ and $\gamma_1=1,\gamma_2=0$, i.e., $\mathcal{F}_1$ and $\mathcal{F}_{N\mathcal{F}}$.
Let $\mathcal{E}=[e_{ij}] \in \mathds{R}^{n \times n}$ be a consensus matrix. 
In lines 2 to 4, $\mathcal{E}$ is defined according to \citet{Lancichinetti2012}: in line 2, $e_{ij}$ receives the number of times that vertices $i$ and $j$ appear in the same cluster in the partitions from $\mathcal{F}'$; in line 3, matrix $\mathcal{E}$ is normalized; and, in line 4, elements from $\mathcal{E}$ below a threshold $\tau$ are set to $0$ to avoid noisy data. \textcolor{black}{In particular, the step described in line 4 is skipped  for  $i',j' \in V$, where $j' = \arg\max_{j} e_{i'j}$ and $e_{i'j'} < \tau$.}

In order to favor the grouping of vertices that are in the same cluster in the majority of the partitions from $\mathcal{F}'$, in line 5, the ensemble algorithm adjusts the original graph by adding the consensus matrix to the adjacency matrix.

The ensemble algorithm calculates the eigenvalues and eigenvectors of the original modularity matrix $B$, in line 6. It estimates the number of clusters, $k$, in lines 7 to 9, according to Section \ref{sec:alg_number_k}. In line 10, the vertex vectors $rp^i$ and $rn^i$, $i \in V$, are created from the eigenvalues and eigenvectors of $B$. Finally, \emph{SpecG-EC} calls Algorithm \ref{alg:memetic} to find the partition $EP$ that maximizes the modularity of the adjusted graph in line 11. The ensemble algorithm returns $EP$.

%%%%%%%%%%%%%%%%%%%%%%%%%%%%%%%%%%%%%%%%%%%%%%%%%%%%%%
\section{Computational Experiments} \label{sec:experiments}
%%%%%%%%%%%%%%%%%%%%%%%%%%%%%%%%%%%%%%%%%%%%%%%%%%%%%%

This section discusses the computational experiments performed with \emph{MOSpecG} in real and artificial networks. In this section, we refer to \emph{MOSpecG} for maximizing modularity, i.e., with $\gamma_1=\gamma_2=0.5$, as \emph{MOSpecG-mod}. \textcolor{black}{Both \emph{SpecG-EC} and \emph{MOSpecG}} were implemented in C++ using the ARPACK++ library\footnote{\textcolor{black}{The source code is available at \url{https://github.com/camilapsan/MOSpecG_SpecG}.}}\textcolor{black}{. The following values of the parameters were defined in the algorithms after preliminary tests, reported in  \ref{sec:appendix}: $N\mathcal{F}=11$,  $\tau=0.5$, $N\mathcal{G}=50$, $N\mathcal{P}=5$, $N\mathcal{O}=40$ and $p=\lfloor 0.1 n \rfloor$. A single parameter was valued differently in experiments with real networks and artificial networks, which is the number of iterations of the local search,  $IT$.  In the experiments with real networks, the value of $IT$ was 5, whereas in the experiments with artificial networks, which are much larger than the real networks, $IT$ was valued 1. All} the experiments were run on a computer with an Intel Core i7-4790S processor with 3.20GHz and 8GB of main memory.

\textcolor{black}{The experiments are divided into two parts, each of them with two experiments.  The first experiment of the first part shows the results obtained by \emph{MOSpecG} with  real networks. In this experiment, we present the results including the dominated solutions obtained by \emph{MOSpecG} because some of them had good NMI values. Therefore, we  refer to the sets of solutions found by  \emph{MOSpecG} as \textit{solution sets} rather than Pareto frontier approximations. In the second experiment of the first part, also with real networks,  we contrasted the results achieved by \emph{SpecG-EC} and \emph{MOSpecG-mod} with those found by the reference graph clustering algorithms: Moga-Net, OSLOM and Infomap.  Artificial networks were used in the second part of the computational tests. In the first experiment of the second part,  we again present the results  achieved by \emph{MOSpecG}. In the second experiment, we  compared the results achieved by \emph{SpecG-EC} and \emph{MOSpecG-mod} with those obtained by  Moga-Net, OSLOM and Infomap.  The codes of the reference algorithms used in the experiments are those provided in the authors' website. }

\textcolor{black}{In all experiments,} the expected partitions of the tested networks are known. Thereby, to evaluate the correlation between the solutions found by the algorithms and the ground truth partitions, we used the measure Normalized Mutual Information (NMI) \citep{Shannon1948}. The NMI values lie in the range $[0,1]$ and the higher they are, the more correlated is the pair of partitions.

%%%%%%%%%%%%%%%%%%%%%%%%%%%%%%%%%%%%%%%%%%%%%%%%%%%%%% 
 \subsection{\textcolor{black}{Experiments with real networks}}
%%%%%%%%%%%%%%%%%%%%%%%%%%%%%%%%%%%%%%%%%%%%%%%%%%%%%% 
 
This section shows the results of the experiments with the real benchmark networks: Karate \citep{Zachary1977}, Dolphins \citep{Lusseau2003}, Polbooks \citep{Krebs2008} and Football \citep{Girvan2002}. \textcolor{black}{Table \ref{tab:exp_mo_real_number} presents the number of vertices and edges in these networks.}

 \begin{table*}[!htb]
      \caption{Number of vertices and edges in the real benchmark networks.}
   \centering
     \setlength{\tabcolsep}{6pt} 
     \begin{tabular}{|c|c|c|c|c|c|}
 \cline{3-6}
 \multicolumn{2}{c}{} & \multicolumn{4}{|c|}{Network}\\ \cline{3-6}
  \multicolumn{2}{c|}{}  & Karate&	Dolphins&	Polbooks&	Football\\   \cline{1-6}
\multicolumn{2}{|c|}{Number of vertices} &	34&	62&	105&	115\\ 
\multicolumn{2}{|c|}{Number of edges} &	78&	159&	441&	613\\  \hline
     \end{tabular}
     \label{tab:exp_mo_real_number}
 \end{table*}

\subsubsection{\textcolor{black}{Solution sets found by MOSpecG}}

\begin{figure}[!htb]
 \subfigure[fig:karatepareto][Karate network.]{\includegraphics[width=0.49\textwidth]{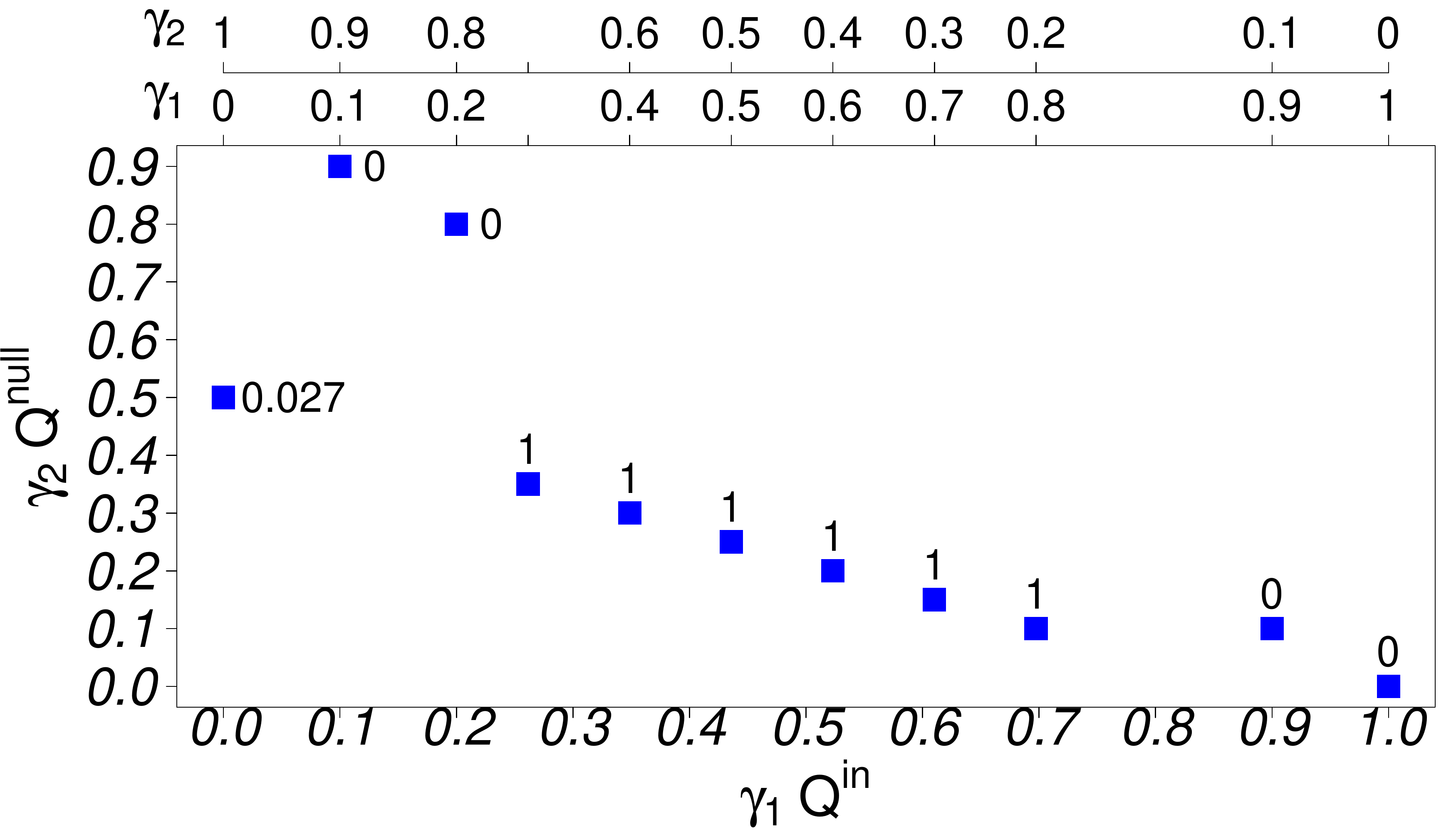}\label{fig:exp_pareto_real_karate}}
 \subfigure[fig:dolphinspareto][Dolphins network.]{\includegraphics[width=0.49\textwidth]{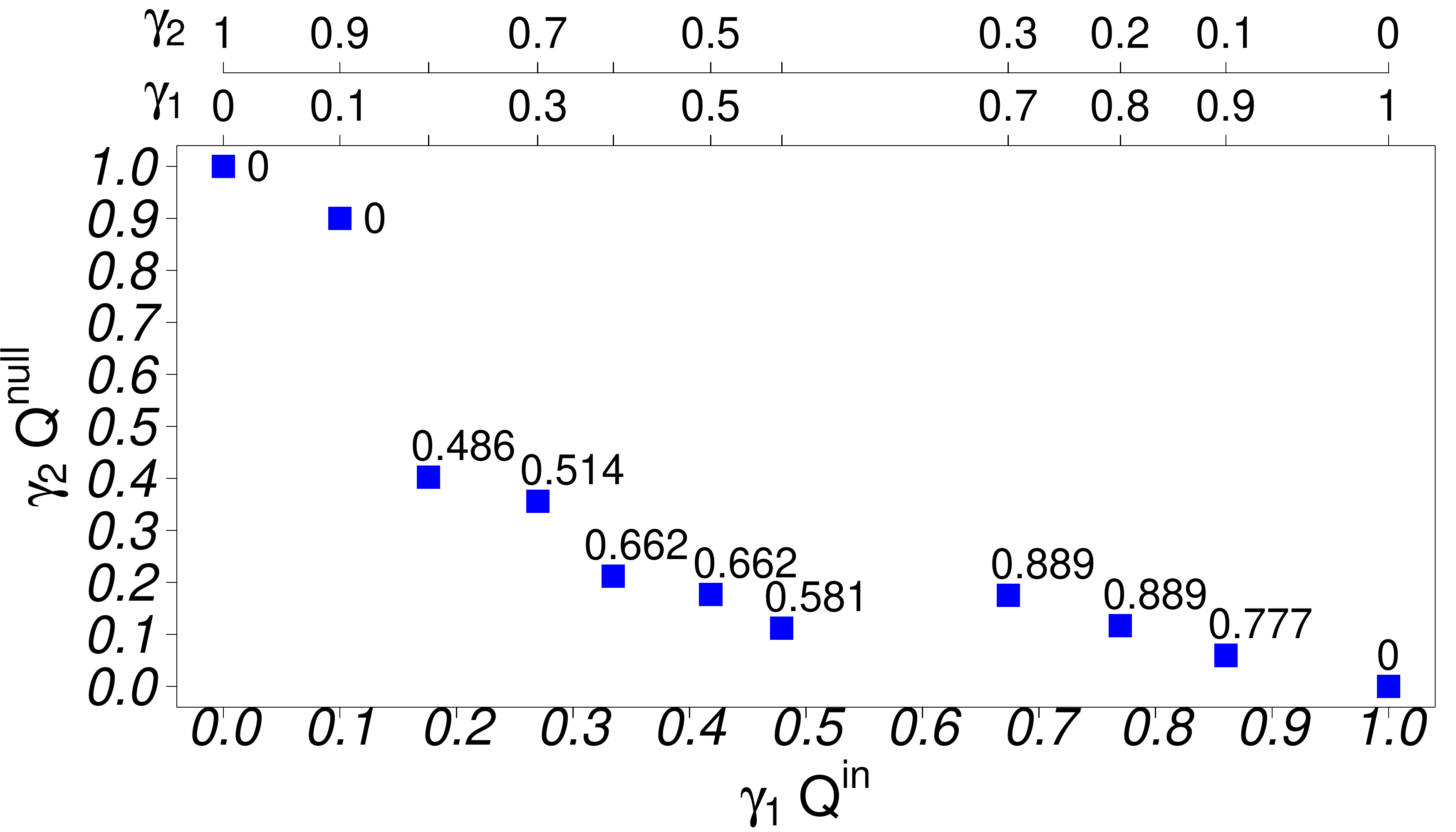}\label{fig:exp_pareto_real_dolphins}}
  \subfigure[fig:polbookspareto][Polbooks network.]{\includegraphics[width=0.49\textwidth]{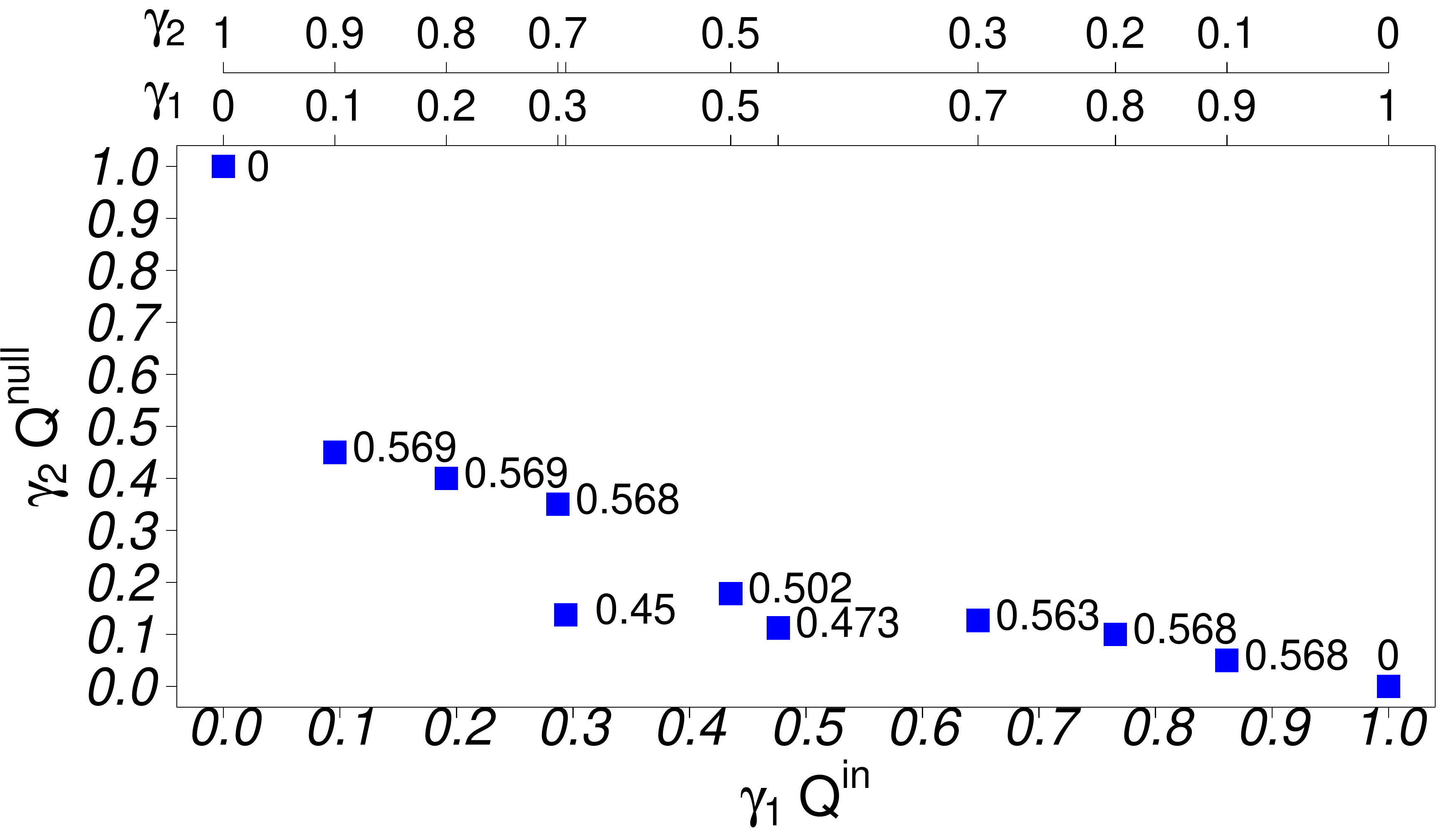}\label{fig:exp_pareto_real_polbooks}}
   \subfigure[fig:footballpareto][Football network.]{\includegraphics[width=0.49\textwidth]{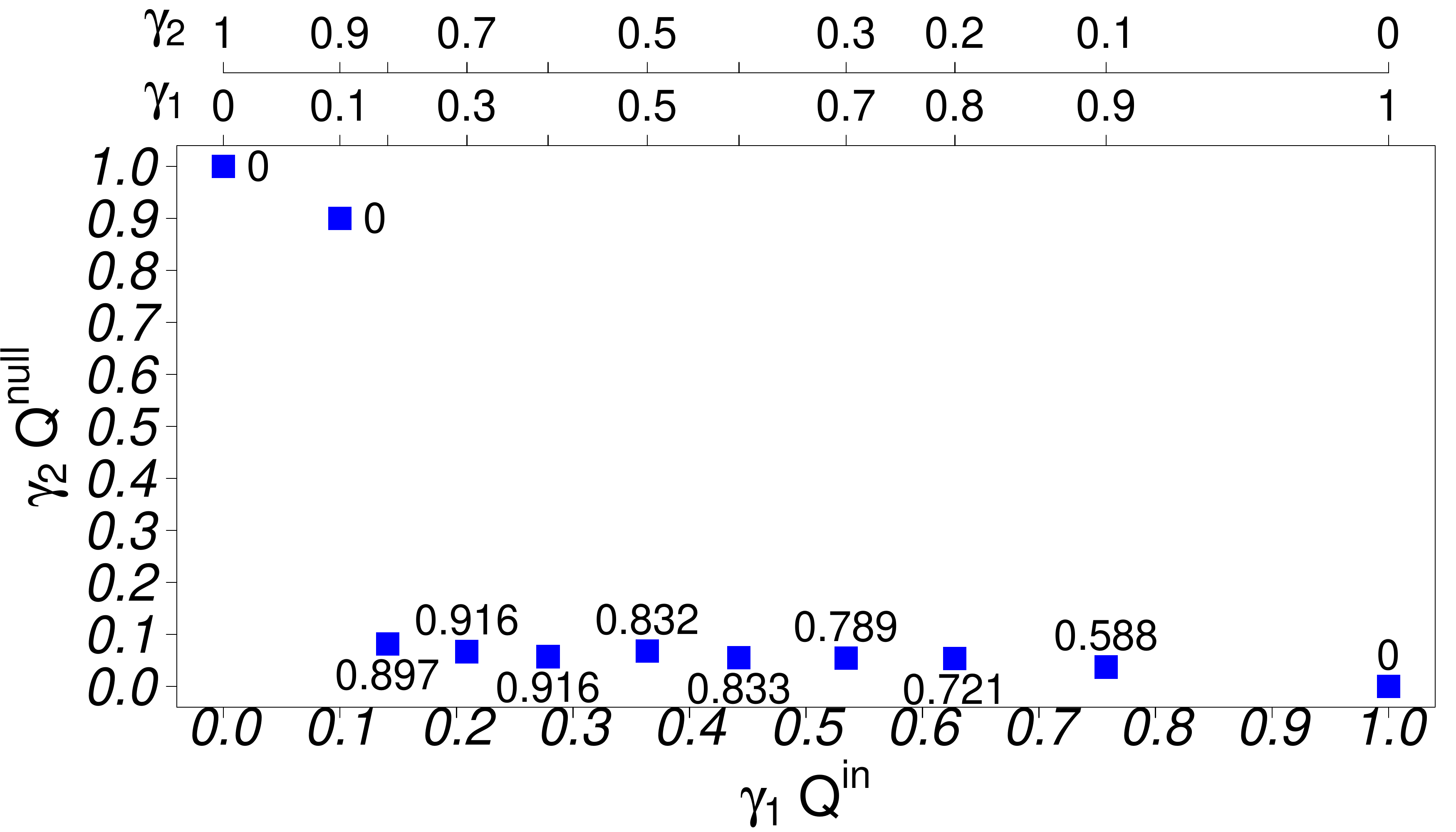}\label{fig:exp_pareto_real_football}}
 \caption{\textit{Solution sets}.}
 \label{fig:exp_pareto_real}
 \end{figure}

Figure \ref{fig:exp_pareto_real} exhibits the \textit{solution sets} achieved by a single execution of \emph{MOSpecG} for the real benchmark networks. This figure illustrates the trade-offs between the two conflicting objectives. 
Each point is labeled with the NMI value achieved by the corresponding partitions.

\emph{MOSpecG} was able to correctly identify the expected partitions of the Karate network for $\gamma_1 \in \{0.3,0.4,\dots,0.8\}$ and $\gamma_2 \in \{0.2,0.3,\dots,0.7\}$. On the one hand, \emph{MOSpecG} achieved the highest NMI values for the Dolphins network when $\gamma_1Q^{IN}>\gamma_2Q^{NULL}$ and $\gamma_1 > \gamma_2$. On the other hand, it achieved the highest NMI values for the Football and Polbooks networks when $\gamma_2Q^{NULL}>\gamma_1Q^{IN}$ and $\gamma_2>\gamma_1$.

%-----------------------------------------------------

\begin{figure}[!htb]
 \center
 \includegraphics[width=0.32\textwidth]{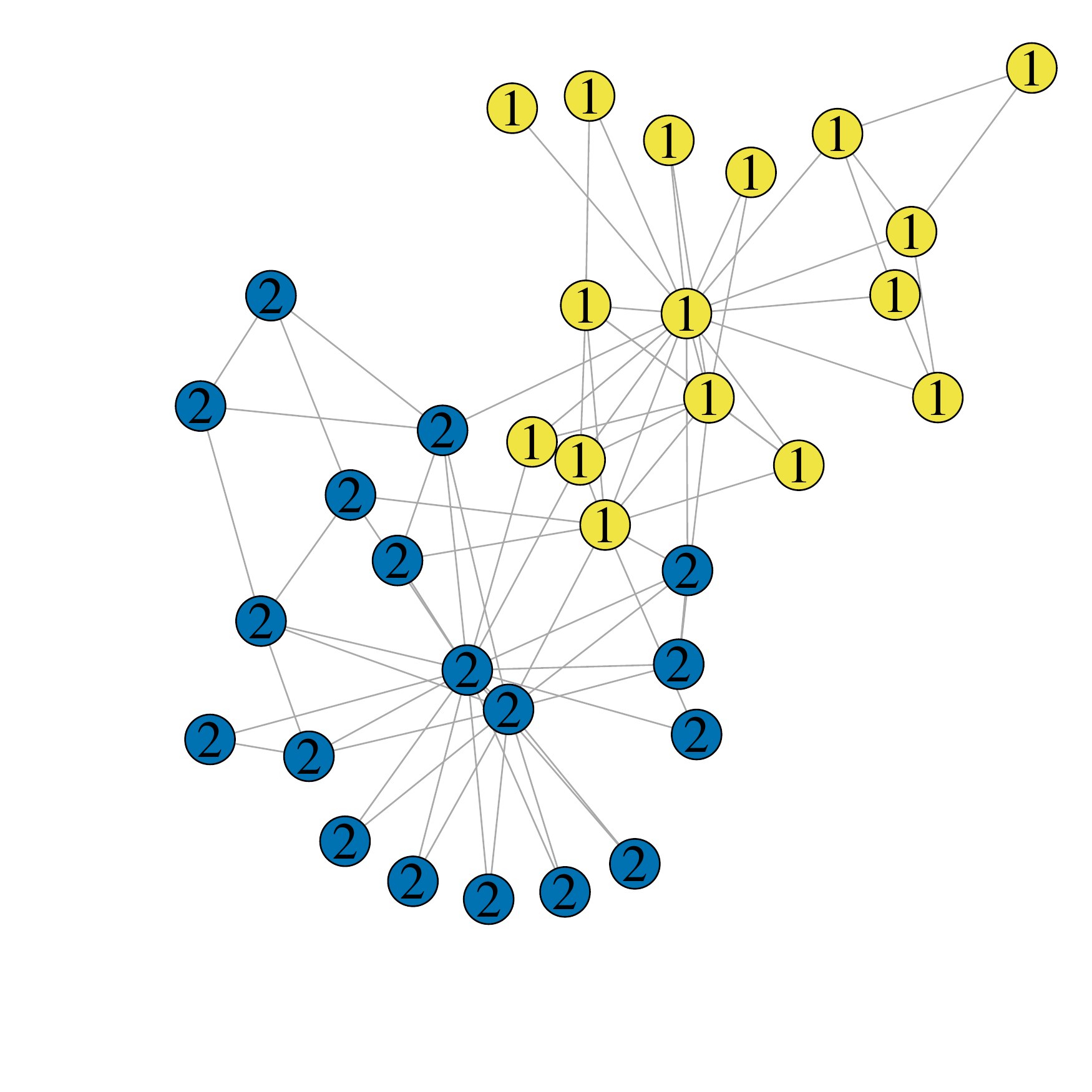}
 \caption{The partition found by \emph{SpecG-EC}, \emph{MOSpecG-mod} and \emph{MOSpecG} with $\gamma_1 \in \{0.3,0.4,\dots,0.8\}$ and $\gamma_2 \in \{0.2,0.3,\dots,0.7\}$ for the Karate network: NMI value of $1$.}
 \label{fig:exp_ens_karate}
 \end{figure}

\begin{figure}[!htb]
  \center
 \subfigure[fig:dolphinsens][The partition found by \emph{SpecG-EC} and the best partition found by \emph{MOSpecG} with $\gamma_1 \in \{0.7,0.8\}, \gamma_2 \in \{0.2,0.3\}$: NMI value of $0.889$.]{\includegraphics[width=0.32\textwidth]{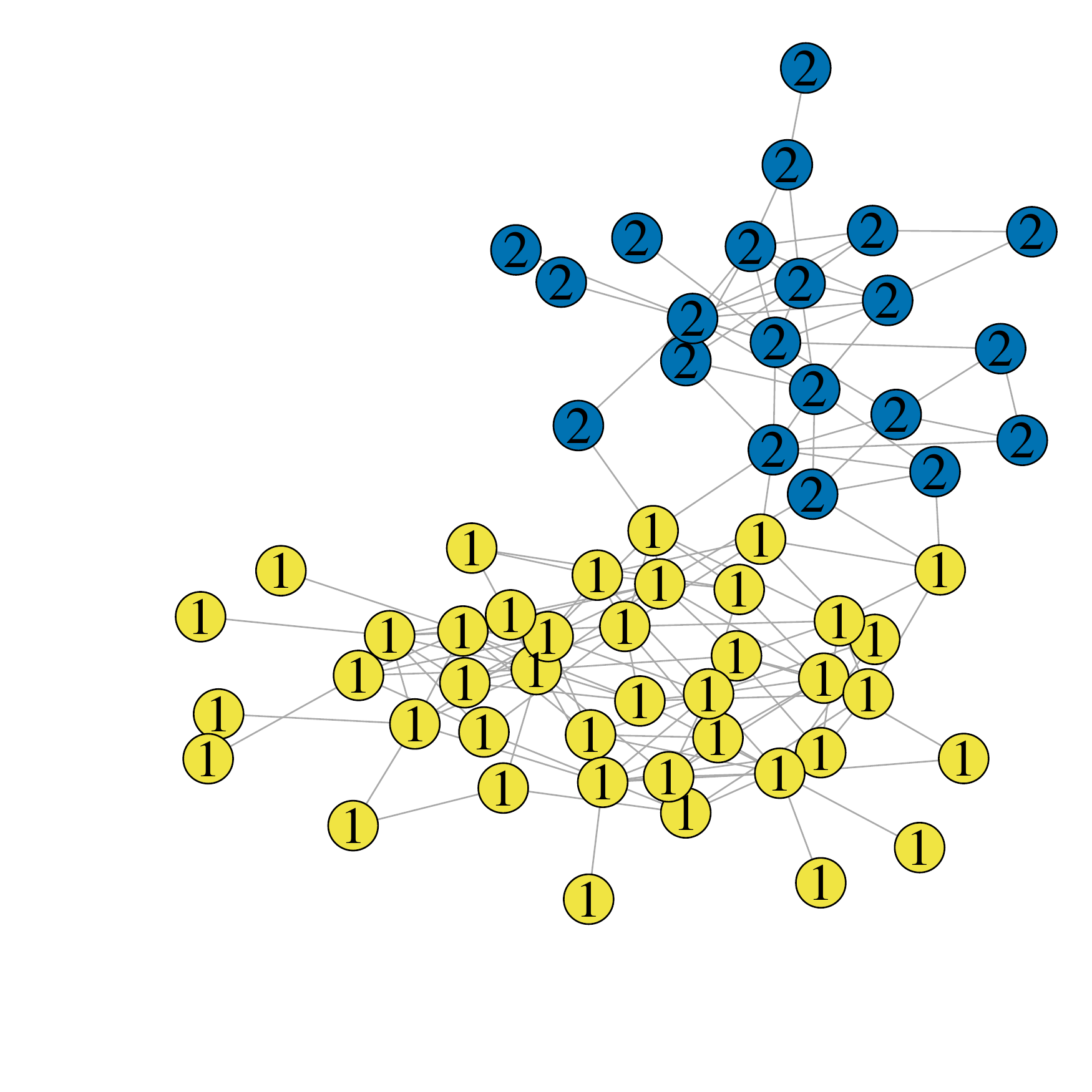}\label{fig:exp_ens_dolphins_ens}}\quad
 \subfigure[fig:dolphinsmod][The partition found by \emph{MOSpecG-mod}: NMI value of $0.662$.]{\includegraphics[width=0.32\textwidth]{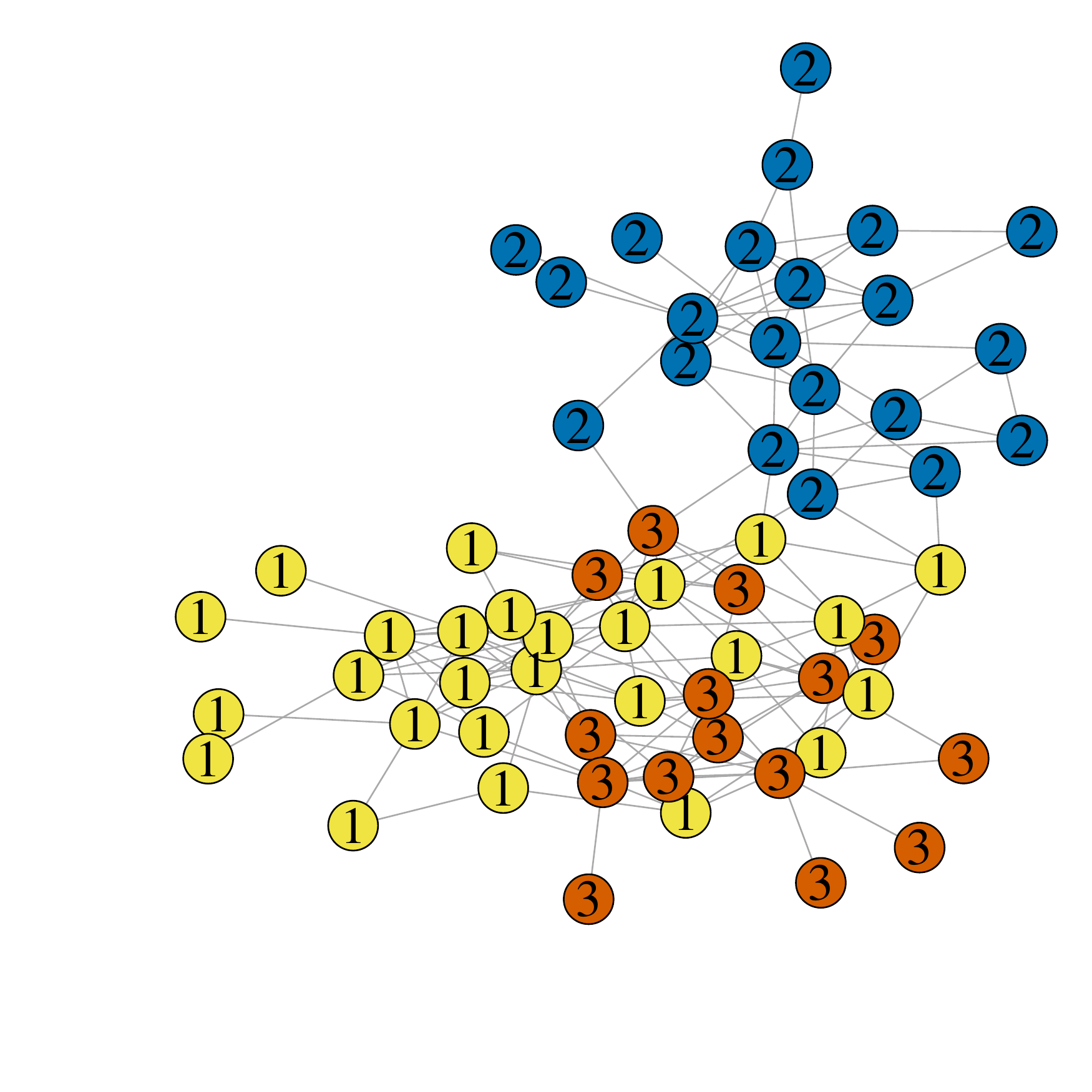}\label{fig:exp_ens_dolphins_mod}}
 \caption{Dolphins network.}
 \label{fig:exp_ens_dolphins}
 \end{figure}

\begin{figure}[!htb]
  \center
 \subfigure[fig:polbookssens][The partition found by \emph{SpecG-EC} and the best partition found by \emph{MOSpecG} with $\gamma_1 \in \{0.1,0.2\}, \gamma_2 \in \{0.8,0.9\}$: NMI value of $0.569$.]{\includegraphics[width=0.32\textwidth]{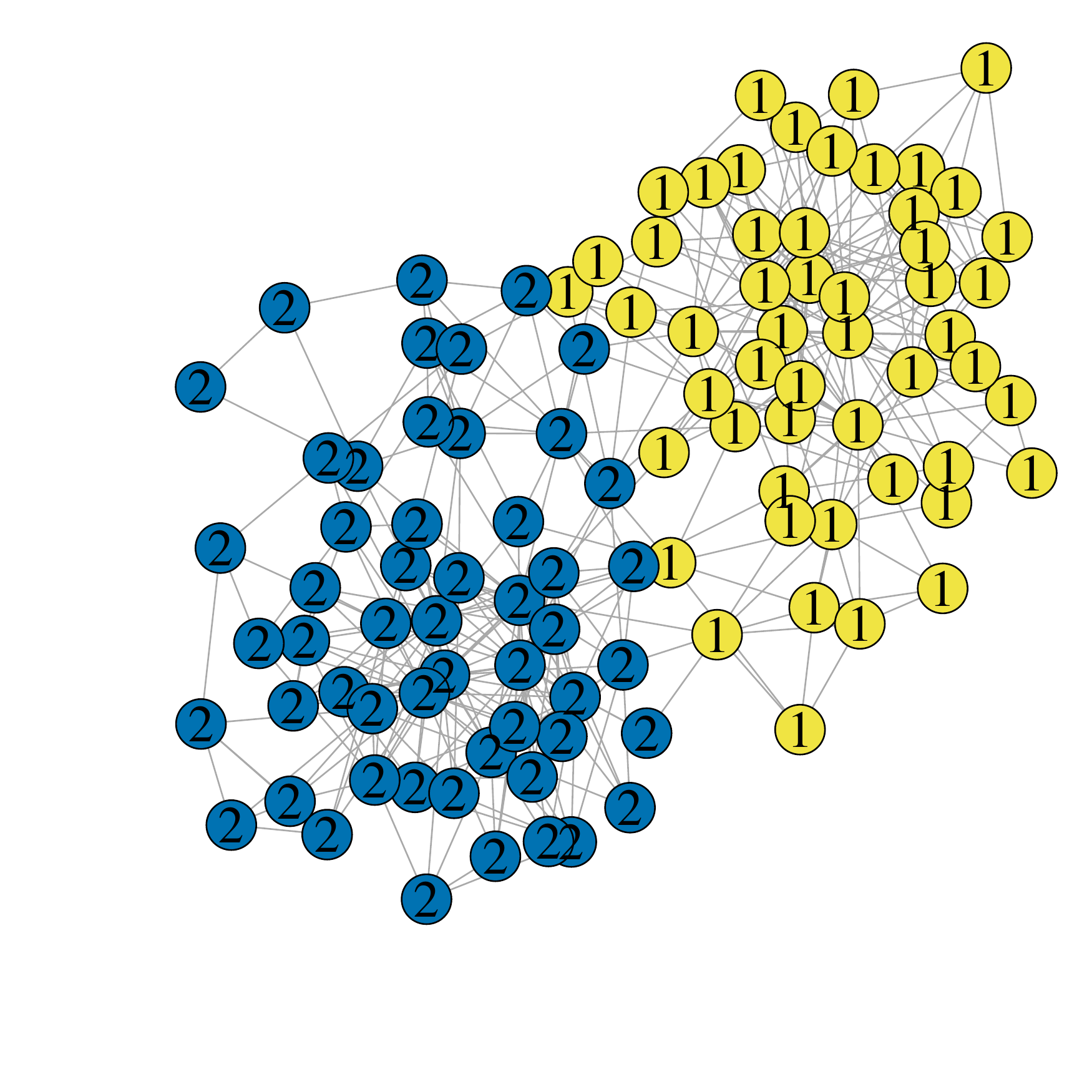}\label{fig:exp_ens_polbooks_ens}}\quad
 \subfigure[fig:polbooksmod][The partition found by \emph{MOSpecG-mod}: NMI value of $0.502$.]{\includegraphics[width=0.32\textwidth]{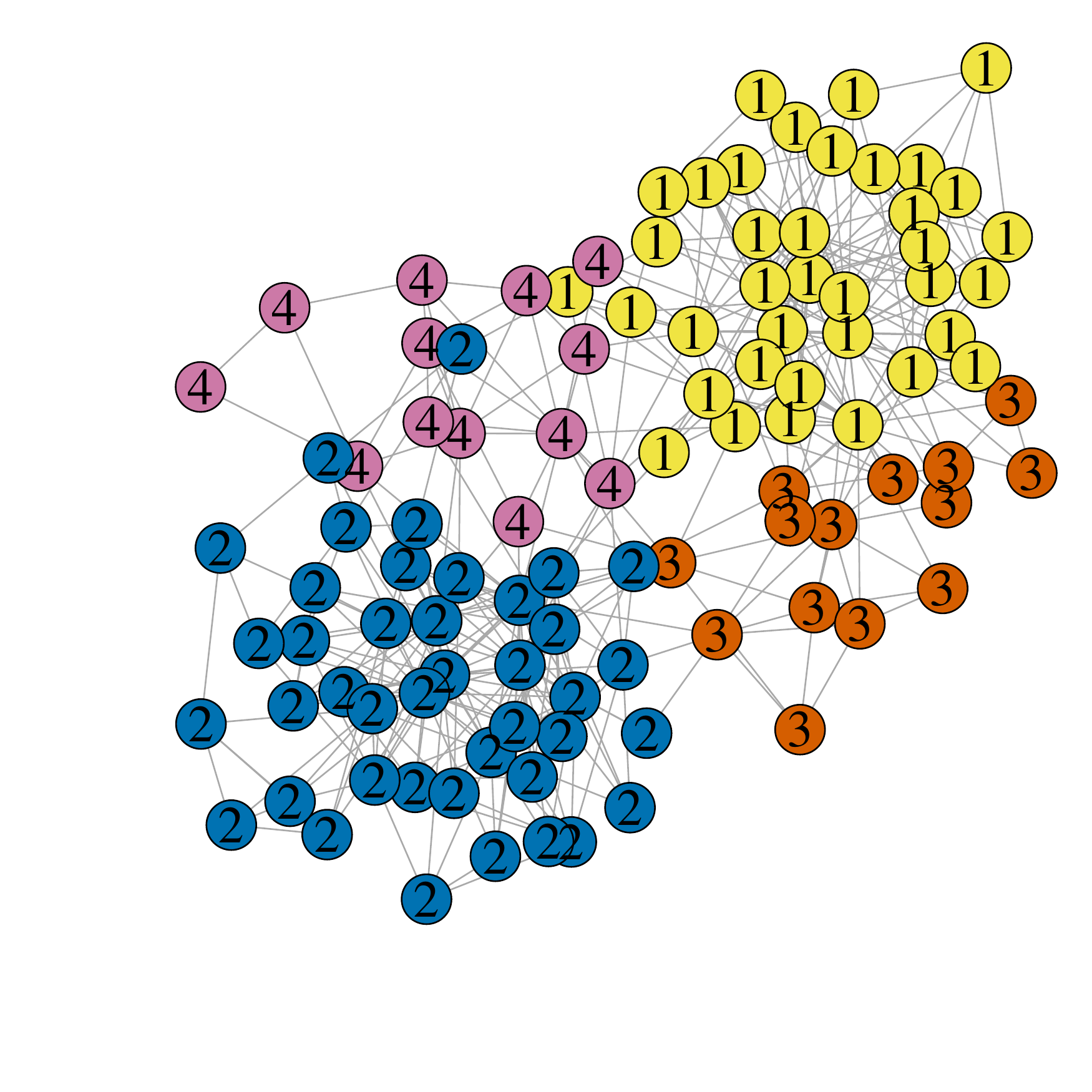}\label{fig:exp_ens_polbooks_mod}}
 \caption{Polbooks network.}
 \label{fig:exp_ens_polbooks}
 \end{figure}

\begin{figure}[!htb]
 \subfigure[fig:footballsens][The partition found by \emph{SpecG-EC}: NMI value of $0.839$.]{\includegraphics[width=0.32\textwidth]{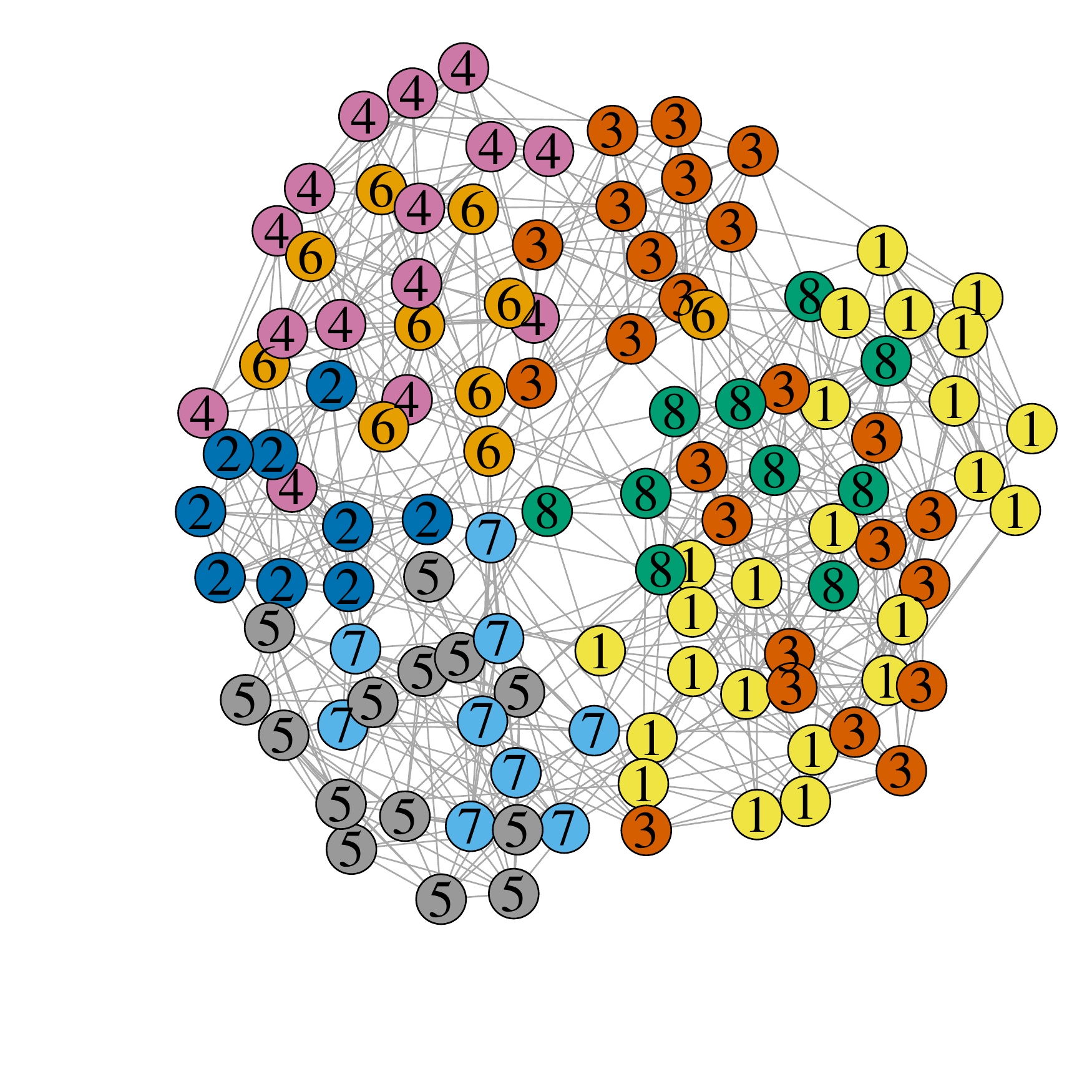}\label{fig:exp_ens_football_ens}}
 \subfigure[fig:footballsmod][The partition found by \emph{MOSpecG-mod}: NMI value of $0.832$.]{\includegraphics[width=0.32\textwidth]{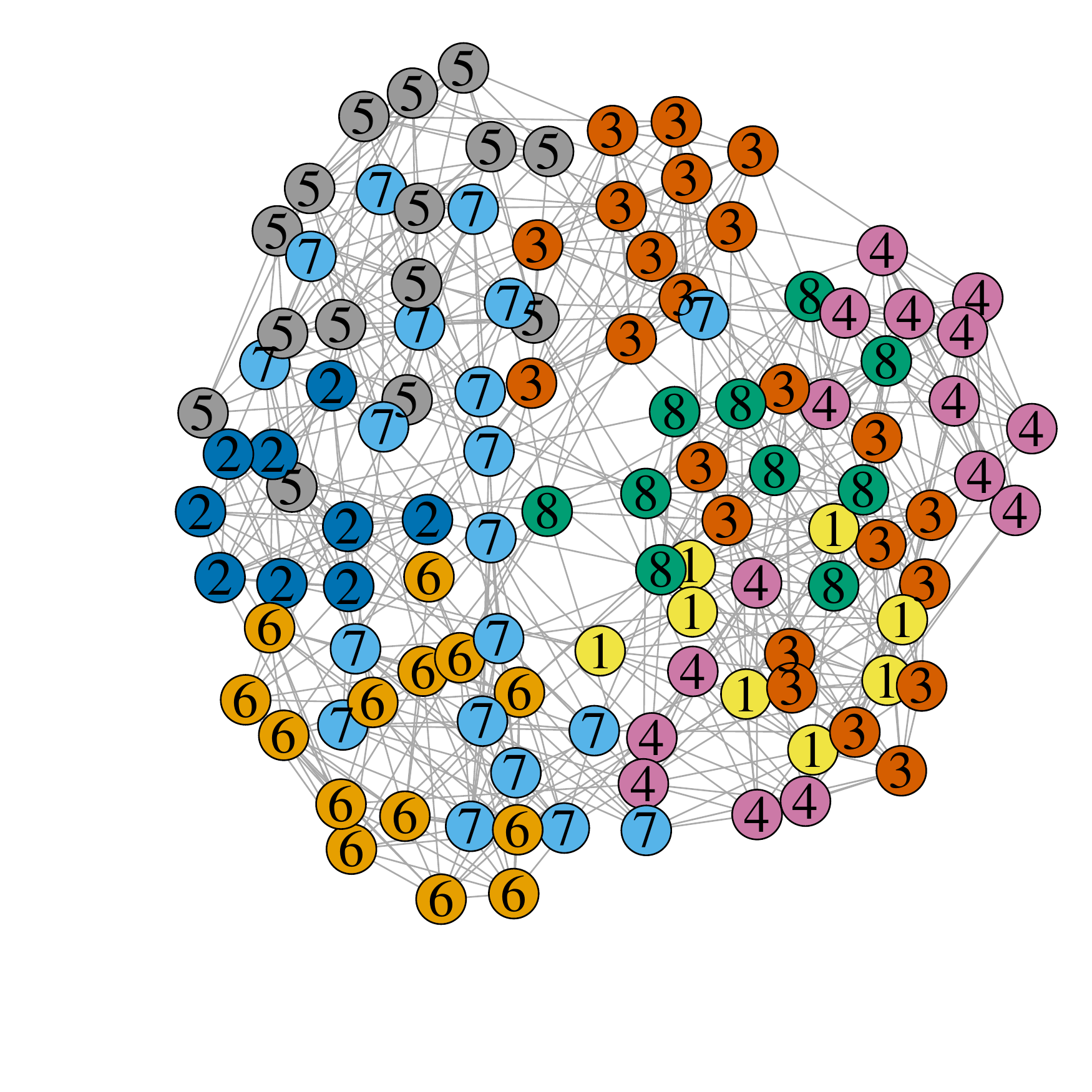}\label{fig:exp_ens_football_mod}}
 \subfigure[fig:footballsmod][The best partition found by \emph{MOSpecG} with $\gamma_1=0.4, \gamma_2=0.6$: NMI value of $0.916$.]{\includegraphics[width=0.32\textwidth]{charts_reais/football_P5.pdf}\label{fig:exp_ens_football_best}}
 \caption{Football network.}
 \label{fig:exp_ens_football}
 \end{figure}

Figures \ref{fig:exp_ens_karate}, \ref{fig:exp_ens_dolphins}, \ref{fig:exp_ens_polbooks} and \ref{fig:exp_ens_football} illustrate the partitions found by \emph{SpecG-EC} and \emph{MOSpecG-mod} for the Karate, Dolphins, Polbooks and Football networks, respectively. These figures also report the partitions found by \emph{MOSpecG} with the $\gamma_1$ and $\gamma_2$ values that resulted in the highest NMI values, here referred to as best partitions. Each vertex is identified by its cluster label in these figures.
 
%Karate

 % Dolphins

 % Polbooks
 
  % Football

Figure \ref{fig:exp_ens_karate} exhibits the expected partition of the Karate network, found by the proposed algorithm.
Figure \ref{fig:exp_ens_dolphins} shows that the ensemble and the best partition obtained by \emph{MOSpecG} for the Dolphins network have the expected number of clusters.
The cluster with label $3$ from the partition returned by \emph{MOSpecG-mod}, in Figure \ref{fig:exp_ens_dolphins_mod}, is merged with the cluster with label $1$ in the ensemble partition in Figure \ref{fig:exp_ens_dolphins_ens}.
Figure \ref{fig:exp_ens_polbooks} shows that most of the vertices from clusters with labels $3$ and $4$ in the partition found by \emph{MOSpecG-mod} for the Polbooks network are merged in, respectively, clusters with labels $1$ and $2$ in the ensemble partition obtained by \emph{SpecG-EC}. 
None of the partitions found for the Football network in Figure 
\ref{fig:exp_ens_football} correctly defined the number of clusters.

%-----------------------------------------------------

\subsubsection{\textcolor{black}{Comparative analysis}}\label{sec:realcomp}

Table \ref{tab:exp_mo_real} reports the average results of ten independent executions of \emph{SpecG-EC}, \emph{MOSpecG-mod}, Moga-Net, OSLOM and Infomap to detect communities in real networks. The results presented are the NMI values, the CPU-times in seconds and number of clusters. The standard deviation of the presented values is shown between parentheses. Table \ref{tab:exp_mo_real} also presents the number of clusters in the expected partitions.

 \begin{table*}[!htb]
   \centering
     \caption{NMI, CPU-times and number of clusters achieved by the algorithms for real graphs.}
     \setlength{\tabcolsep}{3pt} 
     \begin{tabular}{|c|c|c|c|c|c|}
 \cline{3-6}
 \multicolumn{2}{c}{} & \multicolumn{4}{|c|}{Network}\\ \cline{3-6}
  \multicolumn{2}{c|}{}  & Karate&	Dolphins&	Polbooks&	Football\\   \cline{1-6}
\multirow{3}{*}{\emph{SpecG-EC}} 
& NMI&	\textbf{1} (0)&	\textbf{0.889} (0)&	 \textbf{0.565} (0.006)&	0.864 (0.03)\\ 
& CPU-time (s)&	0.172 (0.05)&	0.33 (0.053)&	0.884 (0.131)&	1.377 (0.203)\\ 
& \#Clusters&	2 (0)&	2 (0)&	2.3 (0.483)&	9 (1.054)\\  \hline 
\multirow{3}{*}{\emph{MOSpecG-mod}} 
& NMI&	\textbf{1} (0)&	0.662 (0)&	0.485 (0.026)&	0.876 (0.023)\\ 
& CPU-time (s)&	0.015 (0.003)&	0.025 (0.004)&	0.081 (0.015)&	0.129 (0.028)\\ 
& \#Clusters&	2 (0)&	3 (0)&	4.3 (0.483)&	9.4 (0.699)\\  \hline 
\multirow{3}{*}{Moga-Net}
& NMI&	0.682 (0.047)&	0.538 (0.067)&	0.511 (0.054)&	0.736 (0.048)\\ 
& CPU-time (s)&	7.85 (0.627)&	11.827 (1.255)&	25.231 (2.656)&	28.61 (2.888)\\ 
& \#Clusters&	3.9 (0.316)&	6.1 (1.449)&	5.5 (1.65)&	7.8 (1.033)\\  \hline 
\multirow{3}{*}{OSLOM} 
& NMI&	\textbf{1} (0)&	0.786 (0.11)&	0.558 (0.017)&	0.916 (0)\\ 
& CPU-time (s)&	0.3 (0.483)&	0.9 (0.568)&	1.2 (0.422)&	0.7 (0.483)\\ 
& \#Clusters&	2 (0)&	2 (0)&	3.7 (0.675)&	11 (0)\\  \hline 
\multirow{3}{*}{Infomap} 
& NMI&	0.699 (0)&	0.519 (0)&	0.537 (0)&	\textbf{0.924} (0)\\ 
& CPU-time (s)&	0.2 (0.422)&	0.5 (0.527)&	0.4 (0.516)&	0.4 (0.516)\\ 
& \#Clusters&	3 (0)&	6 (0)&	5 (0)&	12 (0)\\  \hline   \hline 
Expected& \#Clusters &  2&	2&	3&	12\\  \hline 
     \end{tabular}
     \label{tab:exp_mo_real}
 \end{table*}
 
As can be seen in Table~\ref{tab:exp_mo_real}, on the one hand, \emph{SpecG-EC} outperformed Moga-Net in all the networks. On the other hand, \emph{MOSpecG-mod} only found lower NMI values than Moga-Net for the Polbooks network.
Moga-Net and Infomap were the only algorithms which did not obtain the expected partition for the Karate network. \emph{SpecG-EC} achieved higher NMI values than all the reference algorithms, including \emph{MOSpecG-mod}, for the Dolphins and Polbooks networks.
Furthermore, the number of clusters in the partitions obtained by \emph{SpecG-EC} and \emph{MOSpecG-mod} varied at a maximum of $25\%$ and $43.333\%$, respectively and on average, when compared to the expected number of clusters. Thereby, there is empirical evidence suggesting the effectiveness of the proposed algorithm in estimating the number of clusters.

\textcolor{black}{The differences between the NMI values reported in Figures \ref{fig:exp_ens_polbooks} and \ref{fig:exp_ens_football} and those presented in Table \ref{tab:exp_mo_real} are due to the fact that the figures only report the results of a single execution, whereas the table shows the average NMI values of ten executions.
According to Table \ref{tab:exp_mo_real}, the average NMI value of partitions obtained by \emph{SpecG-EC} for the Polbooks network is only 0.703\% lower than the highest NMI value of partitions from the \textit{solution set} presented in Figure \ref{fig:exp_pareto_real_polbooks}. Furthermore, \emph{SpecG-EC} found partitions for Football network whose average NMI value was 5.677\% worse than the highest NMI value of partitions from the \textit{solution set} presented in Figure \ref{fig:exp_pareto_real_football}.}

\textcolor{black}{Table \ref{tab:exp_detail_real} demonstrates details of the experiment performed with the proposed and reference algorithms on network Dolphins\footnote{The table additionally reports the results obtained by \emph{MOSpecG-MO} for each combination of $\gamma_1$ and $\gamma_2$.}. The table presents the NMI and modularity values; the running time in seconds; the number of pairs of vertices which were grouped correctly in the same cluster and incorrectly grouped in the same or different clusters, when compared to the expected partition; and the number and size of the clusters in the partitions obtained by the algorithms. The expected partition of network dolphins has 2 clusters with 42 and 20 vertices.}

\begin{table}[!htb]
\setlength{\tabcolsep}{3pt} 
\caption{Details of the experiment performed using network dolphins.}
\begin{tabular}{|l|l|l|l|l|l|l|l|}
\hline
\multicolumn{1}{|c|}{Algorithm}  & NMI & \multicolumn{1}{|c|}{$Q$} &  \multicolumn{1}{|c|}{CPU-} &  \multicolumn{2}{|c|}{Pairs of vertices} &   \multicolumn{2}{|c|}{Clusters}                                          \\  \cline{5-8}
   &  & &  \multicolumn{1}{|c|}{time (s)} & Correct & Wrong   & \#   & \multicolumn{1}{|c|}{Sizes}                       \\\hline
 \emph{MOSpecG} - $\gamma_1$=0,$\gamma_2$=1     &	0       &	0       &	0.018&  1051    &	840     &		1&	62\\ 
\emph{MOSpecG}  - $\gamma_1$=0.1,$\gamma_2$=0.9 &   0       &	0       &	0.019&	1051    &	840     &	1&	62\\ 
\emph{MOSpecG}  - $\gamma_1$=0.2,$\gamma_2$=0.8 & 	0.486   &	0.378   &	0.025&	731     &	520     &	2&	32, 30\\ 
\emph{MOSpecG}  - $\gamma_1$=0.3,$\gamma_2$=0.7 &	0.514   &	0.391   &	0.025&	754     &	477     &	2&	33, 29\\ 
\emph{MOSpecG}  - $\gamma_1$=0.4,$\gamma_2$=0.6 &	0.662   &	0.483   &	0.024&	620     &	451     &	3&	26, 21, 15\\ 
\emph{MOSpecG}  - $\gamma_1$=0.5,$\gamma_2$=0.5 &	0.662   &	0.483   &	0.022&	620     &	451     &	3&	26, 21, 15\\ 
\emph{MOSpecG} - $\gamma_1$=0.6,$\gamma_2$=0.4  &	0.581   &	0.518   &	0.026&	492     &	579     &	4&	21, 20, 14, 7\\ 
\emph{MOSpecG}  - $\gamma_1$=0.7,$\gamma_2$=0.3 &	0.889   &	0.379   &	0.025&	1010    &	61      &	2&	41, 21\\ 
\emph{MOSpecG}  - $\gamma_1$=0.8,$\gamma_2$=0.2 &	0.889   &	0.379   &	0.025&	1010    &	61      &	2&	41, 21\\ 
\emph{MOSpecG}  - $\gamma_1$=0.9,$\gamma_2$=0.1 &	0.777   &	0.359   &	0.025&	991     &	120     &	2&	42, 20\\ 
\emph{MOSpecG} - $\gamma_1$=1,$\gamma_2$=0      &	0       &	0       &	0.024&	1051    &	840     &	1&	62\\ \hline
\multicolumn{1}{|c|}{\emph{SpecG-EC}}           &	0.889   &	0.379   &	0.281&	1010    &	61      &	2&	41, 21\\ \hline
\multicolumn{1}{|c|}{{Moga-Net}}                &	0.472   &	0.417   &	13.92&	532     &	537     &	7&	31, 10, 6, 6, 4, 3, 2\\ \hline
\multicolumn{1}{|c|}{{Infomap}}                 &	0.519   &	0.523   &	1    &	417     &	654     &	6&	21, 17, 12, 7, 3, 2\\ \hline
\multicolumn{1}{|c|}{{OSLOM}}                   &	0.814   &	0.385   &	1    &	971     &	120     &	2&	40, 22\\ \hline
\end{tabular}
\label{tab:exp_detail_real}
\end{table}

\textcolor{black}{The results in Table \ref{tab:exp_detail_real}  show that the larger and the lower the number of pairs of vertices classified correctly and incorrectly, respectively, the larger the NMI values of the partitions. This table also shows that partitions with higher values of modularity are not necessarily more similar to the expected partitions considering the NMI values, the number and size of  the clusters. The partition obtained by \emph{SpecG-EC} presented the highest NMI value and matched the expected number of clusters, 2. In this partition, exactly one vertex was classified in the wrong cluster. In combining the solutions of \emph{MOSpecG} by the consensus strategy, \emph{SpecG-EC} found the partition with the largest NMI.
Even though \emph{MOSpecG} with $\gamma_1=0.9$ and $\gamma_2=0.1$ and OSLOM identified partitions with 2 clusters and whose numbers of vertices are the expected,  2 vertices were  displaced. Both Moga-Net and Infomap identified a large number of clusters, far from the expected.}

%%%%%%%%%%%%%%%%%%%%%%%%%%%%%%%%%%%%%%%%%%%%%%%%%%%%%% 
\subsection{\textcolor{black}{Experiments with artificial networks}}
%%%%%%%%%%%%%%%%%%%%%%%%%%%%%%%%%%%%%%%%%%%%%%%%%%%%%%

This experiment used 80 undirected LFR networks \citep{Lancichinetti2011} with the following characteristics: 1000 vertices; average degree within the range $[20,50]$; small-sized communities, whose number of vertices are in the interval $[10,50]$; large-sized communities, whose number of vertices are in the interval $[20,100]$; and degree of mixture (mixture coefficient) between groups ($\mu$) with values from the set $\{0.1,0.2,\dots,0.8\}$.

\subsubsection{\textcolor{black}{Solution sets found by MOSpecG}}
 
\begin{figure}[!htb]
 \subfigure[fig:S1000nmi][Small-sized community networks.]{\includegraphics[width=0.49\textwidth]{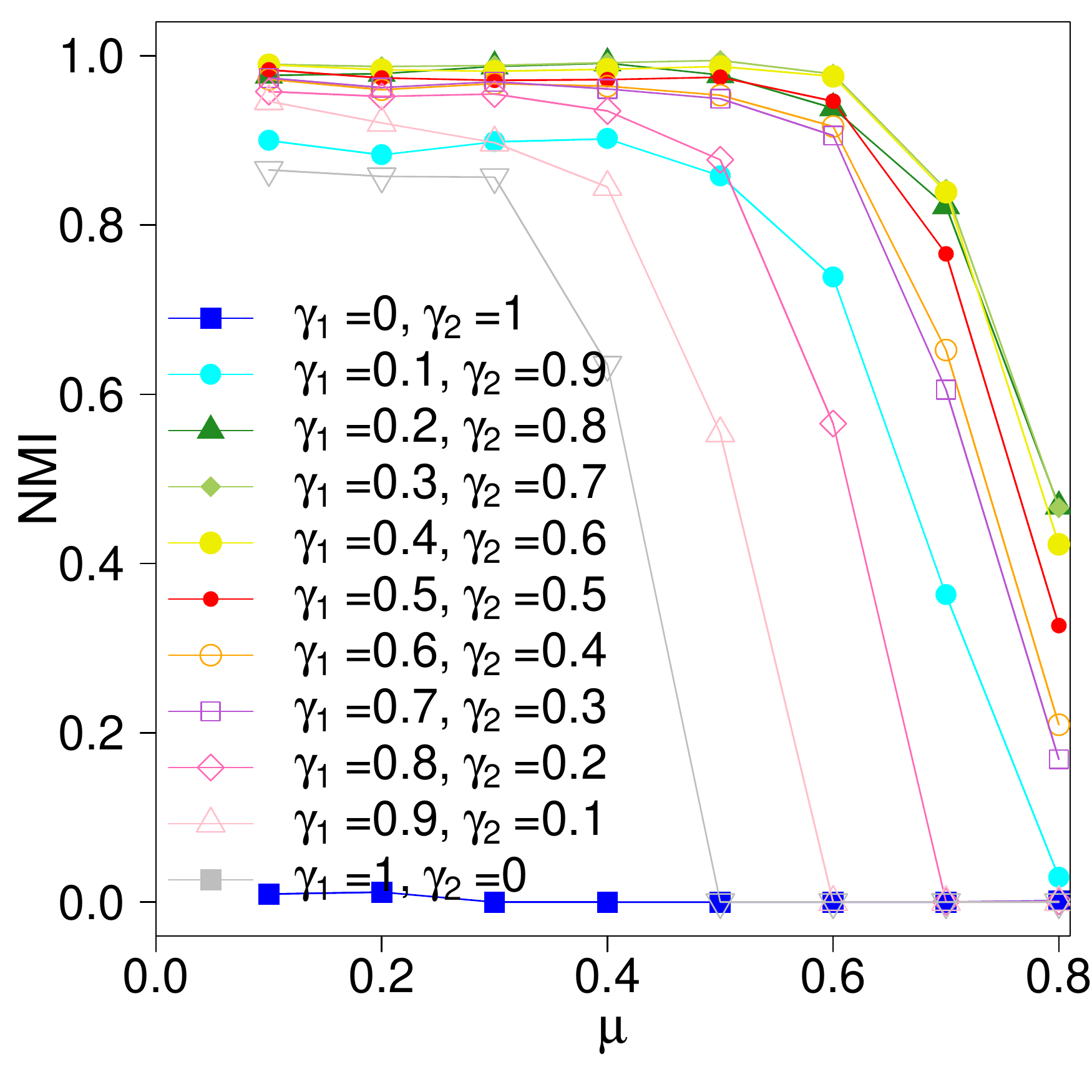}\label{fig:S1000nmi}}
 \subfigure[fig:L1000nmi][Large-sized community networks.]{\includegraphics[width=0.49\textwidth]{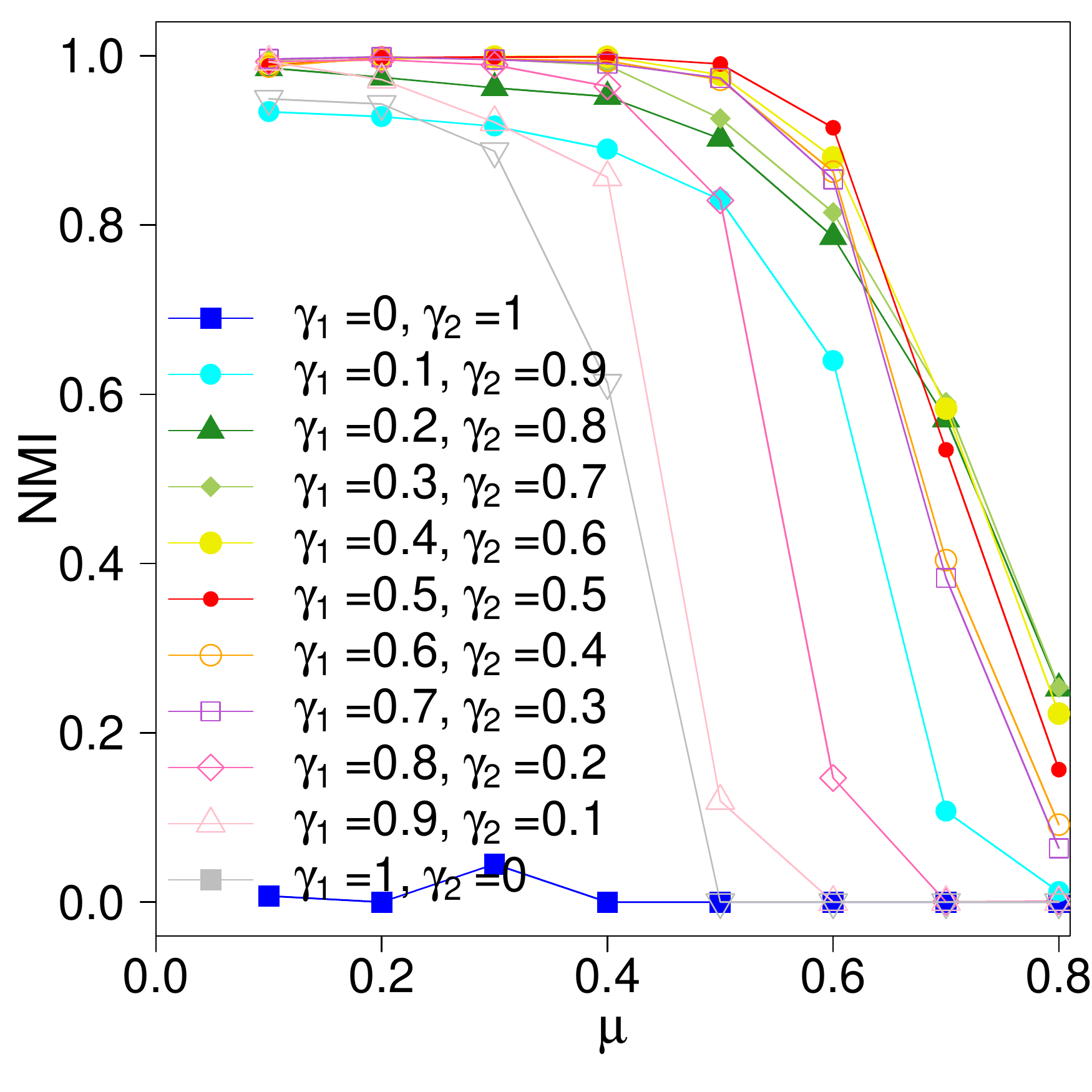}\label{fig:L1000nmi}}
 \caption{Average NMI values of the partitions obtained by \emph{MOSpecG} considering different values of $\gamma_1$ and $\gamma_2$.}
 \label{fig:exp_S1000_nmi}
 \end{figure}

\begin{figure}[!htb]
\subfigure[fig:S1000number][Small-sized community networks.]{\includegraphics[width=0.49\textwidth]{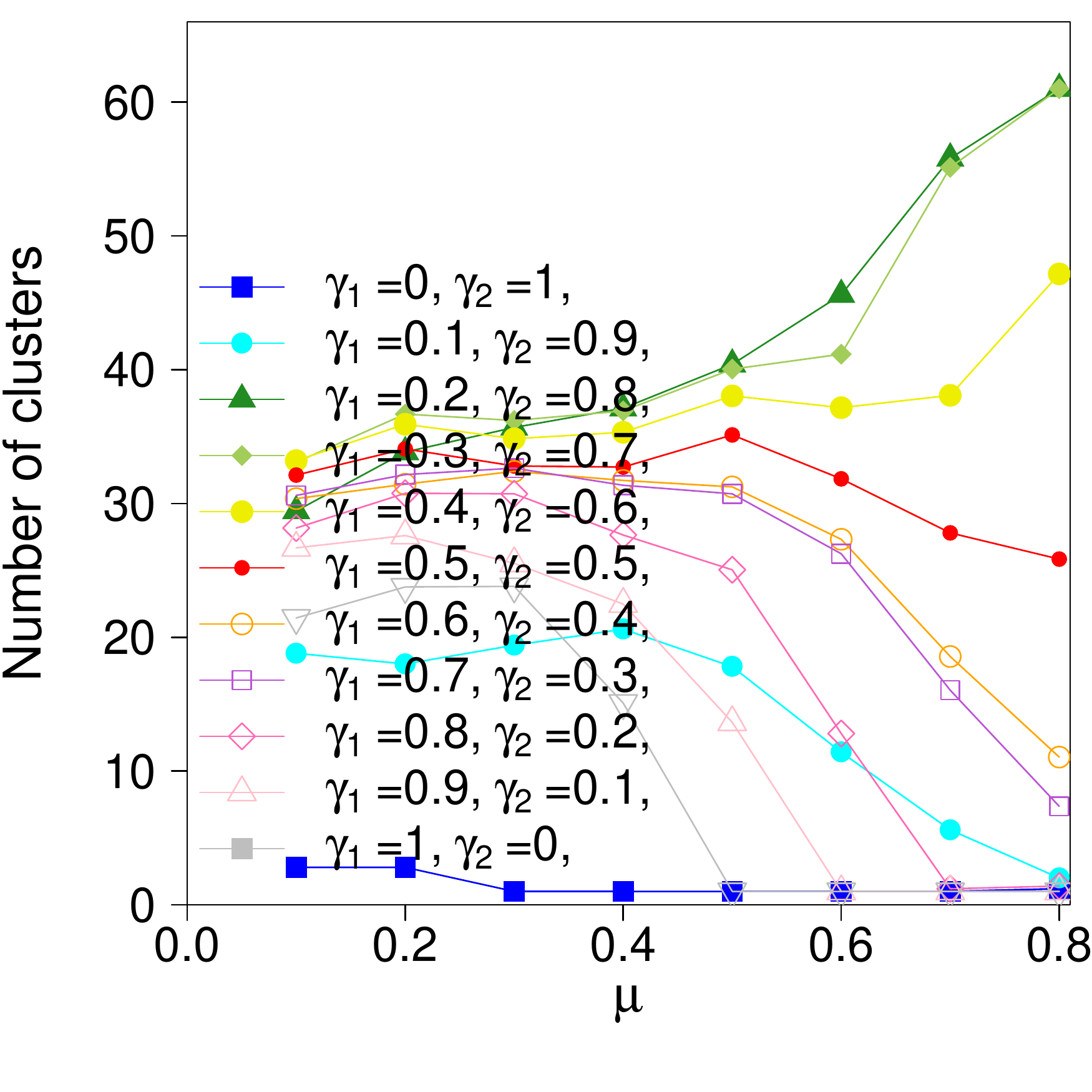}\label{fig:S1000number}}
\subfigure[fig:L1000number][Large-sized community networks.]{\includegraphics[width=0.49\textwidth]{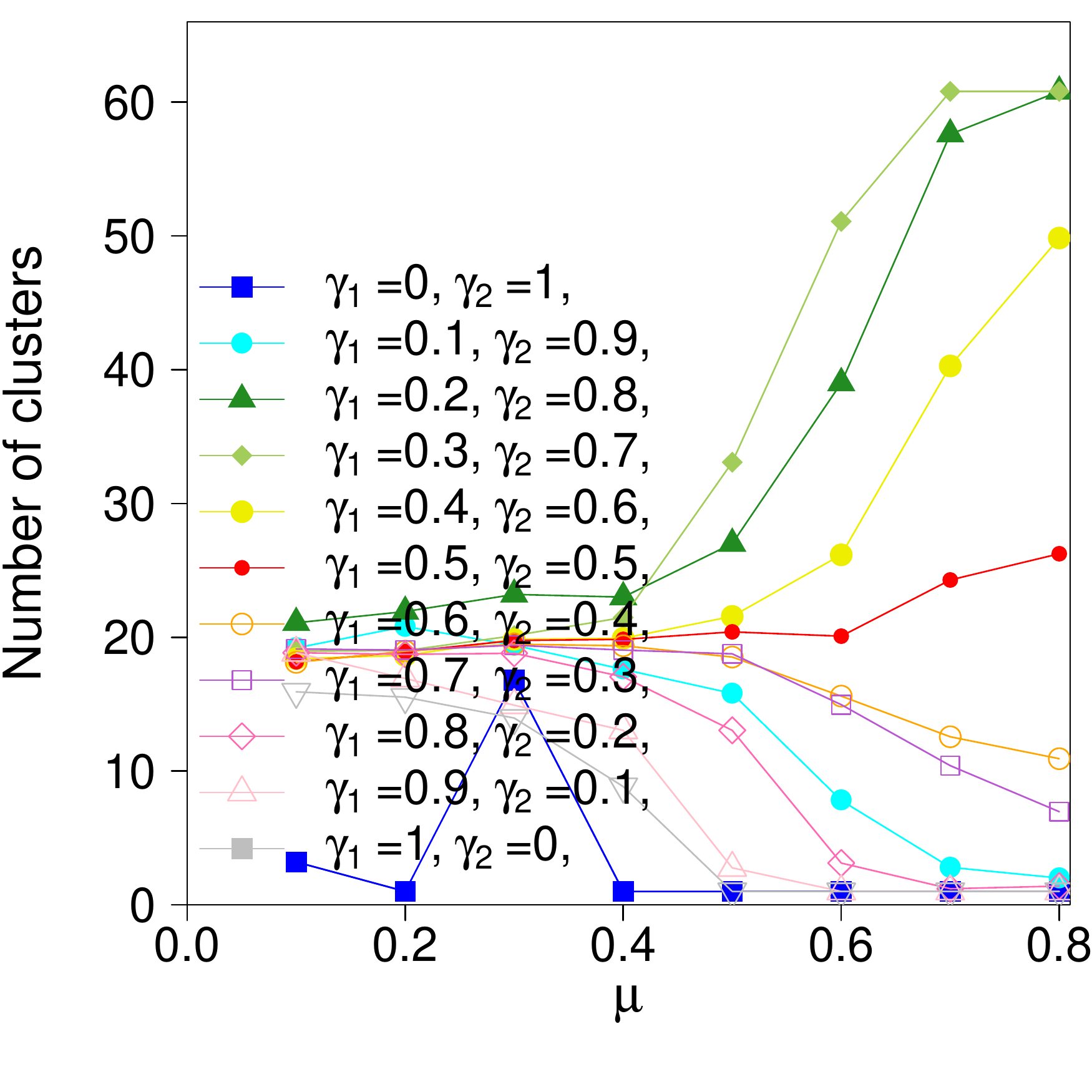}\label{fig:L1000number}}
\caption{Average number of clusters in the partitions found by \emph{MOSpecG} considering different values of $\gamma_1$ and $\gamma_2$.}
\label{fig:exp_S1000_number}
\end{figure}

%-----------------------------------------------------
% MO NMI
Figures \ref{fig:exp_S1000_nmi} to \ref{fig:exp_ens_S1000_number} display the average results of the algorithms applied to the LFR networks (y-axis) by $\mu$ (x-axis). Figures \ref{fig:S1000nmi} and \ref{fig:L1000nmi} present the average NMI values of partitions obtained by \emph{MOSpecG} for, respectively, the small and large-sized community networks considering each combination of weights $\gamma_1$ and $\gamma_2$. 

As can be noted in Figures \ref{fig:S1000nmi} and \ref{fig:L1000nmi}, the values of $\gamma_1 \in \{0.2,0.3\}$ and $\gamma_2 \in \{0.7,0.8\}$ resulted in partitions with the highest average NMI values for the networks with $\mu \geq 0.7$. The proposed heuristic presented competitive results when detecting communities in all networks by optimizing the modularity, i.e.,  when $\gamma_1 = \gamma_2 = 0.5$. The partitions found when considering $\gamma_1=0$ and $\gamma_2=1$ failed to identify good quality clusters.

%-----------------------------------------------------
% MO Number of clusters

Figures \ref{fig:S1000number} and \ref{fig:L1000number} show the average number of clusters in the partitions from the \textit{solution sets} for, respectively, the small and large-sized community networks. Except for the results when $\gamma_1=0,\gamma_2=1$, which misidentified the number of clusters, and when $\gamma_1=0.1,\gamma_2=0.9$, the lower the $\gamma_1$ and the larger the $\gamma_2$, the larger the number of clusters. Thereby, the number of communities grows with $\gamma=\frac{\gamma_2}{\gamma_1}$. 

%-----------------------------------------------------
\subsubsection{\textcolor{black}{Comparative analysis}}
% Ensemble NMI

\begin{figure}[!htb]
 \subfigure[fig:S1000nmi][Small-sized community networks.]{\includegraphics[width=0.49\textwidth]{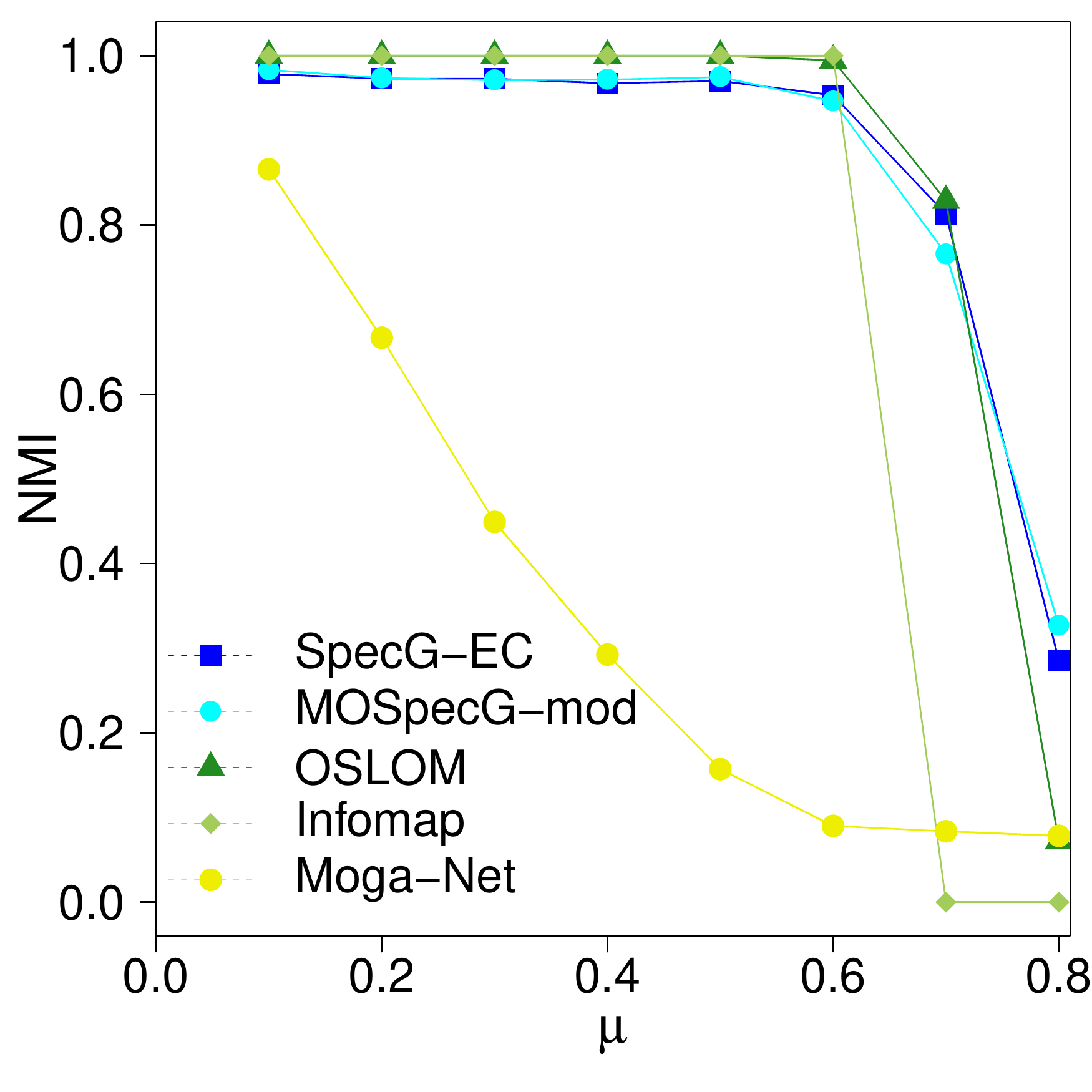}}
 \subfigure[fig:L1000nmi][Large-sized community networks.] {\includegraphics[width=0.49\textwidth]{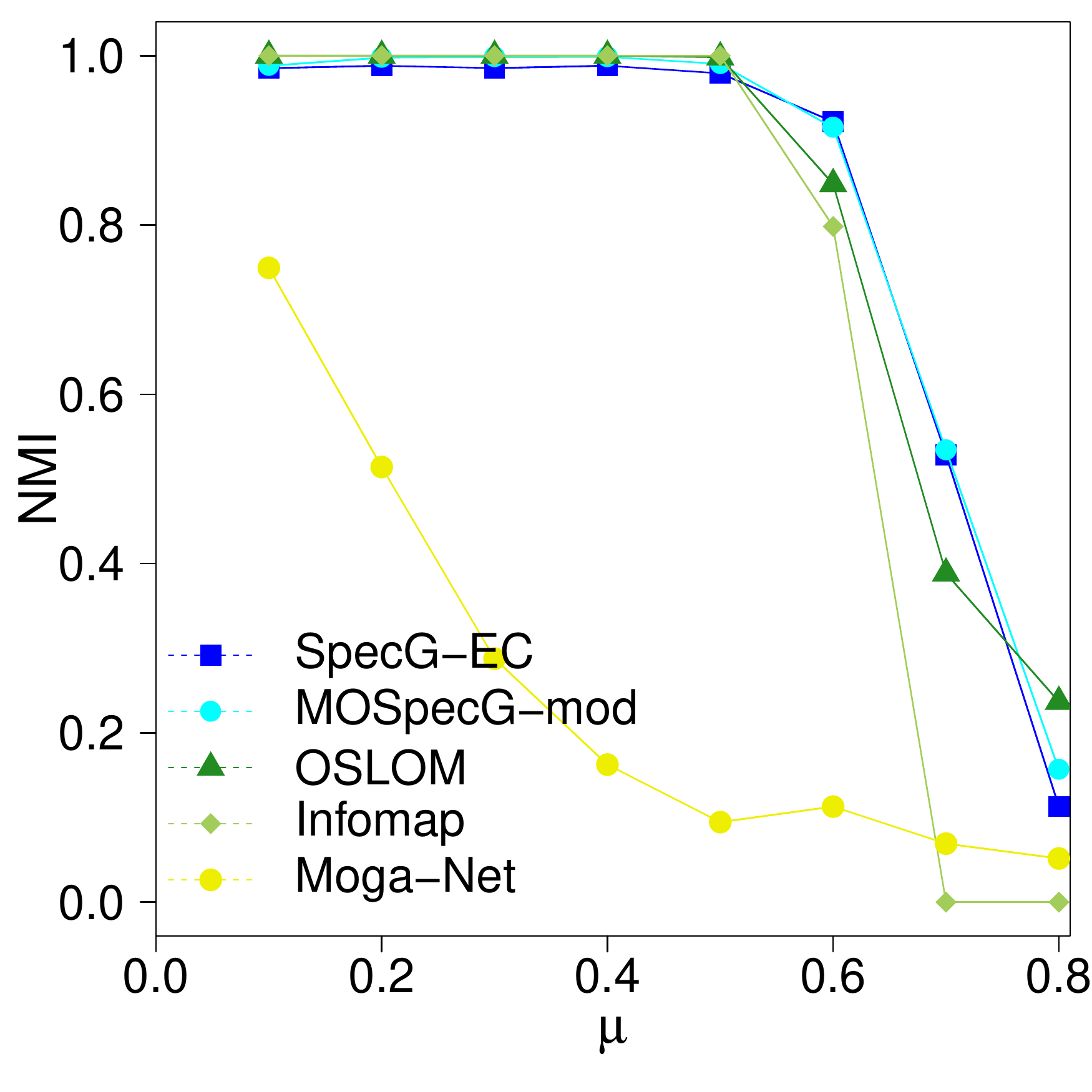}}
 \caption{Average NMI values achieved by the algorithms.}
 \label{fig:exp_ens_S1000_nmi}
 \end{figure}

\begin{figure}[!htb]
 \subfigure[fig:S1000time][Small-sized community networks.]{\includegraphics[width=0.49\textwidth]{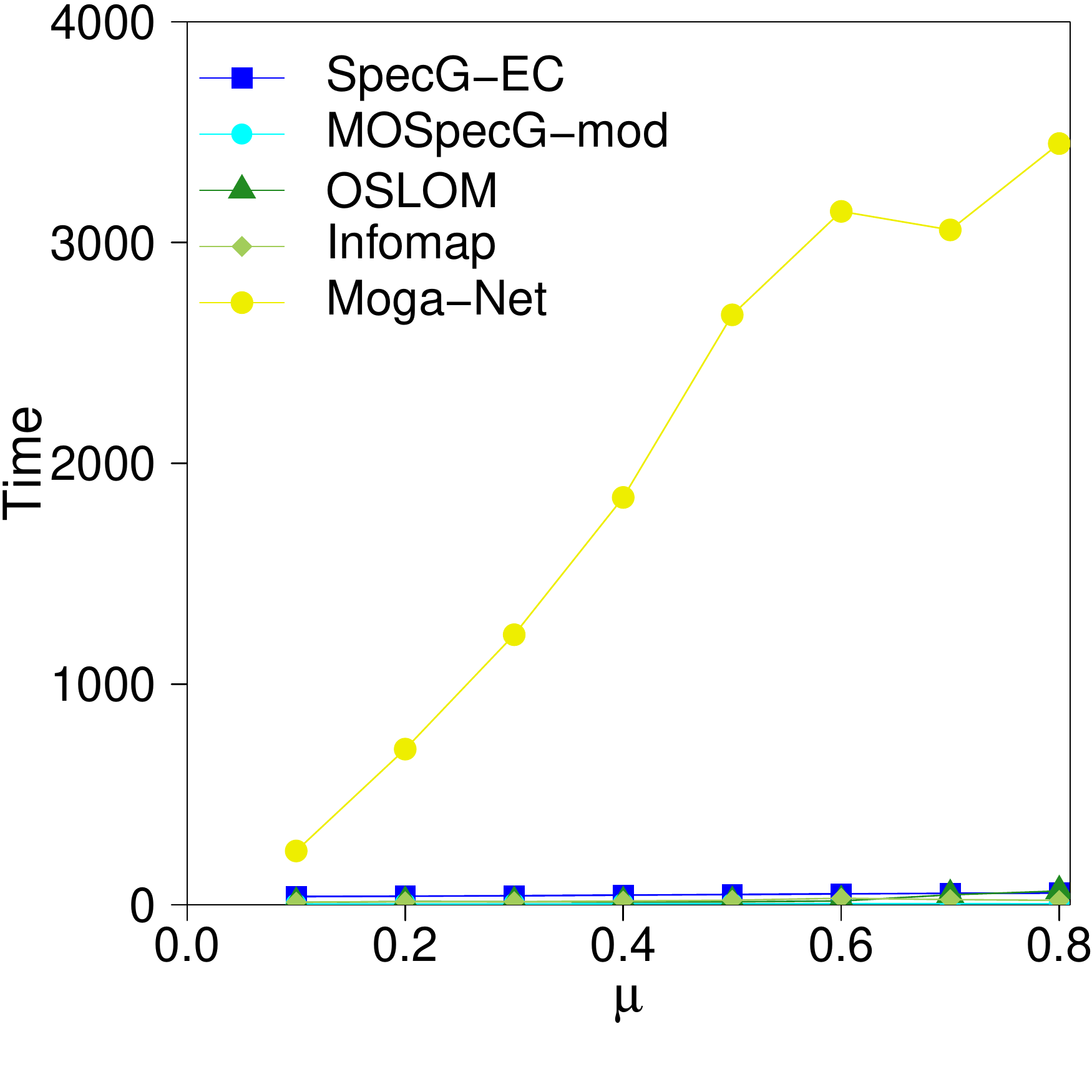}}
 \subfigure[fig:L1000time][Large-sized community networks.] {\includegraphics[width=0.49\textwidth]{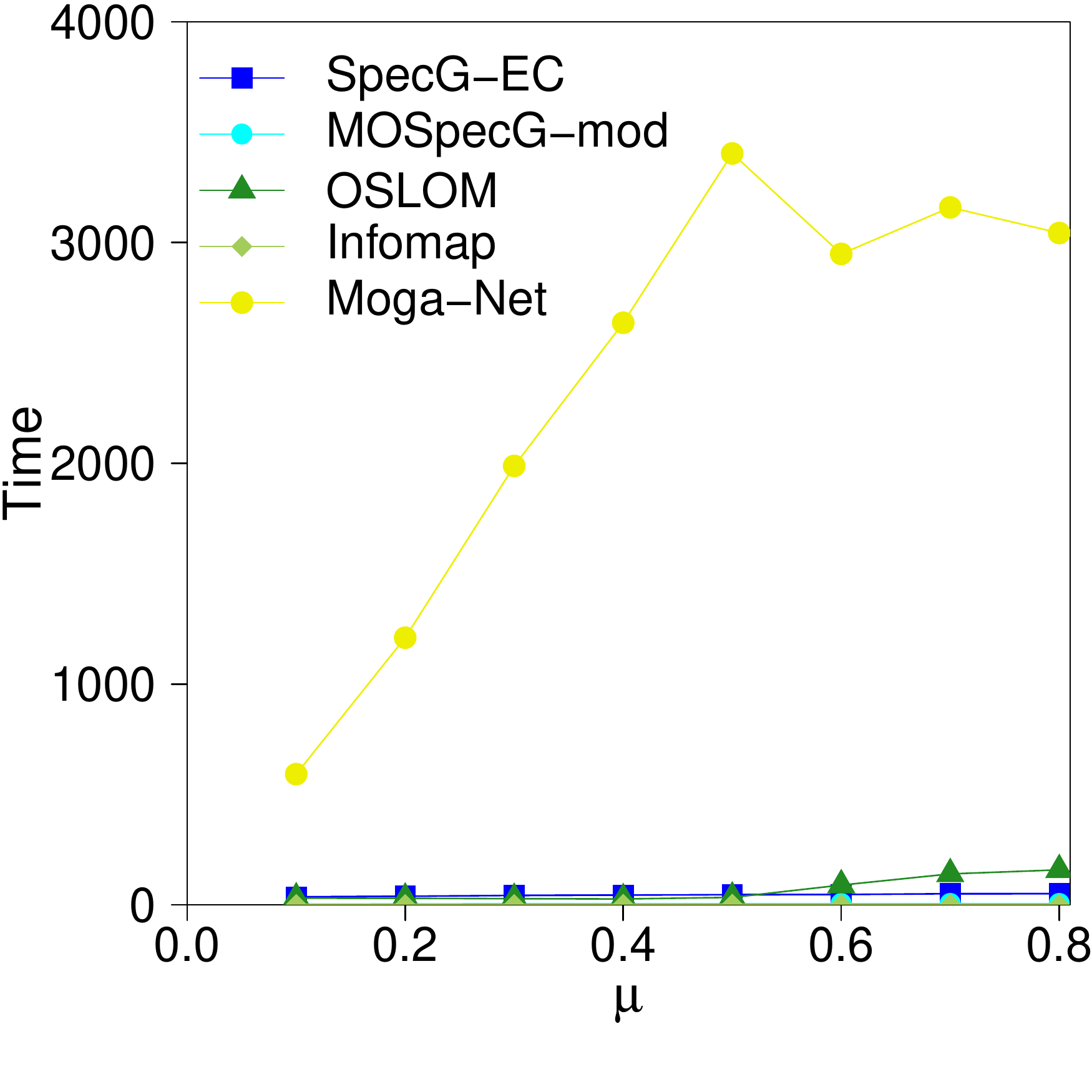}}
  \caption{Average CPU-times (s) required by the algorithms.}
 \label{fig:exp_ens_S1000_time}
 \end{figure}

Figure \ref{fig:exp_ens_S1000_nmi} presents the average NMI values of the partitions found by \emph{SpecG-EC}, \emph{MOSpecG-mod}, OSLOM, Infomap and Moga-Net whereas Figure \ref{fig:exp_ens_S1000_time} shows the respective average CPU-times for the small and large-sized community networks.
 
 Figure \ref{fig:exp_ens_S1000_nmi} shows that the partitions found by \emph{SpecG-EC} and \emph{MOSpecG-mod} had average NMI values higher than those with the largest modularity found by Moga-Net. 
 Moreover, \emph{SpecG-EC} outperformed \emph{MOSpecG-mod}, Infomap and OSLOM in the small-sized community networks with, respectively, $\mu \in \{0.6,0.7\}$, $\mu \geq 0.7$ and $\mu=0.8$. 
 In the small-sized community networks with $\mu\leq 0.6$, \emph{SpecG-EC} obtained partitions whose NMI values were higher or equal to $0.953$.
\emph{SpecG-EC} outperformed \emph{MOSpecG-mod}, Infomap and OSLOM in large-sized community networks with, respectively, $\mu=0.6$, $\mu \geq 0.6$ and $\mu \in \{0.6,0.7\}$, and achieved NMI values of at least $0.979$ in the networks when $\mu \leq 0.5$.

%-----------------------------------------------------
% Ensemble time
 
 \emph{MOSpecG-mod} and Infomap were the algorithms with the lowest CPU-times in networks with, respectively, small and large-sized community networks.
Nonetheless, \emph{SpecG-EC} was from $6$ to $64$ times faster than Moga-Net in all the networks. On the one hand, \emph{SpecG-EC} was faster than OSLOM in the large-sized community networks with $\mu \geq 0.6$. On the other, it required from $1.18$ to $3.056$ times more than the CPU time required by OSLOM in the remaining networks. Because \emph{MOSpecG-mod} was approximately $13.286$ times faster than \emph{SpecG-EC}, it was also faster than OSLOM in all the networks.
 
%-----------------------------------------------------
% Clusters

\begin{figure}[!htb]
 \subfigure[fig:S1000time][Small-sized community networks.]{\includegraphics[width=0.49\textwidth]{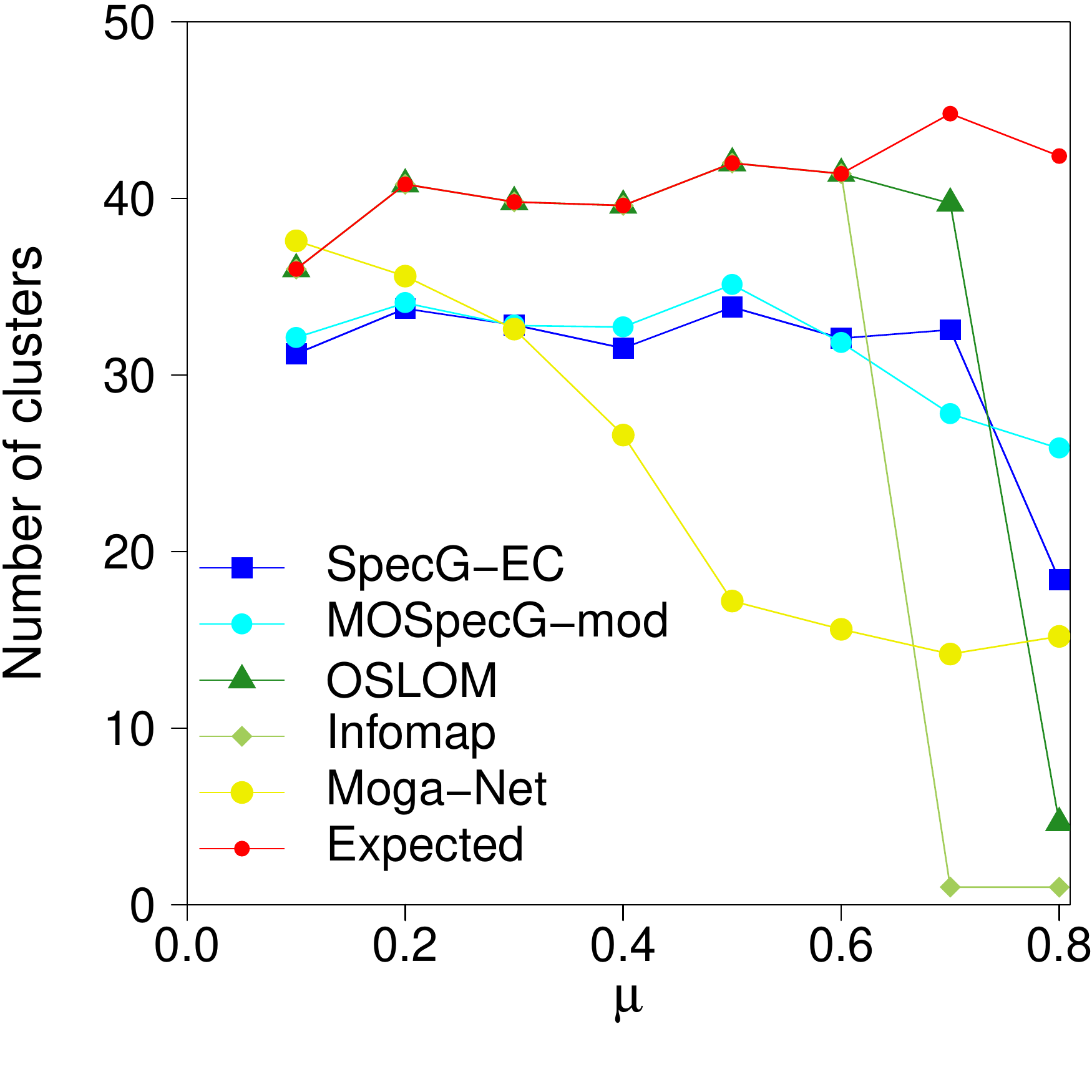}}
 \subfigure[fig:L1000time][Large-sized community networks.] {\includegraphics[width=0.49\textwidth]{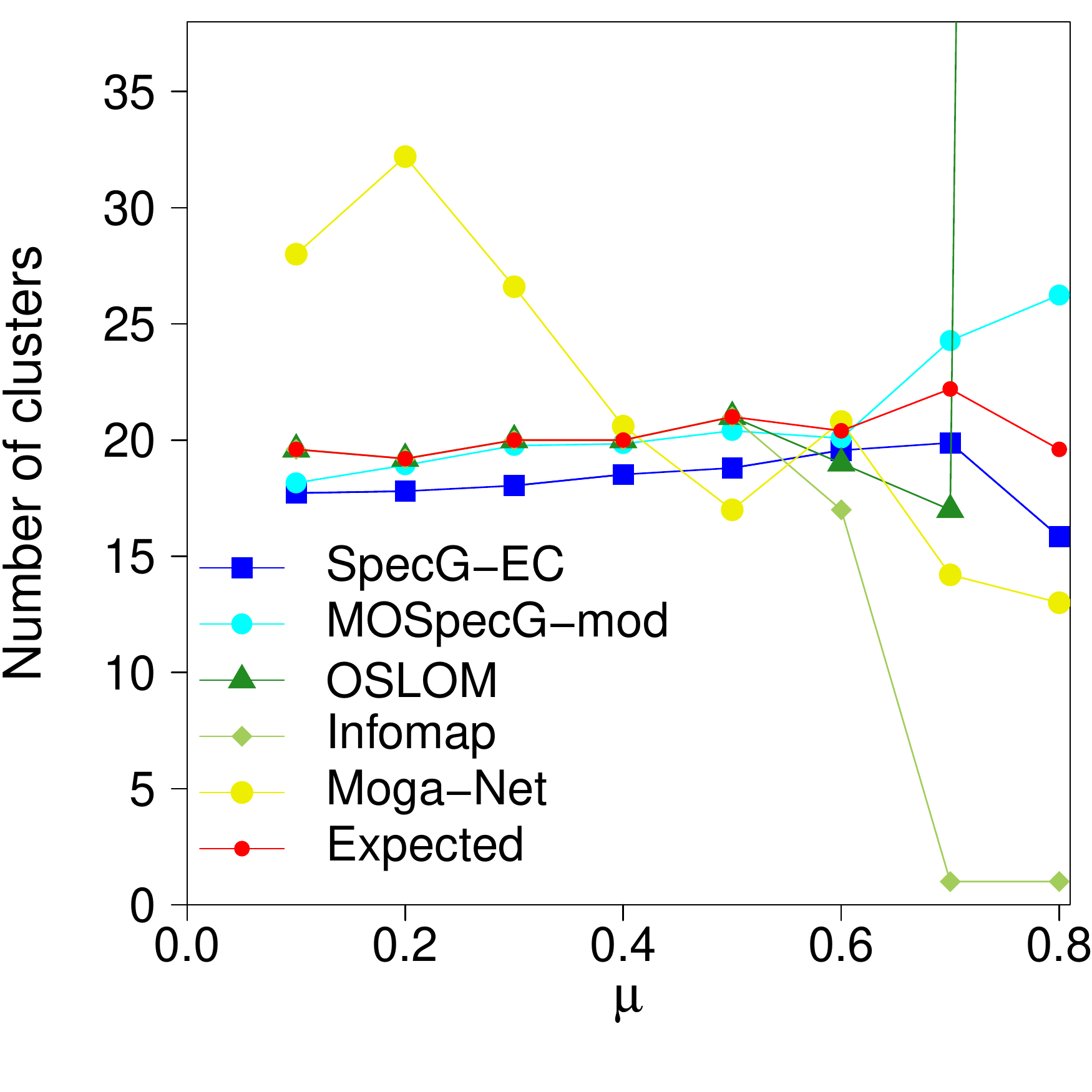}}
  \caption{Average number of clusters found by the algorithms in the partitions and in the expected partitions.}
 \label{fig:exp_ens_S1000_number}
 \end{figure}

Figure \ref{fig:exp_ens_S1000_number} shows the number of clusters obtained by the algorithms in the partitions and in the expected partitions. %\camila{[TIREI PARÁGRAFO AQUII!]}
As can be seen in Figure \ref{fig:exp_ens_S1000_number}, Moga-Net obtained the partitions with the worst estimation of numbers of clusters with regard to the expected partitions.
 On the one hand, OSLOM and Infomap found partitions whose number of clusters is exactly the expected in small and large-sized community networks with, respectively, $\mu \leq 0.6$ and $\mu \leq 0.5$. On the other hand, as opposed to \emph{SpecG-EC} and \emph{MOSpecG-mod}, Infomap failed to identify the number of clusters in the networks with $\mu \geq 0.7$.  OSLOM obtained partitions with worse estimations of the number of clusters with regard to the expected partitions than both versions of the proposed algorithm for small and large-sized community networks with, respectively, $\mu=0.8$ and $\mu \geq 0.7$. In particular, despite presenting slightly better NMI values than \emph{SpecG-EC} and \emph{MOSpecG-mod} for large-sized community network with $\mu=0.8$, OSLOM wrongly identified approximately 381 clusters, on average, more than the expected.

%%%%%%%%%%%%%%%%%%%%%%%%%%%%%%%%%%%%%%%%%%%%%%%%%%%%%%
\section{Final Remarks and Future Works}\label{sec:conclusion}
%%%%%%%%%%%%%%%%%%%%%%%%%%%%%%%%%%%%%%%%%%%%%%%%%%%%%%

This paper presented a novel spectral decomposition of modularity to clustering graphs through a multi-objective memetic algorithm called \emph{MOSpecG}. In addition, it introduced an ensemble algorithm, here called \emph{SpecG-EC}, that combines partitions obtained by \emph{MOSpecG} to provide a single partition.

The results of computational experiments using real and LFR networks showed that \emph{SpecG-EC} and the version of \emph{MOSpecG} that maximizes modularity, named \emph{MOSpecG-mod}, outperformed a multi-objective genetic algorithm found in the literature and presented reasonable running times when compared to reference algorithms. The \emph{SpecG-EC} and \emph{MOSpecG-mod} found partitions more similar to the expected ones than state-of-the-art mono-objective algorithms in artificial networks with higher mixture coefficients and satisfactory results in the remaining artificial networks. In particular, \emph{SpecG-EC} performed better in artificial large-sized community networks and outperformed state-of-the-art mono-objective algorithms in two real networks.

Because \emph{SpecG-EC} obtained better results than \emph{MOSpecG-mod} for most of the networks, we can conclude that the ensemble strategy outperformed the maximization of the classical modularity. Nonetheless, \emph{SpecG-EC} constructs its solution using partitions found by \emph{MOSpecG} and thus is slower than \emph{MOSpecG-mod}.
The experiments also suggested that both the ensemble and the modularity maximization version of the proposed algorithm provide a reasonable number of clusters in real and artificial networks.

The empirical finding  that some partitions  obtained by \emph{MOSpecG}  were more similar to the expected partitions than both the modularity maximization and ensemble partitions suggests advantages of studying the duality between the terms of the modularity using multi-objective graph clustering algorithms. In this sense, as future work, we intend to further improve the results achieved by \emph{SpecG-EC} by studying more effective procedures to select partitions from multi-objective problems for the ensemble.

 \textcolor{black}{To combine pairs of vertex partitions, evolutionary algorithms usually match the communities of the different partitions  to then perform the crossover operator. The matching of the communities is  difficult to establish. In \emph{SpecG-EC}, however, we propose a spectral analysis to this step. Unfortunately, \emph{SpecG-EC} does not scale well due to the computational burden in the  eigenvalues and eigenvectors computation. Therefore, reducing the computational cost of the spectral decomposition would make this algorithm more effective in detecting communities in larger graphs. As in  applications the networks are mostly sparse, a future research direction would be the study of the spectral decomposition of the non-backtracking matrix as the fitness function to reduce the cost of the eigen-decomposition operations of \emph{SpecG-EC}.}

\textcolor{black}{Moreover, it is worth to highlight that  in many case-oriented applications, such as the study of metabolic networks, the specialist who performs the cluster analysis prefers to investigate the results of a set of solutions instead of a unique partition. In fact,   hierarchical clustering algorithms are widely employed in these studies, primarily due to the unclear definition of clustering and the diversified characteristics of the applications. Therefore, in this sense, \emph{MOSpecG} can be a powerful tool, since it  provides  a pool of solutions from the optimization of the bi-objective problem.}

\vspace{-0.05cm}

\section*{Acknowledgments}
%%%%%%%%%%%%%%%%%%%%%%%%%%%%%%%%%%%%%%%%%%%%%%%%%%%%%%
\vspace{-0.14cm}

 The authors would like to acknowledge the foundings provided by São Paulo Research Foundation (FAPESP), grant numbers: 2016/22688-2 and 2015/21660-4; and by Conselho Nacional de Desenvolvimento Científico e Tecnológico (CNPq), grant numbers: 306036/2018-5. The authors would also like to thank the anonymous reviewers for their valuable com ments.
The second author also thanks Leonardo V. Rosset for giving her a hand.
%\section*{References}
\bibliographystyle{model5-names}
\bibliography{Bibliography}

\appendix
\section{Setting up parameters}\label{sec:appendix}

In this appendix, we present the preliminary experiments carried out using LFR and real networks to fine-tune the parameters of the proposed algorithms. First, we identified the best parameters for $\emph{MOSpecG-mod}$, which is $\emph{MOSpecG}$ when maximizing the classical version of the modularity measure.  After these parameters have been defined, we decide the best value of the parameters $N\mathcal{F}$ and $\tau$ of \emph{SpecG-EC}.

Tables \ref{tab:apx_single_S_1}, \ref{tab:apx_single_S_2}, \ref{tab:apx_single_L_1} and \ref{tab:apx_single_L_2} present the results achieved by $\emph{MOSpecG-mod}$ to detect communities in LFR networks. Tables \ref{tab:apx_single_real_1} and \ref{tab:apx_single_real_2} demonstrate the average results of 10 independent executions of $\emph{MOSpecG-mod}$ to find partitions of the real networks Karate \citep{Zachary1977}, Dolphins \citep{Lusseau2003}, Polbooks \citep{Krebs2008} and Football \citep{Girvan2002}. These tables show the NMI values of the partitions with respect to the expected ones  and the respective  execution times in seconds of \emph{MOSpecG-mod}. 
 All possible combinations of the following parameter values were considered in these experiments: $N\mathcal{G}\in \{10,50\}$, $N\mathcal{P}\in\{5,10\}$, $N\mathcal{O} \in \{10,40,70\}$, $p \in \{ \lfloor{0.1n}\rfloor, \lfloor{0.3n}\rfloor, \lfloor{0.5n}\rfloor \}$ and $IT \in \{1,5\}$. In addition, Tables \ref{tab:apx_single_S_1}, \ref{tab:apx_single_S_2}, \ref{tab:apx_single_L_1} and \ref{tab:apx_single_L_2}  report the average (AVG), maximum (MAX) and minimum (MIN) NMI values and execution times in detecting communities in the networks with each mixture coefficient value. The shades of gray used as background colors of the table cells  are in accordance with the NMI values.

In Tables \ref{tab:apx:pearson_S} and \ref{tab:apx:pearson_L}, we show the Pearson correlation coefficients\footnote{The Pearson correlation coefficient assesses the linear correlation between two variables. It is valued from -1 to 1, where -1 and 1 indicate a perfect negative and positive linear correlation, respectively; the value 0 indicates no linear correlation between the pair of variables.} between the  \emph{MOSpecG-mod}  parameters and (i)  the NMI values of the partitions with respect to the expected partitioning of LFR networks considering each possible value of $\mu$, and (ii) the corresponding average (AVG), maximum (MAX) and minimum (MIN) execution times. Table \ref{tab:apx:pearson_real} presents the Pearson correlation coefficients between the \emph{MOSpecG-mod}  parameters and the NMI values and execution time in seconds for real networks.

 As expected, the parameter $p$, which refers to the number of eigenvalues and eigenvectors  used in the spectral relaxation, presented the highest positive correlation coefficient values with respect to the execution times. 
On the one hand, the correlation coefficients between $p$ and the NMI values were lower than $0$  for small- and large-sized community networks with, respectively, $\mu \leq 0.2$ and $\mu\leq 0.5$. On the other, when considering networks with $\mu \geq 0.7$, the correlation coefficients between $p$ and the NMI values ranged from $0.197$ to $0.903$. The Pearson correlations between $p$ and the NMI values of the real network partitions were also conflicting: their values were $-0.803$, $0.406$ and $0.374$ for Karate, Polbooks and Football networks, respectively. Therefore, in spite of being  difficult to draw general conclusions by these results, we remark that the NMI values of partitions of LFR networks increased with $p$ only in networks with high mixture coefficients.

In addition, according to Tables \ref{tab:apx:pearson_S} and \ref{tab:apx:pearson_L}, one can also observe that the correlations between $N\mathcal{P}$ and the NMI values of the partitions of LFR networks with $\mu \leq 0.7$ were at least $0.371$. Nevertheless, the correlations were $-0.109$ and $-0.226$ when considering small- and large-sized community networks with $\mu=0.8$.  One conjecture that might justify the negative correlations in such cases is that \emph{MOSpecG-mod}  has more chance of selecting high-quality partitions more often for the crossover operator when $N\mathcal{P}$ is lower.
 Tables \ref{tab:apx:pearson_S}, \ref{tab:apx:pearson_L} and \ref{tab:apx:pearson_real} also show that regardless the network under consideration the execution time increases alongside $N\mathcal{P}$.

In the parameter-tuning, we selected the set of parameters for \emph{MOSpecG-mod} to obtain  partitions with satisfactory NMI values taking a reasonable running time. In line with this, parameters $p$ and $N\mathcal{P}$ received values $ \lfloor{0.1n}\rfloor$ and $5$, respectively. Nonetheless, we recommend setting higher values for $p$ and  $N\mathcal{P}$ when the networks under study  have high and low mixture coefficients, respectively.

The correlation coefficients between parameter $N\mathcal{O}$ and the NMI values did not present a clear pattern when considering  the LFR networks showed in Tables \ref{tab:apx:pearson_S} and \ref{tab:apx:pearson_L}. On the one hand, these correlations were at least $0.463$ when $\mu\leq 0.7$.  On the other, they were $-0.157$ and $-0.286$ respectively for small- and large-community size networks with $\mu=0.8$. Since increasing $N\mathcal{O}$ does not augment the execution times, we adopted the median value to this parameter, $40$.

$N\mathcal{G}$ and $IT$ did not have a strong correlation with the NMI values for the LFR networks even though the execution times increased alongside $N\mathcal{G}$ and $IT$. 
When considering the real networks, however, the correlations between $N\mathcal{G}$ and the NMI values shown in Table \ref{tab:apx:pearson_real} ranged from $0.034$ to $0.211$. We therefore considered $N\mathcal{G}=50$ to enhance the robustness of \emph{MOSpecG-mod}.
Because the parameter $IT$ only showed a strong correlation with the NMI value of the partition obtained to the Polbooks network, we used the lowest value of $IT$, 1, to carry out the experiments with the LFR networks. Nevertheless, in small networks such as the real networks tested in the experiments presented in this paper, we recommend increasing $IT$ since the computational time is significantly low in practice.

Therefore, we chose the following values of  parameters: $N\mathcal{G}=50, N\mathcal{P}=5$, $N\mathcal{O}=40$, $p=\lfloor{0.1 n}\rfloor$. The parameter  $IT$ was 1 in tests with LFR networks and 5 in the experiments with real networks. 

Tables \ref{tab:apx:MO_S}, \ref{tab:apx:MO_L} and \ref{tab:apx:MO_real} show  the NMI values of the partitions obtained by $\emph{SpecG-EC}$ in  small- and large-sized community networks with different mixture coefficients and in  real networks. This experiment was performed by fixing the values of $N\mathcal{G}, N\mathcal{P}, N\mathcal{O}$ and $p$ at values decided on the previous parameter tuning experiments and considering the respective combination of the remaining parameters: $N\mathcal{F}\in \{6,11\}$ and $\tau \in \{0.1, 0.25, 0.5, 0.75\}$.  In addition, Tables \ref{tab:apx:MO_S} and \ref{tab:apx:MO_L} report the average (AVG), maximum (MAX) and minimum (MIN) values of NMI and execution times in seconds of the consensus step in \emph{SpecG-EC} considering the networks with different mixture coefficients. Even though we did not report the running times of $\emph{SpecG-EC}$ to obtain the $N\mathcal{F}$ partitions required to define the consensus partition, the average execution times increase alongside $N\mathcal{F}$ and $\tau$.

According to Tables \ref{tab:apx:MO_S} and \ref{tab:apx:MO_L},  the highest NMI values were achieved when $N\mathcal{F}=11$ and $\tau=0.75$.   Considering Table \ref{tab:apx:MO_real}, the highest NMI values were obtained when $N\mathcal{F}=11$ and $\tau=0.5$. 
The NMI values of partitions of the LFR networks  with $\mu \leq 0.5$ when $\tau=0.5$ were on average only $2\%$ worse than the NMI values of partitions when $\tau=0.75$. As considering the real networks \emph{SpecG-EC} performed better when  $\tau$ was fixed at 0.5,  we assigned  0.5 to parameter $\tau$.

%------------------------------------------ Tables - MOSpecG-mod

\begin{table}[!htb]
  \setlength{\tabcolsep}{1pt} 
  \footnotesize
    \caption{Experiments to adjust parameters of \emph{MOSpecG-mod} to  small-sized community networks - part 1.}
    \label{tab:apx_single_S_1}
    \begin{center}
\begin{tabular}{|c|c|c|c|c|R|R|R|R|R|R|R|R|RRR|ccc|}
\hline
\multicolumn{5}{|c|}{Parameters}                   & \multicolumn{8}{c|}{$\mu$}                                     & \multicolumn{3}{c|}{NMI}     & \multicolumn{3}{c|}{Time (s)}   \\ \hline
$N\mathcal{G}$ & $N\mathcal{P}$ & $N\mathcal{O}$ & $p$ & $IT$ &
\multicolumn{1}{c|}{0.1} &	\multicolumn{1}{c|}{0.2} & \multicolumn{1}{c|}{0.3} & \multicolumn{1}{c|}{0.4} & \multicolumn{1}{c|}{0.5} &	\multicolumn{1}{c|}{0.6} & \multicolumn{1}{c|}{0.7} &	\multicolumn{1}{c|}{0.8}  & \multicolumn{1}{c}{AVG} & \multicolumn{1}{c}{MAX} & \multicolumn{1}{c|}{MIN} & \multicolumn{1}{c}{AVG} & \multicolumn{1}{c}{MAX} & \multicolumn{1}{c|}{MIN}  \\ \hline
10 & 5 & 10 & $\lfloor{0.1n}\rfloor$ & 10 & 0.901 & 0.882 & 0.887 & 0.907 & 0.853 & 0.774 & 0.55 & 0.323 & 0.76 & 0.907 & 0.323 & 3.939 & 4.563 & 3.396 \\ 
10 & 5 & 10 & $\lfloor{0.1n}\rfloor$ & 50 & 0.886 & 0.882 & 0.897 & 0.896 & 0.844 & 0.769 & 0.55 & 0.327 & 0.756 & 0.897 & 0.327 & 5.86 & 6.632 & 5.321 \\ 
10 & 5 & 10 & $\lfloor{0.3n}\rfloor$ & 10 & 0.849 & 0.838 & 0.851 & 0.831 & 0.781 & 0.697 & 0.589 & 0.469 & 0.738 & 0.851 & 0.469 & 11.346 & 13.567 & 7.686 \\ 
10 & 5 & 10 & $\lfloor{0.3n}\rfloor$ & 50 & 0.841 & 0.838 & 0.858 & 0.83 & 0.781 & 0.69 & 0.597 & 0.465 & 0.738 & 0.858 & 0.465 & 28.864 & 36.643 & 13.642 \\ 
10 & 5 & 10 & $\lfloor{0.5n}\rfloor$ & 10 & 0.821 & 0.81 & 0.822 & 0.798 & 0.761 & 0.685 & 0.634 & 0.548 & 0.735 & 0.822 & 0.548 & 20.778 & 28.898 & 12.412 \\ 
10 & 5 & 10 & $\lfloor{0.5n}\rfloor$ & 50 & 0.829 & 0.811 & 0.82 & 0.803 & 0.758 & 0.695 & 0.629 & 0.542 & 0.736 & 0.829 & 0.542 & 60.065 & 91.883 & 22.256 \\ 
10 & 5 & 40 & $\lfloor{0.1n}\rfloor$ & 10 & 0.985 & 0.98 & 0.98 & 0.969 & 0.968 & 0.943 & 0.767 & 0.331 & 0.865 & 0.985 & 0.331 & 1.865 & 2.118 & 1.689 \\ 
10 & 5 & 40 & $\lfloor{0.1n}\rfloor$ & 50 & 0.979 & 0.977 & 0.979 & 0.972 & 0.969 & 0.952 & 0.79 & 0.339 & 0.87 & 0.979 & 0.339 & 4.136 & 4.53 & 3.498 \\ 
10 & 5 & 40 & $\lfloor{0.3n}\rfloor$ & 10 & 0.981 & 0.98 & 0.999 & 0.999 & 0.996 & 0.99 & 0.919 & 0.409 & 0.909 & 0.999 & 0.409 & 10.05 & 13.067 & 5.876 \\ 
10 & 5 & 40 & $\lfloor{0.3n}\rfloor$ & 50 & 0.98 & 0.983 & 0.997 & 0.999 & 0.999 & 0.998 & 0.937 & 0.411 & 0.913 & 0.999 & 0.411 & 29.572 & 38.344 & 11.395 \\ 
10 & 5 & 40 & $\lfloor{0.5n}\rfloor$ & 10 & 0.988 & 0.987 & 0.998 & 0.999 & 1 & 0.997 & 0.903 & 0.469 & 0.918 & 1 & 0.469 & 20.374 & 27.569 & 12.813 \\ 
10 & 5 & 40 & $\lfloor{0.5n}\rfloor$ & 50 & 0.986 & 0.987 & 1 & 1 & 1 & 0.999 & 0.959 & 0.456 & 0.923 & 1 & 0.456 & 61.71 & 100.439 & 22.056 \\ 
10 & 5 & 70 & $\lfloor{0.1n}\rfloor$ & 10 & 0.981 & 0.981 & 0.973 & 0.969 & 0.969 & 0.953 & 0.749 & 0.33 & 0.863 & 0.981 & 0.33 & 2.335 & 2.622 & 1.896 \\ 
10 & 5 & 70 & $\lfloor{0.1n}\rfloor$ & 50 & 0.984 & 0.981 & 0.98 & 0.969 & 0.973 & 0.953 & 0.779 & 0.334 & 0.869 & 0.984 & 0.334 & 4.728 & 5.086 & 3.979 \\ 
10 & 5 & 70 & $\lfloor{0.3n}\rfloor$ & 10 & 0.983 & 0.98 & 0.997 & 1 & 0.997 & 0.99 & 0.916 & 0.414 & 0.91 & 1 & 0.414 & 10.37 & 12.769 & 6.580 \\ 
10 & 5 & 70 & $\lfloor{0.3n}\rfloor$ & 50 & 0.981 & 0.983 & 0.999 & 1 & 0.996 & 0.996 & 0.942 & 0.419 & 0.915 & 1 & 0.419 & 27.99 & 40.331 & 11.746 \\ 
10 & 5 & 70 & $\lfloor{0.5n}\rfloor$ & 10 & 0.988 & 0.99 & 0.998 & 0.999 & 0.999 & 0.997 & 0.901 & 0.464 & 0.917 & 0.999 & 0.464 & 19.707 & 26.755 & 13.802 \\ 
10 & 5 & 70 & $\lfloor{0.5n}\rfloor$ & 50 & 0.983 & 0.988 & 0.999 & 0.999 & 1 & 0.999 & 0.962 & 0.447 & 0.922 & 1 & 0.447 & 57.508 & 92.233 & 20.299 \\ 
10 & 10 & 10 & $\lfloor{0.1n}\rfloor$ & 10 & 0.987 & 0.986 & 0.98 & 0.973 & 0.97 & 0.948 & 0.749 & 0.329 & 0.865 & 0.987 & 0.329 & 2.466 & 2.823 & 2.217 \\ 
10 & 10 & 10 & $\lfloor{0.1n}\rfloor$ & 50 & 0.988 & 0.984 & 0.982 & 0.976 & 0.977 & 0.958 & 0.792 & 0.339 & 0.875 & 0.988 & 0.339 & 6.916 & 7.535 & 5.380 \\ 
10 & 10 & 10 & $\lfloor{0.3n}\rfloor$ & 10 & 0.987 & 0.985 & 0.998 & 0.999 & 0.996 & 0.992 & 0.908 & 0.418 & 0.91 & 0.999 & 0.418 & 13.816 & 17.218 & 7.192 \\ 
10 & 10 & 10 & $\lfloor{0.3n}\rfloor$ & 50 & 0.985 & 0.984 & 0.998 & 1 & 1 & 0.996 & 0.95 & 0.419 & 0.917 & 1 & 0.419 & 49.406 & 66.762 & 18.639 \\ 
10 & 10 & 10 & $\lfloor{0.5n}\rfloor$ & 10 & 0.985 & 0.99 & 0.999 & 0.998 & 0.999 & 0.997 & 0.887 & 0.47 & 0.916 & 0.999 & 0.47 & 28.757 & 42.609 & 13.662 \\ 
10 & 10 & 10 & $\lfloor{0.5n}\rfloor$ & 50 & 0.987 & 0.99 & 1 & 1 & 1 & 1 & 0.947 & 0.461 & 0.923 & 1 & 0.461 & 108.069 & 175.175 & 36.357 \\ 
10 & 10 & 40 & $\lfloor{0.1n}\rfloor$ & 10 & 0.986 & 0.98 & 0.982 & 0.976 & 0.973 & 0.95 & 0.78 & 0.331 & 0.87 & 0.986 & 0.331 & 2.391 & 2.648 & 2.184 \\ 
10 & 10 & 40 & $\lfloor{0.1n}\rfloor$ & 50 & 0.989 & 0.983 & 0.983 & 0.982 & 0.973 & 0.95 & 0.793 & 0.337 & 0.874 & 0.989 & 0.337 & 7.228 & 8.476 & 5.596 \\ 
10 & 10 & 40 & $\lfloor{0.3n}\rfloor$ & 10 & 0.986 & 0.985 & 0.999 & 0.999 & 1 & 0.992 & 0.919 & 0.411 & 0.911 & 1 & 0.411 & 14.935 & 18.538 & 8.249 \\ 
10 & 10 & 40 & $\lfloor{0.3n}\rfloor$ & 50 & 0.984 & 0.983 & 1 & 0.999 & 0.999 & 0.999 & 0.942 & 0.411 & 0.915 & 1 & 0.411 & 49.835 & 64.276 & 19.583 \\ 
10 & 10 & 40 & $\lfloor{0.5n}\rfloor$ & 10 & 0.987 & 0.99 & 1 & 0.999 & 0.999 & 0.999 & 0.929 & 0.482 & 0.923 & 1 & 0.482 & 28.679 & 44.118 & 12.843 \\ 
10 & 10 & 40 & $\lfloor{0.5n}\rfloor$ & 50 & 0.989 & 0.99 & 1 & 1 & 1 & 0.999 & 0.969 & 0.46 & 0.926 & 1 & 0.46 & 106.041 & 175.882 & 33.539 \\ 
10 & 10 & 70 & $\lfloor{0.1n}\rfloor$ & 10 & 0.992 & 0.979 & 0.979 & 0.976 & 0.975 & 0.955 & 0.754 & 0.326 & 0.867 & 0.992 & 0.326 & 2.527 & 3.088 & 2.236 \\ 
10 & 10 & 70 & $\lfloor{0.1n}\rfloor$ & 50 & 0.984 & 0.982 & 0.982 & 0.973 & 0.972 & 0.956 & 0.788 & 0.338 & 0.872 & 0.984 & 0.338 & 6.912 & 7.577 & 5.488 \\ 
10 & 10 & 70 & $\lfloor{0.3n}\rfloor$ & 10 & 0.986 & 0.981 & 0.999 & 0.999 & 0.998 & 0.995 & 0.919 & 0.411 & 0.911 & 0.999 & 0.411 & 13.736 & 17.072 & 6.914 \\ 
10 & 10 & 70 & $\lfloor{0.3n}\rfloor$ & 50 & 0.985 & 0.986 & 0.999 & 1 & 0.999 & 0.997 & 0.954 & 0.429 & 0.919 & 1 & 0.429 & 48.612 & 64.408 & 18.673 \\ 
10 & 10 & 70 & $\lfloor{0.5n}\rfloor$ & 10 & 0.99 & 0.99 & 0.996 & 1 & 0.999 & 0.998 & 0.911 & 0.446 & 0.916 & 1 & 0.446 & 28.189 & 42.39 & 13.094 \\ 
10 & 10 & 70 & $\lfloor{0.5n}\rfloor$ & 50 & 0.986 & 0.993 & 0.999 & 1 & 1 & 0.999 & 0.957 & 0.453 & 0.923 & 1 & 0.453 & 106.077 & 175.132 & 32.210
 \\ \hline
\end{tabular}
\end{center}
\end{table}

\begin{table}[!htb]
  \setlength{\tabcolsep}{1pt} 
  \footnotesize
  \caption{Experiments to adjust parameters of \emph{MOSpecG-mod} to  small-sized community networks  - part 2.}
  \label{tab:apx_single_S_2}
  \begin{center}
\begin{tabular}{|c|c|c|c|c|R|R|R|R|R|R|R|R|RRR|ccc|}
\hline
\multicolumn{5}{|c|}{Parameters}                   & \multicolumn{8}{c|}{$\mu$}                                     & \multicolumn{3}{c|}{NMI}     & \multicolumn{3}{c|}{Time (s)}   \\ \hline
$N\mathcal{G}$ & $N\mathcal{P}$ & $N\mathcal{O}$ & $p$ & $IT$ &
\multicolumn{1}{c|}{0.1} &	\multicolumn{1}{c|}{0.2} & \multicolumn{1}{c|}{0.3} & \multicolumn{1}{c|}{0.4} & \multicolumn{1}{c|}{0.5} &	\multicolumn{1}{c|}{0.6} & \multicolumn{1}{c|}{0.7} &	\multicolumn{1}{c|}{0.8}  & \multicolumn{1}{c}{AVG} & \multicolumn{1}{c}{MAX} & \multicolumn{1}{c|}{MIN} & \multicolumn{1}{c}{AVG} & \multicolumn{1}{c}{MAX} & \multicolumn{1}{c|}{MIN}  \\ \hline
50 & 5 & 10 & $\lfloor{0.1n}\rfloor$ & 10 & 0.899 & 0.885 & 0.896 & 0.892 & 0.849 & 0.776 & 0.542 & 0.326 & 0.758 & 0.899 & 0.326 & 4.241 & 4.554 & 3.764 \\ 
50 & 5 & 10 & $\lfloor{0.1n}\rfloor$ & 50 & 0.885 & 0.888 & 0.901 & 0.892 & 0.852 & 0.783 & 0.562 & 0.315 & 0.76 & 0.901 & 0.315 & 15.326 & 16.689 & 11.897 \\ 
50 & 5 & 10 & $\lfloor{0.3n}\rfloor$ & 10 & 0.849 & 0.851 & 0.859 & 0.819 & 0.786 & 0.695 & 0.595 & 0.463 & 0.74 & 0.859 & 0.463 & 26.935 & 34.837 & 11.574 \\ 
50 & 5 & 10 & $\lfloor{0.3n}\rfloor$ & 50 & 0.851 & 0.841 & 0.845 & 0.823 & 0.777 & 0.689 & 0.597 & 0.467 & 0.736 & 0.851 & 0.467 & 114.172 & 153.981 & 39.929 \\ 
50 & 5 & 10 & $\lfloor{0.5n}\rfloor$ & 10 & 0.826 & 0.824 & 0.818 & 0.795 & 0.754 & 0.693 & 0.633 & 0.549 & 0.737 & 0.826 & 0.549 & 57.523 & 91.836 & 21.432 \\ 
50 & 5 & 10 & $\lfloor{0.5n}\rfloor$ & 50 & 0.824 & 0.809 & 0.822 & 0.792 & 0.751 & 0.689 & 0.633 & 0.545 & 0.733 & 0.824 & 0.545 & 250.024 & 423.033 & 68.129 \\ 
50 & 5 & 40 & $\lfloor{0.1n}\rfloor$ & 10 & 0.984 & 0.98 & 0.98 & 0.971 & 0.975 & 0.95 & 0.779 & 0.336 & 0.869 & 0.984 & 0.336 & 4.101 & 4.582 & 3.415 \\ 
50 & 5 & 40 & $\lfloor{0.1n}\rfloor$ & 50 & 0.986 & 0.987 & 0.976 & 0.977 & 0.973 & 0.957 & 0.812 & 0.343 & 0.876 & 0.987 & 0.343 & 15.084 & 16.476 & 11.360 \\ 
50 & 5 & 40 & $\lfloor{0.3n}\rfloor$ & 10 & 0.979 & 0.98 & 0.997 & 0.999 & 0.996 & 0.991 & 0.909 & 0.412 & 0.908 & 0.999 & 0.412 & 27.033 & 36.392 & 11.138 \\ 
50 & 5 & 40 & $\lfloor{0.3n}\rfloor$ & 50 & 0.985 & 0.983 & 0.998 & 1 & 0.999 & 0.996 & 0.963 & 0.4 & 0.916 & 1 & 0.4 & 113.672 & 152.987 & 39.445 \\ 
50 & 5 & 40 & $\lfloor{0.5n}\rfloor$ & 10 & 0.988 & 0.99 & 0.999 & 0.998 & 1 & 0.998 & 0.928 & 0.465 & 0.921 & 1 & 0.465 & 56.769 & 91.557 & 20.047 \\ 
50 & 5 & 40 & $\lfloor{0.5n}\rfloor$ & 50 & 0.989 & 0.99 & 0.999 & 1 & 1 & 1 & 0.982 & 0.436 & 0.925 & 1 & 0.436 & 250.195 & 421.961 & 67.939 \\ 
50 & 5 & 70 & $\lfloor{0.1n}\rfloor$ & 10 & 0.979 & 0.978 & 0.98 & 0.975 & 0.967 & 0.948 & 0.767 & 0.336 & 0.866 & 0.98 & 0.336 & 4.082 & 4.389 & 3.398 \\ 
50 & 5 & 70 & $\lfloor{0.1n}\rfloor$ & 50 & 0.981 & 0.975 & 0.977 & 0.971 & 0.968 & 0.957 & 0.806 & 0.342 & 0.872 & 0.981 & 0.342 & 15.157 & 16.679 & 11.699 \\ 
50 & 5 & 70 & $\lfloor{0.3n}\rfloor$ & 10 & 0.98 & 0.981 & 0.998 & 0.998 & 0.996 & 0.996 & 0.922 & 0.413 & 0.911 & 0.998 & 0.413 & 26.693 & 34.541 & 11.423 \\ 
50 & 5 & 70 & $\lfloor{0.3n}\rfloor$ & 50 & 0.983 & 0.983 & 0.996 & 0.999 & 0.998 & 0.995 & 0.958 & 0.425 & 0.917 & 0.999 & 0.425 & 113.984 & 152.902 & 39.230 \\ 
50 & 5 & 70 & $\lfloor{0.5n}\rfloor$ & 10 & 0.984 & 0.982 & 0.998 & 0.999 & 0.997 & 0.999 & 0.928 & 0.478 & 0.921 & 0.999 & 0.478 & 56.74 & 91.630 & 20.113 \\ 
50 & 5 & 70 & $\lfloor{0.5n}\rfloor$ & 50 & 0.984 & 0.986 & 0.998 & 1 & 1 & 0.999 & 0.983 & 0.427 & 0.922 & 1 & 0.427 & 250.17 & 421.768 & 68.330 \\ 
50 & 10 & 10 & $\lfloor{0.1n}\rfloor$ & 10 & 0.988 & 0.98 & 0.984 & 0.977 & 0.974 & 0.957 & 0.754 & 0.332 & 0.868 & 0.988 & 0.332 & 6.882 & 7.759 & 5.594 \\ 
50 & 10 & 10 & $\lfloor{0.1n}\rfloor$ & 50 & 0.987 & 0.982 & 0.982 & 0.978 & 0.978 & 0.957 & 0.811 & 0.344 & 0.877 & 0.987 & 0.344 & 28.632 & 31.543 & 21.711 \\ 
50 & 10 & 10 & $\lfloor{0.3n}\rfloor$ & 10 & 0.986 & 0.985 & 0.997 & 0.999 & 0.997 & 0.991 & 0.917 & 0.416 & 0.911 & 0.999 & 0.416 & 48.794 & 64.591 & 18.255 \\ 
50 & 10 & 10 & $\lfloor{0.3n}\rfloor$ & 50 & 0.986 & 0.987 & 1 & 1 & 0.999 & 0.998 & 0.962 & 0.424 & 0.92 & 1 & 0.424 & 222.775 & 300.265 & 74.748 \\ 
50 & 10 & 10 & $\lfloor{0.5n}\rfloor$ & 10 & 0.991 & 0.989 & 0.998 & 1 & 1 & 0.996 & 0.937 & 0.468 & 0.922 & 1 & 0.468 & 105.128 & 174.042 & 32.190 \\ 
50 & 10 & 10 & $\lfloor{0.5n}\rfloor$ & 50 & 0.989 & 0.989 & 0.998 & 1 & 1 & 0.999 & 0.98 & 0.442 & 0.925 & 1 & 0.442 & 490.752 & 833.401 & 129.655 \\ 
50 & 10 & 40 & $\lfloor{0.1n}\rfloor$ & 10 & 0.987 & 0.983 & 0.979 & 0.981 & 0.973 & 0.951 & 0.761 & 0.332 & 0.868 & 0.987 & 0.332 & 6.809 & 7.695 & 5.380 \\ 
50 & 10 & 40 & $\lfloor{0.1n}\rfloor$ & 50 & 0.981 & 0.983 & 0.981 & 0.985 & 0.972 & 0.958 & 0.803 & 0.346 & 0.876 & 0.985 & 0.346 & 28.649 & 31.526 & 21.613 \\ 
50 & 10 & 40 & $\lfloor{0.3n}\rfloor$ & 10 & 0.987 & 0.981 & 0.998 & 1 & 0.998 & 0.996 & 0.937 & 0.416 & 0.914 & 1 & 0.416 & 48.478 & 64.132 & 18.186 \\ 
50 & 10 & 40 & $\lfloor{0.3n}\rfloor$ & 50 & 0.989 & 0.988 & 0.998 & 1 & 0.999 & 0.996 & 0.96 & 0.41 & 0.918 & 1 & 0.41 & 222.641 & 301 & 74.162 \\ 
50 & 10 & 40 & $\lfloor{0.5n}\rfloor$ & 10 & 0.987 & 0.99 & 0.998 & 0.999 & 0.998 & 0.998 & 0.942 & 0.467 & 0.922 & 0.999 & 0.467 & 105.672 & 174.514 & 32.057 \\ 
50 & 10 & 40 & $\lfloor{0.5n}\rfloor$ & 50 & 0.99 & 0.993 & 1 & 1 & 1 & 0.999 & 0.983 & 0.425 & 0.924 & 1 & 0.425 & 490.804 & 834.429 & 128.175 \\ 
50 & 10 & 70 & $\lfloor{0.1n}\rfloor$ & 10 & 0.985 & 0.985 & 0.985 & 0.978 & 0.977 & 0.95 & 0.772 & 0.33 & 0.87 & 0.985 & 0.33 & 6.867 & 7.601 & 5.289 \\ 
50 & 10 & 70 & $\lfloor{0.1n}\rfloor$ & 50 & 0.988 & 0.984 & 0.983 & 0.975 & 0.972 & 0.956 & 0.805 & 0.341 & 0.876 & 0.988 & 0.341 & 28.624 & 31.634 & 21.670 \\ 
50 & 10 & 70 & $\lfloor{0.3n}\rfloor$ & 10 & 0.984 & 0.984 & 0.998 & 0.999 & 0.998 & 0.995 & 0.939 & 0.418 & 0.914 & 0.999 & 0.418 & 48.673 & 64.261 & 18.514 \\ 
50 & 10 & 70 & $\lfloor{0.3n}\rfloor$ & 50 & 0.986 & 0.985 & 0.999 & 1 & 1 & 0.998 & 0.96 & 0.416 & 0.918 & 1 & 0.416 & 223.151 & 300.679 & 75.510 \\ 
50 & 10 & 70 & $\lfloor{0.5n}\rfloor$ & 10 & 0.988 & 0.99 & 0.999 & 1 & 0.999 & 0.997 & 0.953 & 0.459 & 0.923 & 1 & 0.459 & 105.182 & 174.151 & 31.984 \\ 
50 & 10 & 70 & $\lfloor{0.5n}\rfloor$ & 50 & 0.988 & 0.99 & 0.998 & 1 & 0.999 & 1 & 0.983 & 0.433 & 0.924 & 1 & 0.433 & 490.98 & 837.837 & 128.176 \\ \hline
\end{tabular}
\end{center}
\end{table}

\begin{table}[!htb]
  \setlength{\tabcolsep}{1pt} 
  \footnotesize
  \caption{Experiments to adjust parameters of \emph{MOSpecG-mod} to  large-sized community networks  - part 1.}
  \label{tab:apx_single_L_1}
  \begin{center}
\begin{tabular}{|c|c|c|c|c|R|R|R|R|R|R|R|R|RRR|ccc|}
\hline
\multicolumn{5}{|c|}{Parameters}                   & \multicolumn{8}{c|}{$\mu$}                                     & \multicolumn{3}{c|}{NMI}     & \multicolumn{3}{c|}{Time (s)}   \\ \hline
$N\mathcal{G}$ & $N\mathcal{P}$ & $N\mathcal{O}$ & $p$ & $IT$ &
\multicolumn{1}{c|}{0.1} &	\multicolumn{1}{c|}{0.2} & \multicolumn{1}{c|}{0.3} & \multicolumn{1}{c|}{0.4} & \multicolumn{1}{c|}{0.5} &	\multicolumn{1}{c|}{0.6} & \multicolumn{1}{c|}{0.7} &	\multicolumn{1}{c|}{0.8}  & \multicolumn{1}{c}{AVG} & \multicolumn{1}{c}{MAX} & \multicolumn{1}{c|}{MIN} & \multicolumn{1}{c}{AVG} & \multicolumn{1}{c}{MAX} & \multicolumn{1}{c|}{MIN}  \\ \hline
10 & 5 & 10 & $\lfloor{0.1n}\rfloor$ & 10 & 0.926 & 0.931 & 0.92 & 0.895 & 0.795 & 0.595 & 0.369 & 0.188 & 0.702 & 0.931 & 0.188 & 4.15 & 4.754 & 3.774 \\ 
10 & 5 & 10 & $\lfloor{0.1n}\rfloor$ & 50 & 0.927 & 0.921 & 0.91 & 0.886 & 0.783 & 0.606 & 0.354 & 0.188 & 0.697 & 0.927 & 0.188 & 5.878 & 6.711 & 5.152 \\ 
10 & 5 & 10 & $\lfloor{0.3n}\rfloor$ & 10 & 0.833 & 0.794 & 0.748 & 0.699 & 0.625 & 0.507 & 0.419 & 0.329 & 0.619 & 0.833 & 0.329 & 12.368 & 13.562 & 10.294 \\ 
10 & 5 & 10 & $\lfloor{0.3n}\rfloor$ & 50 & 0.839 & 0.789 & 0.738 & 0.699 & 0.624 & 0.513 & 0.419 & 0.327 & 0.619 & 0.839 & 0.327 & 33.63 & 38.342 & 21.380 \\ 
10 & 5 & 10 & $\lfloor{0.5n}\rfloor$ & 10 & 0.803 & 0.755 & 0.719 & 0.662 & 0.605 & 0.527 & 0.484 & 0.412 & 0.621 & 0.803 & 0.412 & 23.246 & 28.721 & 15.972 \\ 
10 & 5 & 10 & $\lfloor{0.5n}\rfloor$ & 50 & 0.81 & 0.757 & 0.719 & 0.668 & 0.608 & 0.528 & 0.481 & 0.415 & 0.623 & 0.81 & 0.415 & 70.536 & 99.119 & 35.598 \\ 
10 & 5 & 40 & $\lfloor{0.1n}\rfloor$ & 10 & 1 & 1 & 1 & 1 & 0.993 & 0.915 & 0.537 & 0.147 & 0.824 & 1 & 0.147 & 2.629 & 3.3 & 2.094 \\ 
10 & 5 & 40 & $\lfloor{0.1n}\rfloor$ & 50 & 1 & 0.999 & 0.999 & 1 & 0.994 & 0.923 & 0.584 & 0.151 & 0.831 & 1 & 0.151 & 4.989 & 5.562 & 4.291 \\ 
10 & 5 & 40 & $\lfloor{0.3n}\rfloor$ & 10 & 1 & 1 & 0.999 & 1 & 0.996 & 0.911 & 0.524 & 0.227 & 0.832 & 1 & 0.227 & 11.791 & 12.949 & 9.636 \\ 
10 & 5 & 40 & $\lfloor{0.3n}\rfloor$ & 50 & 1 & 1 & 1 & 1 & 1 & 0.965 & 0.576 & 0.212 & 0.844 & 1 & 0.212 & 34.373 & 38.461 & 21.337 \\ 
10 & 5 & 40 & $\lfloor{0.5n}\rfloor$ & 10 & 1 & 1 & 1 & 1 & 0.994 & 0.879 & 0.528 & 0.288 & 0.836 & 1 & 0.288 & 22.75 & 28.699 & 15.740 \\ 
10 & 5 & 40 & $\lfloor{0.5n}\rfloor$ & 50 & 1 & 1 & 1 & 1 & 1 & 0.94 & 0.585 & 0.273 & 0.85 & 1 & 0.273 & 71.339 & 95.149 & 36.073 \\ 
10 & 5 & 70 & $\lfloor{0.1n}\rfloor$ & 10 & 0.998 & 1 & 0.998 & 0.997 & 0.991 & 0.909 & 0.534 & 0.156 & 0.823 & 1 & 0.156 & 2.429 & 3 & 2.041 \\ 
10 & 5 & 70 & $\lfloor{0.1n}\rfloor$ & 50 & 1 & 1 & 0.999 & 0.998 & 0.993 & 0.92 & 0.562 & 0.147 & 0.827 & 1 & 0.147 & 5.268 & 6.615 & 4.113 \\ 
10 & 5 & 70 & $\lfloor{0.3n}\rfloor$ & 10 & 1 & 1 & 1 & 1 & 0.998 & 0.926 & 0.528 & 0.228 & 0.835 & 1 & 0.228 & 11.001 & 11.555 & 10.506 \\ 
10 & 5 & 70 & $\lfloor{0.3n}\rfloor$ & 50 & 1 & 1 & 1 & 1 & 1 & 0.962 & 0.595 & 0.213 & 0.846 & 1 & 0.213 & 31.367 & 34.998 & 19.831 \\ 
10 & 5 & 70 & $\lfloor{0.5n}\rfloor$ & 10 & 1 & 1 & 0.999 & 1 & 0.992 & 0.875 & 0.549 & 0.293 & 0.839 & 1 & 0.293 & 20.65 & 26.147 & 13.595 \\ 
10 & 5 & 70 & $\lfloor{0.5n}\rfloor$ & 50 & 1 & 1 & 1 & 1 & 0.998 & 0.957 & 0.587 & 0.265 & 0.851 & 1 & 0.265 & 67.941 & 93.517 & 33.107 \\ 
10 & 10 & 10 & $\lfloor{0.1n}\rfloor$ & 10 & 1 & 1 & 1 & 1 & 0.993 & 0.917 & 0.527 & 0.156 & 0.824 & 1 & 0.156 & 2.487 & 2.936 & 2.125 \\ 
10 & 10 & 10 & $\lfloor{0.1n}\rfloor$ & 50 & 0.999 & 1 & 1 & 1 & 0.995 & 0.923 & 0.558 & 0.163 & 0.83 & 1 & 0.163 & 6.855 & 7.51 & 6.141 \\ 
10 & 10 & 10 & $\lfloor{0.3n}\rfloor$ & 10 & 1 & 1 & 1 & 1 & 0.991 & 0.907 & 0.516 & 0.232 & 0.831 & 1 & 0.232 & 15.943 & 18.082 & 10.799 \\ 
10 & 10 & 10 & $\lfloor{0.3n}\rfloor$ & 50 & 1 & 1 & 1 & 1 & 1 & 0.952 & 0.579 & 0.217 & 0.844 & 1 & 0.217 & 59.627 & 73.394 & 34.367 \\ 
10 & 10 & 10 & $\lfloor{0.5n}\rfloor$ & 10 & 1 & 1 & 1 & 0.995 & 0.986 & 0.871 & 0.523 & 0.304 & 0.835 & 1 & 0.304 & 32.498 & 43.211 & 18.572 \\ 
10 & 10 & 10 & $\lfloor{0.5n}\rfloor$ & 50 & 1 & 1 & 1 & 1 & 1 & 0.936 & 0.564 & 0.282 & 0.848 & 1 & 0.282 & 128.304 & 179.584 & 57.375 \\ 
10 & 10 & 40 & $\lfloor{0.1n}\rfloor$ & 10 & 1 & 1 & 0.999 & 1 & 0.993 & 0.914 & 0.546 & 0.152 & 0.826 & 1 & 0.152 & 2.584 & 2.807 & 2.286 \\ 
10 & 10 & 40 & $\lfloor{0.1n}\rfloor$ & 50 & 1 & 1 & 1 & 1 & 0.995 & 0.924 & 0.557 & 0.15 & 0.828 & 1 & 0.15 & 6.831 & 7.496 & 6.279 \\ 
10 & 10 & 40 & $\lfloor{0.3n}\rfloor$ & 10 & 1 & 1 & 1 & 1 & 1 & 0.924 & 0.55 & 0.228 & 0.838 & 1 & 0.228 & 15.832 & 17.616 & 11.098 \\ 
10 & 10 & 40 & $\lfloor{0.3n}\rfloor$ & 50 & 1 & 1 & 1 & 1 & 1 & 0.958 & 0.581 & 0.22 & 0.845 & 1 & 0.22 & 57.065 & 64.895 & 33.826 \\ 
10 & 10 & 40 & $\lfloor{0.5n}\rfloor$ & 10 & 1 & 1 & 1 & 1 & 0.998 & 0.894 & 0.534 & 0.288 & 0.839 & 1 & 0.288 & 32.421 & 43.250 & 18.655 \\ 
10 & 10 & 40 & $\lfloor{0.5n}\rfloor$ & 50 & 1 & 1 & 1 & 1 & 1 & 0.955 & 0.552 & 0.267 & 0.847 & 1 & 0.267 & 126.972 & 178.203 & 59.322 \\ 
10 & 10 & 70 & $\lfloor{0.1n}\rfloor$ & 10 & 1 & 1 & 0.998 & 1 & 0.993 & 0.916 & 0.55 & 0.163 & 0.828 & 1 & 0.163 & 2.474 & 2.855 & 2.129 \\ 
10 & 10 & 70 & $\lfloor{0.1n}\rfloor$ & 50 & 1 & 1 & 0.997 & 1 & 0.994 & 0.923 & 0.562 & 0.162 & 0.83 & 1 & 0.162 & 6.802 & 7.603 & 6.094 \\ 
10 & 10 & 70 & $\lfloor{0.3n}\rfloor$ & 10 & 1 & 1 & 1 & 1 & 1 & 0.951 & 0.562 & 0.224 & 0.842 & 1 & 0.224 & 15.567 & 17.461 & 10.675 \\ 
10 & 10 & 70 & $\lfloor{0.3n}\rfloor$ & 50 & 1 & 1 & 1 & 1 & 1 & 0.957 & 0.604 & 0.218 & 0.847 & 1 & 0.218 & 56.963 & 64.597 & 33.589 \\ 
10 & 10 & 70 & $\lfloor{0.5n}\rfloor$ & 10 & 1 & 1 & 1 & 1 & 1 & 0.908 & 0.549 & 0.288 & 0.843 & 1 & 0.288 & 33.021 & 43.464 & 19.996 \\ 
10 & 10 & 70 & $\lfloor{0.5n}\rfloor$ & 50 & 1 & 1 & 1 & 1 & 1 & 0.958 & 0.586 & 0.264 & 0.851 & 1 & 0.264 & 125.699 & 177.726 & 58.092  
\\ \hline
\end{tabular}
\end{center}
\end{table}

\begin{table}[!htb]
  \setlength{\tabcolsep}{1pt} 
  \footnotesize
  \caption{Experiments to adjust parameters of \emph{MOSpecG-mod} to  large-sized community networks  - part 2.}
  \label{tab:apx_single_L_2}
  \begin{center}
\begin{tabular}{|c|c|c|c|c|R|R|R|R|R|R|R|R|RRR|ccc|}
\hline
\multicolumn{5}{|c|}{Parameters}                   & \multicolumn{8}{c|}{$\mu$}                                     & \multicolumn{3}{c|}{NMI}     & \multicolumn{3}{c|}{Time (s)}   \\ \hline
$N\mathcal{G}$ & $N\mathcal{P}$ & $N\mathcal{O}$ & $p$ & $IT$ &
\multicolumn{1}{c|}{0.1} &	\multicolumn{1}{c|}{0.2} & \multicolumn{1}{c|}{0.3} & \multicolumn{1}{c|}{0.4} & \multicolumn{1}{c|}{0.5} &	\multicolumn{1}{c|}{0.6} & \multicolumn{1}{c|}{0.7} &	\multicolumn{1}{c|}{0.8}  & \multicolumn{1}{c}{AVG} & \multicolumn{1}{c}{MAX} & \multicolumn{1}{c|}{MIN} & \multicolumn{1}{c}{AVG} & \multicolumn{1}{c}{MAX} & \multicolumn{1}{c|}{MIN}  \\ \hline
50 & 5 & 10 & $\lfloor{0.1n}\rfloor$ & 10 & 0.932 & 0.926 & 0.915 & 0.881 & 0.779 & 0.592 & 0.358 & 0.186 & 0.696 & 0.932 & 0.186 & 4.224 & 4.759 & 3.710 \\ 
50 & 5 & 10 & $\lfloor{0.1n}\rfloor$ & 50 & 0.93 & 0.931 & 0.914 & 0.893 & 0.792 & 0.591 & 0.366 & 0.181 & 0.7 & 0.931 & 0.181 & 14.998 & 16.184 & 13.520 \\ 
50 & 5 & 10 & $\lfloor{0.3n}\rfloor$ & 10 & 0.844 & 0.793 & 0.753 & 0.684 & 0.626 & 0.524 & 0.423 & 0.325 & 0.622 & 0.844 & 0.325 & 31.105 & 35.189 & 19.478 \\ 
50 & 5 & 10 & $\lfloor{0.3n}\rfloor$ & 50 & 0.843 & 0.789 & 0.754 & 0.702 & 0.622 & 0.509 & 0.426 & 0.322 & 0.621 & 0.843 & 0.322 & 135.181 & 153.748 & 76.298 \\ 
50 & 5 & 10 & $\lfloor{0.5n}\rfloor$ & 10 & 0.808 & 0.756 & 0.718 & 0.666 & 0.609 & 0.528 & 0.479 & 0.412 & 0.622 & 0.808 & 0.412 & 67.111 & 93.828 & 33.073 \\ 
50 & 5 & 10 & $\lfloor{0.5n}\rfloor$ & 50 & 0.807 & 0.745 & 0.716 & 0.669 & 0.602 & 0.532 & 0.478 & 0.409 & 0.62 & 0.807 & 0.409 & 301.115 & 428.742 & 131.385 \\ 
50 & 5 & 40 & $\lfloor{0.1n}\rfloor$ & 10 & 1 & 0.998 & 1 & 0.998 & 0.99 & 0.911 & 0.547 & 0.163 & 0.826 & 1 & 0.163 & 4.087 & 4.635 & 3.712 \\ 
50 & 5 & 40 & $\lfloor{0.1n}\rfloor$ & 50 & 1 & 0.999 & 0.997 & 0.997 & 0.99 & 0.921 & 0.579 & 0.158 & 0.83 & 1 & 0.158 & 14.683 & 16.143 & 13.091 \\ 
50 & 5 & 40 & $\lfloor{0.3n}\rfloor$ & 10 & 1 & 1 & 1 & 1 & 1 & 0.937 & 0.563 & 0.23 & 0.841 & 1 & 0.23 & 31.282 & 34.875 & 20.368 \\ 
50 & 5 & 40 & $\lfloor{0.3n}\rfloor$ & 50 & 1 & 1 & 1 & 1 & 1 & 0.974 & 0.641 & 0.201 & 0.852 & 1 & 0.201 & 134.389 & 153.39 & 75.743 \\ 
50 & 5 & 40 & $\lfloor{0.5n}\rfloor$ & 10 & 1 & 1 & 1 & 1 & 0.999 & 0.937 & 0.529 & 0.301 & 0.846 & 1 & 0.301 & 67.367 & 94.62 & 32.814 \\ 
50 & 5 & 40 & $\lfloor{0.5n}\rfloor$ & 50 & 1 & 1 & 1 & 1 & 1 & 0.985 & 0.644 & 0.229 & 0.857 & 1 & 0.229 & 300.352 & 428.367 & 131.312 \\ 
50 & 5 & 70 & $\lfloor{0.1n}\rfloor$ & 10 & 1 & 0.999 & 0.997 & 1 & 0.99 & 0.918 & 0.527 & 0.162 & 0.824 & 1 & 0.162 & 4.027 & 4.447 & 3.703 \\ 
50 & 5 & 70 & $\lfloor{0.1n}\rfloor$ & 50 & 1 & 1 & 1 & 1 & 0.989 & 0.922 & 0.58 & 0.164 & 0.832 & 1 & 0.164 & 14.66 & 16.232 & 13.067 \\ 
50 & 5 & 70 & $\lfloor{0.3n}\rfloor$ & 10 & 1 & 1 & 1 & 1 & 1 & 0.936 & 0.535 & 0.226 & 0.837 & 1 & 0.226 & 30.878 & 34.783 & 19.224 \\ 
50 & 5 & 70 & $\lfloor{0.3n}\rfloor$ & 50 & 1 & 1 & 1 & 1 & 1 & 0.976 & 0.656 & 0.204 & 0.855 & 1 & 0.204 & 135.167 & 155.312 & 75.903 \\ 
50 & 5 & 70 & $\lfloor{0.5n}\rfloor$ & 10 & 0.999 & 1 & 1 & 1 & 1 & 0.927 & 0.541 & 0.293 & 0.845 & 1 & 0.293 & 67.147 & 93.092 & 33.515 \\ 
50 & 5 & 70 & $\lfloor{0.5n}\rfloor$ & 50 & 1 & 1 & 1 & 1 & 1 & 0.981 & 0.643 & 0.233 & 0.857 & 1 & 0.233 & 299.192 & 427.945 & 130.060 \\ 
50 & 10 & 10 & $\lfloor{0.1n}\rfloor$ & 10 & 1 & 1 & 1 & 1 & 0.993 & 0.911 & 0.545 & 0.15 & 0.825 & 1 & 0.150 & 6.723 & 7.645 & 5.909 \\ 
50 & 10 & 10 & $\lfloor{0.1n}\rfloor$ & 50 & 1 & 1 & 1 & 1 & 0.992 & 0.921 & 0.584 & 0.149 & 0.831 & 1 & 0.149 & 28.019 & 30.141 & 25.396 \\ 
50 & 10 & 10 & $\lfloor{0.3n}\rfloor$ & 10 & 1 & 1 & 1 & 1 & 1 & 0.936 & 0.56 & 0.228 & 0.841 & 1 & 0.228 & 56.882 & 64.804 & 33.292 \\ 
50 & 10 & 10 & $\lfloor{0.3n}\rfloor$ & 50 & 1 & 1 & 1 & 1 & 1 & 0.964 & 0.626 & 0.201 & 0.849 & 1 & 0.201 & 264.048 & 302.558 & 146.594 \\ 
50 & 10 & 10 & $\lfloor{0.5n}\rfloor$ & 10 & 1 & 1 & 1 & 1 & 0.996 & 0.907 & 0.534 & 0.292 & 0.841 & 1 & 0.292 & 125.298 & 176.993 & 57.505 \\ 
50 & 10 & 10 & $\lfloor{0.5n}\rfloor$ & 50 & 1 & 1 & 1 & 1 & 1 & 0.982 & 0.646 & 0.24 & 0.859 & 1 & 0.24 & 591.621 & 848.621 & 253.270 \\ 
50 & 10 & 40 & $\lfloor{0.1n}\rfloor$ & 10 & 1 & 1 & 0.999 & 1 & 0.992 & 0.919 & 0.55 & 0.159 & 0.827 & 1 & 0.159 & 6.754 & 7.557 & 6.089 \\ 
50 & 10 & 40 & $\lfloor{0.1n}\rfloor$ & 50 & 1 & 1 & 1 & 1 & 0.994 & 0.924 & 0.574 & 0.157 & 0.831 & 1 & 0.157 & 27.789 & 30.046 & 25.031 \\ 
50 & 10 & 40 & $\lfloor{0.3n}\rfloor$ & 10 & 1 & 1 & 1 & 1 & 1 & 0.943 & 0.574 & 0.222 & 0.842 & 1 & 0.222 & 56.830 & 64.593 & 33.236 \\ 
50 & 10 & 40 & $\lfloor{0.3n}\rfloor$ & 50 & 1 & 1 & 1 & 1 & 1 & 0.978 & 0.657 & 0.202 & 0.855 & 1 & 0.202 & 264.380 & 304.638 & 146.540 \\ 
50 & 10 & 40 & $\lfloor{0.5n}\rfloor$ & 10 & 1 & 1 & 1 & 1 & 1 & 0.915 & 0.529 & 0.289 & 0.842 & 1 & 0.289 & 125.720 & 176.901 & 58.693 \\ 
50 & 10 & 40 & $\lfloor{0.5n}\rfloor$ & 50 & 1 & 1 & 1 & 1 & 1 & 0.985 & 0.638 & 0.228 & 0.856 & 1 & 0.228 & 590.405 & 846.248 & 251.792 \\ 
50 & 10 & 70 & $\lfloor{0.1n}\rfloor$ & 10 & 1 & 1 & 1 & 1 & 0.994 & 0.923 & 0.554 & 0.153 & 0.828 & 1 & 0.153 & 6.721 & 7.487 & 5.951 \\ 
50 & 10 & 70 & $\lfloor{0.1n}\rfloor$ & 50 & 1 & 1 & 1 & 1 & 0.994 & 0.917 & 0.574 & 0.16 & 0.831 & 1 & 0.160 & 27.836 & 30.097 & 25.076 \\ 
50 & 10 & 70 & $\lfloor{0.3n}\rfloor$ & 10 & 1 & 1 & 1 & 1 & 1 & 0.955 & 0.574 & 0.213 & 0.843 & 1 & 0.213 & 56.987 & 64.592 & 33.512 \\ 
50 & 10 & 70 & $\lfloor{0.3n}\rfloor$ & 50 & 1 & 1 & 1 & 1 & 1 & 0.97 & 0.657 & 0.206 & 0.854 & 1 & 0.206 & 267.935 & 310.419 & 146.354 \\ 
50 & 10 & 70 & $\lfloor{0.5n}\rfloor$ & 10 & 1 & 1 & 1 & 1 & 1 & 0.943 & 0.555 & 0.29 & 0.849 & 1 & 0.290 & 127.842 & 177.911 & 57.568 \\ 
50 & 10 & 70 & $\lfloor{0.5n}\rfloor$ & 50 & 1 & 1 & 1 & 1 & 1 & 0.987 & 0.636 & 0.229 & 0.857 & 1 & 0.229 & 605.327 & 927.813 & 255.824 
 \\ \hline
\end{tabular}
\end{center}
\end{table}

\begin{table}[!htb]
  \setlength{\tabcolsep}{4pt} 
  \footnotesize
    \caption{Experiments to adjust parameters of \emph{MOSpecG-mod} to real networks - part 1.}
    \label{tab:apx_single_real_1}
    \begin{center}
\begin{tabular}{|c|c|c|c|c|Rc|Rc|Rc|Rc|}
\hline
\multicolumn{5}{|c|}{Parameters}    &         
 \multicolumn{2}{c|}{Karate} & \multicolumn{2}{c|}{Dolphins} & \multicolumn{2}{c|}{Polbooks} & \multicolumn{2}{c|}{Football}  \\ \hline 
 $N\mathcal{G}$ & $N\mathcal{P}$ & $N\mathcal{O}$ & $p$ & $IT$ & \multicolumn{1}{c}{NMI} & \multicolumn{1}{c|}{Time (s)}  & \multicolumn{1}{c}{NMI} & \multicolumn{1}{c|}{Time (s)} & \multicolumn{1}{c}{NMI} & \multicolumn{1}{c|}{Time (s)} & \multicolumn{1}{c}{NMI} & \multicolumn{1}{c|}{Time (s)} \\ \hline
 10 & 5  & 10 & $\lfloor{0.1n}\rfloor$ & 10 & 0.828 & 0.002 & 0.429 & 0.007 & 0.446 & 0.028 & 0.83  & 0.039 \\
10 & 5  & 10 & $\lfloor{0.1n}\rfloor$ & 50 & 0.795 & 0.005 & 0.448 & 0.022 & 0.441 & 0.073 & 0.828 & 0.074 \\
10 & 5  & 10 & $\lfloor{0.3n}\rfloor$ & 10 & 0.589 & 0.014 & 0.445 & 0.017 & 0.409 & 0.042 & 0.831 & 0.058 \\
10 & 5  & 10 & $\lfloor{0.3n}\rfloor$ & 50 & 0.55  & 0.014 & 0.44  & 0.023 & 0.413 & 0.066 & 0.834 & 0.091 \\
10 & 5  & 10 & $\lfloor{0.5n}\rfloor$ & 10 & 0.489 & 0.012 & 0.42  & 0.012 & 0.419 & 0.042 & 0.829 & 0.06  \\
10 & 5  & 10 & $\lfloor{0.5n}\rfloor$ & 50 & 0.54  & 0.012 & 0.439 & 0.027 & 0.426 & 0.081 & 0.828 & 0.124 \\
10 & 5  & 40 & $\lfloor{0.1n}\rfloor$ & 10 & 1     & 0.002 & 0.541 & 0.005 & 0.454 & 0.018 & 0.863 & 0.017 \\
10 & 5  & 40 & $\lfloor{0.1n}\rfloor$ & 50 & 1     & 0.004 & 0.57  & 0.01  & 0.477 & 0.028 & 0.866 & 0.039 \\
10 & 5  & 40 & $\lfloor{0.3n}\rfloor$ & 10 & 0.699 & 0.011 & 0.523 & 0.013 & 0.524 & 0.024 & 0.89  & 0.041 \\
10 & 5  & 40 & $\lfloor{0.3n}\rfloor$ & 50 & 0.675 & 0.013 & 0.533 & 0.029 & 0.527 & 0.056 & 0.875 & 0.079 \\
10 & 5  & 40 & $\lfloor{0.5n}\rfloor$ & 10 & 0.675 & 0.011 & 0.534 & 0.014 & 0.503 & 0.033 & 0.897 & 0.053 \\
10 & 5  & 40 & $\lfloor{0.5n}\rfloor$ & 50 & 0.699 & 0.013 & 0.573 & 0.029 & 0.536 & 0.097 & 0.892 & 0.121 \\
10 & 5  & 70 & $\lfloor{0.1n}\rfloor$ & 10 & 1     & 0.002 & 0.561 & 0.004 & 0.456 & 0.014 & 0.865 & 0.02  \\
10 & 5  & 70 & $\lfloor{0.1n}\rfloor$ & 50 & 1     & 0.004 & 0.561 & 0.01  & 0.463 & 0.026 & 0.852 & 0.031 \\
10 & 5  & 70 & $\lfloor{0.3n}\rfloor$ & 10 & 0.699 & 0.015 & 0.53  & 0.015 & 0.498 & 0.038 & 0.872 & 0.043 \\
10 & 5  & 70 & $\lfloor{0.3n}\rfloor$ & 50 & 0.699 & 0.016 & 0.529 & 0.025 & 0.524 & 0.06  & 0.891 & 0.082 \\
10 & 5  & 70 & $\lfloor{0.5n}\rfloor$ & 10 & 0.7   & 0.011 & 0.554 & 0.012 & 0.5   & 0.049 & 0.903 & 0.049 \\
10 & 5  & 70 & $\lfloor{0.5n}\rfloor$ & 50 & 0.697 & 0.015 & 0.574 & 0.037 & 0.508 & 0.09  & 0.888 & 0.124 \\
10 & 10 & 10 & $\lfloor{0.1n}\rfloor$ & 10 & 1     & 0.002 & 0.56  & 0.005 & 0.457 & 0.02  & 0.867 & 0.028 \\
10 & 10 & 10 & $\lfloor{0.1n}\rfloor$ & 50 & 1     & 0.006 & 0.561 & 0.017 & 0.477 & 0.05  & 0.866 & 0.069 \\
10 & 10 & 10 & $\lfloor{0.3n}\rfloor$ & 10 & 0.699 & 0.01  & 0.52  & 0.02  & 0.542 & 0.043 & 0.883 & 0.051 \\
10 & 10 & 10 & $\lfloor{0.3n}\rfloor$ & 50 & 0.699 & 0.017 & 0.533 & 0.043 & 0.541 & 0.086 & 0.885 & 0.128 \\
10 & 10 & 10 & $\lfloor{0.5n}\rfloor$ & 10 & 0.699 & 0.013 & 0.58  & 0.017 & 0.533 & 0.044 & 0.895 & 0.102 \\
10 & 10 & 10 & $\lfloor{0.5n}\rfloor$ & 50 & 0.699 & 0.019 & 0.588 & 0.049 & 0.524 & 0.125 & 0.892 & 0.187 \\
10 & 10 & 40 & $\lfloor{0.1n}\rfloor$ & 10 & 1     & 0.003 & 0.561 & 0.006 & 0.47  & 0.024 & 0.88  & 0.024 \\
10 & 10 & 40 & $\lfloor{0.1n}\rfloor$ & 50 & 1     & 0.007 & 0.561 & 0.022 & 0.486 & 0.049 & 0.868 & 0.074 \\
10 & 10 & 40 & $\lfloor{0.3n}\rfloor$ & 10 & 0.699 & 0.011 & 0.524 & 0.018 & 0.533 & 0.037 & 0.903 & 0.054 \\
10 & 10 & 40 & $\lfloor{0.3n}\rfloor$ & 50 & 0.699 & 0.017 & 0.533 & 0.034 & 0.527 & 0.086 & 0.884 & 0.123 \\
10 & 10 & 40 & $\lfloor{0.5n}\rfloor$ & 10 & 0.699 & 0.013 & 0.572 & 0.021 & 0.515 & 0.062 & 0.892 & 0.08  \\
10 & 10 & 40 & $\lfloor{0.5n}\rfloor$ & 50 & 0.699 & 0.018 & 0.587 & 0.051 & 0.547 & 0.135 & 0.9   & 0.181 \\
10 & 10 & 70 & $\lfloor{0.1n}\rfloor$ & 10 & 1     & 0.003 & 0.581 & 0.006 & 0.477 & 0.013 & 0.862 & 0.02  \\
10 & 10 & 70 & $\lfloor{0.1n}\rfloor$ & 50 & 1     & 0.008 & 0.57  & 0.016 & 0.473 & 0.043 & 0.87  & 0.06  \\
10 & 10 & 70 & $\lfloor{0.3n}\rfloor$ & 10 & 0.699 & 0.008 & 0.542 & 0.014 & 0.556 & 0.026 & 0.879 & 0.04  \\
10 & 10 & 70 & $\lfloor{0.3n}\rfloor$ & 50 & 0.699 & 0.008 & 0.533 & 0.034 & 0.541 & 0.084 & 0.884 & 0.125 \\
10 & 10 & 70 & $\lfloor{0.5n}\rfloor$ & 10 & 0.7   & 0.009 & 0.568 & 0.012 & 0.515 & 0.047 & 0.89  & 0.059 \\
10 & 10 & 70 & $\lfloor{0.5n}\rfloor$ & 50 & 0.699 & 0.01  & 0.587 & 0.037 & 0.524 & 0.121 & 0.887 & 0.167 \\
 \hline
 \end{tabular}
 \end{center}
\end{table}

\begin{table}[!htb]
  \setlength{\tabcolsep}{4pt} 
  \footnotesize
    \caption{Experiments to adjust parameters of \emph{MOSpecG-mod} to real networks - part 2.}
    \label{tab:apx_single_real_2}
    \begin{center}
\begin{tabular}{|l|l|l|l|l|Rc|Rc|Rc|Rc|}
\hline
\multicolumn{5}{|c|}{Parameters}    &         
 \multicolumn{2}{c|}{Karate} & \multicolumn{2}{c|}{Dolphins} & \multicolumn{2}{c|}{Polbooks} & \multicolumn{2}{c|}{Football}  \\ \hline 
 $N\mathcal{G}$ & $N\mathcal{P}$ & $N\mathcal{O}$ & $p$ & $IT$ & \multicolumn{1}{c}{NMI} & \multicolumn{1}{c|}{Time (s)}  & \multicolumn{1}{c}{NMI} & \multicolumn{1}{c|}{Time (s)} & \multicolumn{1}{c}{NMI} & \multicolumn{1}{c|}{Time (s)} & \multicolumn{1}{c}{NMI} & \multicolumn{1}{c|}{Time (s)} \\ \hline
50 & 5  & 10 & $\lfloor{0.1n}\rfloor$ & 10 & 0.806 & 0.006 & 0.428 & 0.01  & 0.442 & 0.025 & 0.846 & 0.04  \\
50 & 5  & 10 & $\lfloor{0.1n}\rfloor$ & 50 & 0.808 & 0.014 & 0.425 & 0.044 & 0.44  & 0.095 & 0.844 & 0.137 \\
50 & 5  & 10 & $\lfloor{0.3n}\rfloor$ & 10 & 0.537 & 0.019 & 0.462 & 0.039 & 0.431 & 0.061 & 0.845 & 0.106 \\
50 & 5  & 10 & $\lfloor{0.3n}\rfloor$ & 50 & 0.507 & 0.04  & 0.442 & 0.091 & 0.423 & 0.19  & 0.833 & 0.254 \\
50 & 5  & 10 & $\lfloor{0.5n}\rfloor$ & 10 & 0.59  & 0.025 & 0.426 & 0.035 & 0.411 & 0.108 & 0.844 & 0.162 \\
50 & 5  & 10 & $\lfloor{0.5n}\rfloor$ & 50 & 0.548 & 0.038 & 0.456 & 0.104 & 0.403 & 0.265 & 0.833 & 0.391 \\
50 & 5  & 40 & $\lfloor{0.1n}\rfloor$ & 10 & 1     & 0.008 & 0.581 & 0.015 & 0.47  & 0.034 & 0.855 & 0.082 \\
50 & 5  & 40 & $\lfloor{0.1n}\rfloor$ & 50 & 1     & 0.031 & 0.581 & 0.079 & 0.475 & 0.111 & 0.882 & 0.25  \\
50 & 5  & 40 & $\lfloor{0.3n}\rfloor$ & 10 & 0.699 & 0.032 & 0.52  & 0.055 & 0.529 & 0.128 & 0.873 & 0.176 \\
50 & 5  & 40 & $\lfloor{0.3n}\rfloor$ & 50 & 0.699 & 0.053 & 0.533 & 0.121 & 0.538 & 0.272 & 0.885 & 0.36  \\
50 & 5  & 40 & $\lfloor{0.5n}\rfloor$ & 10 & 0.675 & 0.022 & 0.546 & 0.046 & 0.508 & 0.185 & 0.888 & 0.216 \\
50 & 5  & 40 & $\lfloor{0.5n}\rfloor$ & 50 & 0.699 & 0.077 & 0.575 & 0.156 & 0.531 & 0.389 & 0.89  & 0.559 \\
50 & 5  & 70 & $\lfloor{0.1n}\rfloor$ & 10 & 1     & 0.009 & 0.571 & 0.029 & 0.472 & 0.052 & 0.882 & 0.084 \\
50 & 5  & 70 & $\lfloor{0.1n}\rfloor$ & 50 & 1     & 0.029 & 0.581 & 0.064 & 0.474 & 0.147 & 0.866 & 0.211 \\
50 & 5  & 70 & $\lfloor{0.3n}\rfloor$ & 10 & 0.71  & 0.016 & 0.532 & 0.052 & 0.525 & 0.137 & 0.888 & 0.157 \\
50 & 5  & 70 & $\lfloor{0.3n}\rfloor$ & 50 & 0.699 & 0.049 & 0.533 & 0.114 & 0.538 & 0.291 & 0.889 & 0.378 \\
50 & 5  & 70 & $\lfloor{0.5n}\rfloor$ & 10 & 0.698 & 0.032 & 0.57  & 0.049 & 0.504 & 0.132 & 0.89  & 0.215 \\
50 & 5  & 70 & $\lfloor{0.5n}\rfloor$ & 50 & 0.699 & 0.056 & 0.586 & 0.159 & 0.537 & 0.395 & 0.891 & 0.565 \\
50 & 10 & 10 & $\lfloor{0.1n}\rfloor$ & 10 & 1     & 0.009 & 0.581 & 0.028 & 0.464 & 0.076 & 0.885 & 0.126 \\
50 & 10 & 10 & $\lfloor{0.1n}\rfloor$ & 50 & 1     & 0.097 & 0.581 & 0.121 & 0.486 & 0.24  & 0.875 & 0.325 \\
50 & 10 & 10 & $\lfloor{0.3n}\rfloor$ & 10 & 0.699 & 0.031 & 0.536 & 0.077 & 0.555 & 0.194 & 0.891 & 0.198 \\
50 & 10 & 10 & $\lfloor{0.3n}\rfloor$ & 50 & 0.699 & 0.077 & 0.533 & 0.192 & 0.546 & 0.447 & 0.894 & 0.652 \\
50 & 10 & 10 & $\lfloor{0.5n}\rfloor$ & 10 & 0.699 & 0.033 & 0.58  & 0.078 & 0.536 & 0.264 & 0.904 & 0.284 \\
50 & 10 & 10 & $\lfloor{0.5n}\rfloor$ & 50 & 0.699 & 0.088 & 0.588 & 0.237 & 0.548 & 0.621 & 0.899 & 0.926 \\
50 & 10 & 40 & $\lfloor{0.1n}\rfloor$ & 10 & 1     & 0.011 & 0.581 & 0.036 & 0.476 & 0.095 & 0.873 & 0.128 \\
50 & 10 & 40 & $\lfloor{0.1n}\rfloor$ & 50 & 1     & 0.076 & 0.581 & 0.115 & 0.497 & 0.24  & 0.875 & 0.297 \\
50 & 10 & 40 & $\lfloor{0.3n}\rfloor$ & 10 & 0.699 & 0.034 & 0.529 & 0.07  & 0.533 & 0.17  & 0.904 & 0.239 \\
50 & 10 & 40 & $\lfloor{0.3n}\rfloor$ & 50 & 0.699 & 0.085 & 0.533 & 0.19  & 0.543 & 0.408 & 0.894 & 0.646 \\
50 & 10 & 40 & $\lfloor{0.5n}\rfloor$ & 10 & 0.699 & 0.027 & 0.58  & 0.094 & 0.52  & 0.278 & 0.893 & 0.275 \\
50 & 10 & 40 & $\lfloor{0.5n}\rfloor$ & 50 & 0.699 & 0.118 & 0.588 & 0.209 & 0.533 & 0.581 & 0.89  & 0.803 \\
50 & 10 & 70 & $\lfloor{0.1n}\rfloor$ & 10 & 1     & 0.009 & 0.581 & 0.026 & 0.479 & 0.054 & 0.879 & 0.083 \\
50 & 10 & 70 & $\lfloor{0.1n}\rfloor$ & 50 & 1     & 0.031 & 0.581 & 0.131 & 0.481 & 0.246 & 0.88  & 0.245 \\
50 & 10 & 70 & $\lfloor{0.3n}\rfloor$ & 10 & 0.699 & 0.03  & 0.532 & 0.058 & 0.531 & 0.189 & 0.878 & 0.195 \\
50 & 10 & 70 & $\lfloor{0.3n}\rfloor$ & 50 & 0.699 & 0.073 & 0.533 & 0.129 & 0.547 & 0.378 & 0.885 & 0.588 \\
50 & 10 & 70 & $\lfloor{0.5n}\rfloor$ & 10 & 0.699 & 0.028 & 0.568 & 0.06  & 0.521 & 0.293 & 0.896 & 0.218 \\
50 & 10 & 70 & $\lfloor{0.5n}\rfloor$ & 50 & 0.699 & 0.072 & 0.587 & 0.262 & 0.537 & 0.526 & 0.894 & 0.803 \\
\hline
 \end{tabular}
 \end{center}
\end{table}

%------------------------------------------ Tables - Pearson correlations

\begin{table}[!htb]
  \setlength{\tabcolsep}{4pt} 
\footnotesize
\caption{Pearson correlation coefficients between  \emph{MOSpecG-mod}  parameters and results for small-sized networks.}
\label{tab:apx:pearson_S}
\begin{center}
\begin{tabular}{|c|PPPPPPPP|PPP|}
\hline
\multicolumn{1}{|c|}{\multirow{2}{*}{Parameters}} &\multicolumn{8}{c|}{NMI value achieved over each $\mu$}  & \multicolumn{3}{c|}{Time (s)}  \\  \cline{2-12}
 & \multicolumn{1}{c}{0.1} &	\multicolumn{1}{c}{0.2} & \multicolumn{1}{c}{0.3} & \multicolumn{1}{c}{0.4} & \multicolumn{1}{c}{0.5} &	\multicolumn{1}{c}{0.6} & \multicolumn{1}{c}{0.7} &	\multicolumn{1}{c}{0.8|}  & \multicolumn{1}{c}{AVG} & \multicolumn{1}{c}{MAX} & \multicolumn{1}{c|}{MIN}\\  \hline
 $N\mathcal{G}$  &  0.004 & 0.011 & 0.001 & -0.004 & 0 & 0.006 & 0.056 & -0.016 & 0.397 & 0.376 & 0.428 \\ 
 $N\mathcal{P}$ & 0.463 & 0.456 & 0.441 & 0.435 & 0.44 & 0.438 & 0.381 & -0.109 & 0.198 & 0.188 & 0.215 \\ 
 $N\mathcal{O}$ & 0.518 & 0.522 & 0.52 & 0.51 & 0.523 & 0.531 & 0.463 & -0.157 & -0.002 & 0 & -0.009 \\ 
$p$ & -0.076 & -0.041 & 0.022 & 0.025 & 0.038 & 0.096 & 0.467 & 0.903 & 0.492 & 0.5 & 0.489 \\ 
$IT$ & -0.01 & 0.004 & 0.007 & 0.005 & 0.002 & 0.014 & 0.124 & -0.029 & 0.403 & 0.38 & 0.444 \\ 
  \hline
\end{tabular}
\end{center}
\end{table}

\begin{table}[!htb]
  \setlength{\tabcolsep}{4pt} 
\footnotesize
\caption{ Pearson correlation coefficients between  \emph{MOSpecG-mod}  parameters and results for large-sized networks.}
\label{tab:apx:pearson_L}
\begin{center}
\begin{tabular}{|c|PPPPPPPP|PPP|}
\hline
\multicolumn{1}{|c|}{\multirow{2}{*}{Parameters}} &\multicolumn{8}{c|}{NMI value achieved over each $\mu$}  & \multicolumn{3}{c|}{Time (s)}  \\  \cline{2-12}
 & \multicolumn{1}{c}{0.1} &	\multicolumn{1}{c}{0.2} & \multicolumn{1}{c}{0.3} & \multicolumn{1}{c}{0.4} & \multicolumn{1}{c}{0.5} &	\multicolumn{1}{c}{0.6} & \multicolumn{1}{c}{0.7} &	\multicolumn{1}{c}{0.8|}  & \multicolumn{1}{c}{AVG} & \multicolumn{1}{c}{MAX} & \multicolumn{1}{c|}{MIN}\\  \hline
 $N\mathcal{G}$  & 0.007 & -0.002 & 0.003 & -0.001 & 0.002 & 0.05 & 0.175 & -0.055 & 0.393 & 0.372 & 0.420 \\ 
 $N\mathcal{P}$ & 0.416 & 0.407 & 0.408 & 0.411 & 0.434 & 0.443 & 0.371 & -0.225 & 0.197 & 0.189 & 0.209 \\ 
 $N\mathcal{O}$ & 0.508 & 0.497 & 0.496 & 0.503 & 0.531 & 0.557 & 0.49 & -0.286 & 0 & 0.005 & -0.002 \\ 
$p$ & -0.145 & -0.162 & -0.155 & -0.149 & -0.083 & 0.009 & 0.197 & 0.797 & 0.492 & 0.496 & 0.478 \\ 
$IT$ & 0.003 & -0.004 & -0.003 & 0.005 & 0.005 & 0.095 & 0.34 & -0.119 & 0.398 & 0.376 & 0.428 \\ 
  \hline
\end{tabular}
\end{center}
\end{table}

\begin{table}[!htb]
  \setlength{\tabcolsep}{4pt} 
\begin{center}
  \footnotesize
  \caption{Pearson correlation coefficients between the parameters and results obtained by \emph{MOSpecG-mod} for real networks.}
  \label{tab:apx:pearson_real}
\begin{tabular}{|c|PP|PP|PP|PP|}
\hline
\multicolumn{1}{|c|}{\multirow{2}{*}{Parameters}}                   & \multicolumn{2}{c|}{Karate} & \multicolumn{2}{c|}{Dolphins} & \multicolumn{2}{c|}{Polbooks} & \multicolumn{2}{c|}{Football}  \\ \cline{2-9}
                  & \multicolumn{1}{c}{NMI} & \multicolumn{1}{c|}{Time (s)}  & \multicolumn{1}{c}{NMI} & \multicolumn{1}{c|}{Time (s)} & \multicolumn{1}{c}{NMI} & \multicolumn{1}{c|}{Time (s)} & \multicolumn{1}{c}{NMI} & \multicolumn{1}{c|}{Time (s)} \\ \hline
 $N\mathcal{G}$  & 0.034      & 0.582           & 0.154        & 0.605             & 0.150        & 0.614             & 0.211        & 0.569             \\
 $N\mathcal{P}$  & 0.186      & 0.292           & 0.499        & 0.315             & 0.476        & 0.337             & 0.507        & 0.294             \\
 $N\mathcal{O}$ & 0.181      & -0.071          & 0.414        & -0.006            & 0.276        & 0.001             & 0.313        & -0.026            \\
$p$     & -0.803     & 0.245           & 0.021        & 0.302             & 0.406        & 0.411             & 0.374        & 0.377             \\
$IT$    & -0.006     & 0.487           & 0.077        & 0.509             & 0.115        & 0.436             & -0.040       & 0.485             \\ \hline
\end{tabular}
\end{center}
\end{table}

%------------------------------------------ Tables - SpecG-EC

\begin{table}[!htb]
\begin{center}
  \setlength{\tabcolsep}{2pt} 
  \footnotesize
  \caption{Parameter-tuning experiments for \emph{SpecG-EC} with small-sized community networks.}
  \label{tab:apx:MO_S}
\begin{tabular}{|c|c|R|R|R|R|R|R|R|R|RRR|ccc|}
\hline
\multicolumn{2}{|c|}{Parameters}                   & \multicolumn{8}{c|}{$\mu$}                                     & \multicolumn{3}{c|}{NMI}     & \multicolumn{3}{c|}{Time (s)}   \\ \hline
$N\mathcal{F}$ & $\tau$ & \multicolumn{1}{c|}{0.1} &	\multicolumn{1}{c|}{0.2} & \multicolumn{1}{c}{0.3} & \multicolumn{1}{c|}{0.4} & \multicolumn{1}{c|}{0.5} &	\multicolumn{1}{c|}{0.6} & \multicolumn{1}{c|}{0.7} &	\multicolumn{1}{c|}{0.8}  & \multicolumn{1}{c}{AVG} & \multicolumn{1}{c}{MAX} & \multicolumn{1}{c|}{MIN} & \multicolumn{1}{c}{AVG} & \multicolumn{1}{c}{MAX} & \multicolumn{1}{c|}{MIN}  \\ \hline
6 & 0.1 & 0.339 & 0.296 & 0.007 & 0.222 & 0.097 & 0 & 0 & 0.001 & 0.12 & 0.339 & 0 & 3.189 & 3.637 & 2.030 \\ 
6 & 0.25 & 0.905 & 0.889 & 0.869 & 0.733 & 0.008 & 0.007 & 0 & 0 & 0.426 & 0.905 & 0 & 3.354 & 4.165 & 1.855 \\ 
6 & 0.5 & 0.968 & 0.96 & 0.954 & 0.927 & 0.833 & 0.588 & 0.011 & 0.01 & 0.656 & 0.968 & 0.01 & 3.381 & 4.068 & 2.283 \\ 
6 & 0.75 & 0.982 & 0.972 & 0.976 & 0.977 & 0.973 & 0.97 & 0.842 & 0.397 & 0.886 & 0.982 & 0.397 & 4.434 & 7.283 & 3.163 \\ 
11 & 0.1 & 0.918 & 0.916 & 0.912 & 0.9 & 0.73 & 0 & 0 & 0 & 0.547 & 0.918 & 0 & 3.825 & 4.978 & 2.999 \\ 
11 & 0.25 & 0.966 & 0.958 & 0.953 & 0.961 & 0.958 & 0.878 & 0.33 & 0.032 & 0.755 & 0.966 & 0.032 & 3.403 & 3.614 & 3.146 \\ 
11 & 0.5 & 0.98 & 0.978 & 0.974 & 0.972 & 0.965 & 0.962 & 0.822 & 0.284 & 0.867 & 0.98 & 0.284 & 4.903 & 6.938 & 4.029 \\ 
11 & 0.75 & 0.982 & 0.974 & 0.976 & 0.976 & 0.973 & 0.961 & 0.871 & 0.402 & 0.889 & 0.982 & 0.402 & 5.251 & 8.905 & 3.247 \\  \hline
\end{tabular}
\end{center}
\end{table}

\begin{table}[!htb]
\begin{center}
  \setlength{\tabcolsep}{2pt} 
  \footnotesize
  \caption{Parameter-tuning experiments for \emph{SpecG-EC} with large-sized community networks.}
  \label{tab:apx:MO_L}
\begin{tabular}{|c|c|R|R|R|R|R|R|R|R|RRR|ccc|}
\hline
\multicolumn{2}{|c|}{Parameters}                   & \multicolumn{8}{c|}{$\mu$}                                     & \multicolumn{3}{c|}{NMI}     & \multicolumn{3}{c|}{Time (s)}   \\ \hline
$N\mathcal{F}$ & $\tau$ & \multicolumn{1}{c|}{0.1} &	\multicolumn{1}{c|}{0.2} & \multicolumn{1}{c}{0.3} & \multicolumn{1}{c|}{0.4} & \multicolumn{1}{c|}{0.5} &	\multicolumn{1}{c|}{0.6} & \multicolumn{1}{c|}{0.7} &	\multicolumn{1}{c|}{0.8}  & \multicolumn{1}{c}{AVG} & \multicolumn{1}{c}{MAX} & \multicolumn{1}{c|}{MIN} & \multicolumn{1}{c}{AVG} & \multicolumn{1}{c}{MAX} & \multicolumn{1}{c|}{MIN}  \\ \hline
6 & 0.1 & 0.783 & 0.815 & 0.765 & 0.42 & 0 & 0.002 & 0 & 0 & 0.348 & 0.815 & 0 & 3.56 & 4.313 & 2.825 \\ 
6 & 0.25 & 0.947 & 0.941 & 0.953 & 0.705 & 0.005 & 0 & 0 & 0 & 0.444 & 0.953 & 0 & 3.587 & 4.446 & 2.094 \\ 
6 & 0.5 & 0.986 & 0.997 & 0.988 & 0.965 & 0.771 & 0.112 & 0.006 & 0.006 & 0.604 & 0.997 & 0.006 & 4.059 & 5.211 & 3.152 \\ 
6 & 0.75 & 0.982 & 0.994 & 0.985 & 0.989 & 0.976 & 0.903 & 0.575 & 0.196 & 0.825 & 0.994 & 0.196 & 5.379 & 9.376 & 3.570 \\ 
11 & 0.1 & 0.957 & 0.939 & 0.942 & 0.915 & 0.811 & 0 & 0 & 0 & 0.571 & 0.957 & 0 & 3.794 & 4.447 & 3.218 \\ 
11 & 0.25 & 0.989 & 0.986 & 0.991 & 0.986 & 0.963 & 0.693 & 0.106 & 0.011 & 0.716 & 0.991 & 0.011 & 2.787 & 3.276 & 2.360 \\ 
11 & 0.5 & 0.992 & 0.985 & 0.984 & 0.987 & 0.978 & 0.917 & 0.523 & 0.108 & 0.809 & 0.992 & 0.108 & 4.349 & 7.4 & 3.007 \\ 
11 & 0.75 & 0.98 & 0.99 & 0.979 & 0.991 & 0.979 & 0.931 & 0.594 & 0.201 & 0.831 & 0.991 & 0.201 & 5.54 & 9.455 & 3.908 \\  \hline
\end{tabular}
\end{center}
\end{table}

\begin{table}[!htb]
  \setlength{\tabcolsep}{4pt} 
  \footnotesize
    \caption{Parameter-tuning experiments for \emph{SpecG-EC} with real networks.}
    \label{tab:apx:MO_real}
    \begin{center}
\begin{tabular}{|c|c|Rc|Rc|Rc|Rc|}
\hline
\multicolumn{2}{|c|}{Parameters}                   & \multicolumn{2}{c|}{Karate} & \multicolumn{2}{c|}{Dolphins} & \multicolumn{2}{c|}{Polbooks} & \multicolumn{2}{c|}{Football}  \\ \hline 
$N\mathcal{F}$ & $\tau$ &  \multicolumn{1}{c}{NMI} & \multicolumn{1}{c|}{Time (s)}  & \multicolumn{1}{c}{NMI} & \multicolumn{1}{c|}{Time (s)} & \multicolumn{1}{c}{NMI} & \multicolumn{1}{c|}{Time (s)} & \multicolumn{1}{c}{NMI} & \multicolumn{1}{c|}{Time (s)} \\ \hline
6 & 0.1 & 1 & 0.012 & 0.889 & 0.028 & 0.432 & 0.079 & 0.327 & 0.115 \\ 
6 & 0.25 & 1 & 0.013 & 0.889 & 0.028 & 0.546 & 0.079 & 0.55 & 0.123 \\ 
6 & 0.5 & 1 & 0.013 & 0.889 & 0.035 & 0.552 & 0.084 & 0.748 & 0.135 \\ 
6 & 0.75 & 1 & 0.015 & 0.581 & 0.028 & 0.536 & 0.097 & 0.877 & 0.13 \\ 
11 & 0.1 & 1 & 0.013 & 0.889 & 0.026 & 0.569 & 0.077 & 0.564 & 0.116 \\ 
11 & 0.25 & 1 & 0.011 & 0.889 & 0.028 & 0.569 & 0.079 & 0.717 & 0.121 \\ 
11 & 0.5 & 1 & 0.014 & 0.889 & 0.029 & 0.561 & 0.119 & 0.883 & 0.161 \\ 
11 & 0.75 & 1 & 0.011 & 0.581 & 0.027 & 0.554 & 0.074 & 0.872 & 0.116 \\ \hline
\end{tabular}
\end{center}
\end{table}



% The paper BibTeX style support page is at:
% http://www.michaelshell.org/tex/paper/bibtex/
%\bibliographystyle{paper}

\end{document}